\documentclass[a4paper, amsfonts, amssymb, amsmath, reprint, showkeys, nofootinbib, twoside]{revtex4-1}

\def\mnras{{MNRAS}}             
\def\prc{{Phys.~Rev.~C}}        
\def\prd{{Phys.~Rev.~D}}        
\def\prl{{Phys.~Rev.~Lett.}}    
\usepackage{float}
\usepackage{amsmath,amstext}
\usepackage[T1]{fontenc}
\usepackage{amssymb}
\input{epsf}
\usepackage{graphicx}
\usepackage{ae,aecompl}
\usepackage{hyperref}
\usepackage{amsmath}
\usepackage{amssymb}
\usepackage{mathtools}
\usepackage{bm}
\usepackage{cleveref}
\usepackage{tensor}
\usepackage{braket}
\usepackage{enumitem}
\usepackage{color}

\newcommand{\spc}{\quad \quad \quad}

\def\be{\begin{equation}}
\def\ee{\end{equation}}
\def\beq{\begin{eqnarray}}
\def\eeq{\end{eqnarray}}

\begin{document}


\title{Symmetric-hyperbolic quasi-hydrodynamics}

\author{L.~Gavassino$^1$, M.~Antonelli$^{2,1}$, B.~Haskell$^1$}

\affiliation{$^1$Nicolaus Copernicus Astronomical Center, Polish Academy of Sciences, Warszawa, Poland
\\
$^2$CNRS/in2p3, Laboratoire de Physique Corpusculaire de Caen, 14050 Caen, France
}

\begin{abstract}
We set up a general framework for systematically building and classifying, in the linear regime, causal and stable dissipative hydrodynamic theories that, alongside with the usual hydrodynamic modes, also allow for an  arbitrary number of non-hydrodynamic modes with complex dispersion relation (such theories are often referred to as ``quasi-hydrodynamic'').
To increase the number of non-hydrodynamic modes one needs to add more effective fields to the model. The system of equations governing this class of quasi-hydrodynamic theories is symmetric hyperbolic, thermodynamically consistent (i.e. the entropy is a Lyapunov function) and can be derived from an action principle. 
As a first application of the formalism, we prove that, in the linear regime, the Israel-Stewart theory in the Eckart frame and the Israel-Stewart theory in the Landau frame are exactly the same theory. In addition, with an Onsager-Casimir analysis, we show that in strongly coupled plasmas the non-equilibrium degrees of freedom typically appear in pairs, whose members acquire opposite phase under time reversal. We use this insight to modify Cattaneo's model for diffusion, in a way to make its initial transient consistent with the transient dynamics of holographic plasmas. 
\end{abstract} 

\maketitle

\section{Introduction} 

When the field equations of a relativistic hydrodynamic theory are linearised around a homogeneous static state of global thermodynamic equilibrium, we obtain a system of linear partial differential equations with constant coefficients. If we work in the Fourier space ($t,x \rightarrow \omega,k$), this system becomes algebraic, and its solutions are a set of dispersion relations $\omega_n = \omega_n (k)$, describing the 
\textit{modes} of the theory. Such dispersion relations can then be compared with those computed from statistical mechanics by means of the linear response theory, or other microscopic approaches that allow to derive the spectrum of the system's collective excitations \citep{Kadanoff1963,peliti_book,Kubos2017}.

The dispersion relations can be divided into two broad classes: the modes such that $\omega_n(0)=0$, which constitute the \textit{hydrodynamic sector}, and the modes such that $\omega_n(0)\neq 0$, which constitute the \textit{non-hydrodynamic sector}. It is found that the structure of the modes belonging to the hydrodynamic sector is essentially universal (at least for small $k$) across most fluids and hydrodynamic theories, since they happen to be governed by some Navier-Stokes-type dynamics \cite{Geroch1995,LindblomRelaxation1996,Policastro2002,kovtun_lectures_2012,Glorioso2018}. 
On the other hand, the structure of modes in the non-hydrodynamic sector is different for different fluids and for different hydrodynamic theories, even at small $k$ \cite{Denicol_Relaxation_2011,BulkGavassino,GavassinoFronntiers2021}.

For example, let's consider the Israel-Stewart theory \cite{Hishcock1983}, which approximately describes the non-hydrodynamic sector of ideal relativistic gases \cite{Israel_Stewart_1979,DMNR2012,Chab2021}. Building on the ideas of Cattaneo \cite{cattaneo1958}, the Israel-Stewart theory posits that the shear stress $\Pi_{ab}$ evolves (in the fluid's rest frame, for $k=0$) according to the equation 
\begin{equation}\label{MIUSZ}
\tau \, \partial_t \Pi_{ab} + \Pi_{ab}=0 \, .
\end{equation}
This gives rise to a non-hydrodynamic mode with purely imaginary frequency: $\omega(k=0) = -i/\tau$. 

Now, let's compare the Israel-Stewart theory with the HJSW theory \cite{Heller2014}, which has been constructed to reproduce the dynamics of strongly coupled plasmas \cite{KovtunHolography2005}.
Building on the holographic AdS/CFT description of $\mathcal{N}=4$ super-symmetric Yang-Mills theory, the HJSW theory posits that the viscous stress evolves according to the equation below:
\begin{equation}\label{nonhydroadscft}
  \dfrac{1}{2} \, \chi  \,\tau^2 \, \partial_t^2 \Pi_{ab} + \tau \, \partial_t \Pi_{ab} + \Pi_{ab}=0 \, .
\end{equation}
This equation gives rise to two non-hydrodynamic modes, with frequencies
\begin{equation}\label{twofrequncies}
\omega_{\pm}(k=0) = \dfrac{-i \pm \sqrt{\, 2 \, \chi-1}}{\chi \, \tau} \, .
\end{equation}
For $\chi>1/2$ (in $\mathcal{N}=4$ SYM one finds $1 \lesssim \chi \lesssim 3$ \cite{KovtunHolography2005}), these frequencies have also a real part. This implies that the Israel-Stewart theory \textit{cannot} reproduce them!

This raises the following question: given an arbitrary finite set of dispersion relations\footnote{
    We consider dispersion relations arising from simple poles of the linear-response Green function, leaving out the possibility of having modes related to branch cuts. We briefly comment on this at the end of Sec. \ref{concussioni}.
}
$\omega_n (k)$, is there a systematic technique for building an effective hydrodynamic theory which correctly reproduces both the hydrodynamic and the non-hydrodynamic sector, at least in the small $k$ limit? In fact, given the universality of the hydrodynamic sector, we already know that it should arise from some effective Navier-Stokes kind of dynamics, that is the expected limit of the equations of motion in the low frequency regime (i.e., for small Knudsen number). 
However, if we want to also reproduce some non-hydrodynamic modes, there in no universal answer like Navier-Stokes hydrodynamics \cite{BulkGavassino}. 

Following \citet{Heller2014}, the simplest way of including additional non-hydrodynamic modes in a fluid model is to add higher-order derivatives in time, as is done in going from \eqref{MIUSZ} to \eqref{nonhydroadscft}, fixing the prefactors in a way to generate the desired dispersion relations. The problem is that, by adding higher derivatives by hand, we completely change the mathematical structure of the field equations, and it becomes hard to predict in advance whether the resulting theory will be
\begin{itemize}
\item[(i) -] hyperbolic and causal,
\item[(ii) -] stable,
\item[(iii) -] consistent with the second law of thermodynamics.
\end{itemize}
Of course, if the equations are simple enough (as in the case of the HJSW theory) we can still try to make some adjustment afterwards, ``forcing the theory to work''. But if one aims to include several dissipative phenomena and non-hydrodynamic frequencies, the number of possible couplings may become very large, and fixing everything by hand seems to be a formidable task. We definitely need a more systematic approach, which is what this paper provides.


Here, we develop a construction technique to build symmetric-hyperbolic, causal, and Lyapunov-stable fluid theories in the linear regime, which can reproduce an (almost) arbitrarily assigned set of dispersion relations $\omega_n(k)$, in the limit of small $k$. All these linear theories are consistent with the Onsager-Casimir principle \cite{Onsager_1931,Onsager_Casimir}, the Gibbs stability criterion \cite{GavassinoGibbs2021}, and the principles of Unified Extended Irreversible Thermodynamics (UEIT) described in \cite{GavassinoFronntiers2021}.

Throughout the paper we adopt the signature $(-,+,+,+)$ and work with natural units $c=k_B=\hbar=1$. We follow the same notation of \citet{Geroch_Lindblom_1991_causal}: $a,b,c,m$ are space-time indices, while $A,B,C,D$ are field multi-indices, defined below. The indices $j,k,l$ are pure space indices running from 1 to 3. When two indices are symmetrised, e.g. $\Xi_{(AB)}$, or anti-symmetrised, e.g. $ \Xi_{[AB]}$, we adopt the prefactor $1/2$.

\section{Geroch-Lindblom theories}\label{sec1}

To address the main question outlined in the introduction, we need to identify an appropriate set of fluid theories to work with. It needs to be large enough to accommodate an arbitrary number   of both hydrodynamic and gapped modes $\omega_n(k)$, randomly scattered on the complex plane. 
On the other hand, the mathematical structure of all these theories should be ``elegant enough'', so that we can identify some simple criteria for hyperbolicity, causality and stability, which are valid for the whole set at once. 
In a seminal paper, \citet{Geroch_Lindblom_1991_causal} identified a space of theories that is just right for our aims. In this section, we briefly review their approach.

\subsection{Assumptions and regime of validity of the construction}

Before moving to outline the mathematical details of the Geroch-Lindblom theories, it is important to comment on the regime of validity of our method to construct the hydrodynamic equations from the dispersion relations $\omega_n(k)$.  


We assume that the spatial gradients, as measured in the global rest frame of the fluid, are small (i.e., we work at small Knudsen number). 
This also implies that each fluid element, whose size has to be smaller than the length scale of the gradients, can be taken to be big enough that fluctuations of its average properties are practically negligible. 
Therefore, we will deal only with non-stochastic hydrodynamic models. 

As a consequence of these assumptions, we will be content to correctly reproduce only the first orders of an expansion of the dispersion relations for $k \rightarrow 0$. 
More specifically, if $\omega_n(k)$ belongs to the hydrodynamic sector, then the theory should give the correct predictions up to second order in $k$,
\begin{equation}
\omega_n = \omega_n'(0) \, k + \dfrac{1}{2} \omega_n''(0) \, k^2 + \mathcal{O}(k^3) \, ,
\end{equation}
because the expansion coefficients $\omega_n'(0)$ and $\omega_n''(0)$ carry crucial information about the transport properties of the fluid \cite{LindblomRelaxation1996}. 

On the other hand, if $\omega_n(k)$ is a non-hydrodynamic mode, it is sufficient that the model correctly reproduces the value of the gap $\omega_n(0)$, as this contains most of the relevant information about the initial transient evolution \cite{Denicol_Relaxation_2011,Heller2014}.

Finally, we make an important remark about the physical meaning of the non-hydrodynamic sector. 
The theories that we will construct are consistent with the principles of (Unified) Extended Irreversible Thermodynamics \cite{Jou_Extended,GavassinoFronntiers2021}, according to which the non-hydrodynamic modes describe the evolution of some additional non-equilibrium thermodynamic variables, which are \textit{degrees of freedom} of their own right\footnote{
    Adopting the terminology introduced by \citet{Grozdanov2019}, we may say that the theories considered in this paper are ``quasi-hydrodynamic'', because they treat weakly non-conserved quantities on the same footing as exactly conserved quantities.
    }. 
    
An important example of a non-hydrodynamic mode of this kind is given by chemical reactions \cite{BulkGavassino}: a reaction in a fluid that is prepared out of chemical equilibrium can give rise to a thermodynamic relaxation of the system, which survives in the homogeneous limit (hence $\omega_n(0)\neq 0$), and which involves a measurable change of a physical observable (the reaction coordinate, or the chemical fractions), see e.g., Fig.~1 in \citep{camelio2022arXiv}.

It is, therefore, important to keep in mind that the terms proportional to $\partial_t\Pi_{ab}$ and $\partial_t^2\Pi_{ab}$ in equations \eqref{MIUSZ} and \eqref{nonhydroadscft} are \textit{not} subsequent terms in a derivative expansion close to $\omega =0$ (such a construction is probably meaningless above first order in 3+1 dimensions \cite{ErnstLetters1970,Deschepper1974,KovtunStickiness2011}), but they model the effect of singularities of the retarded correlators in the complex frequency plane \cite{Denicol_Relaxation_2011,Heller2014,Grozdanov2019}.

\subsection{The mathematical construction}\label{Assumpio}

We assume that the macroscopic state of the fluid can be completely characterized by finite set of fields $\varphi^A$, possibly subject to some algebraic constraints\footnote{
    For example, if the fluid's four-velocity $u^a$ is included in the set $\varphi^A$, it will be subject to the algebraic constraint $u^b u_b=-1$. 
    }, see also the more general discussion in \citep{GavassinoFronntiers2021}.
The number and physical meaning of the hydrodynamic fields $\varphi^A$ depends on the particular substance one may want to model and needs to be fixed depending on the structure of the non-hydrodynamic sector that should be reproduced. 
The label $A$ is a multi-index, which contains both abstract indices labelling each tensor field, as well as the space-time indices pertaining to each tensor, if any. 


We now have to choose a class of hydrodynamic equations for the $\varphi^A$ that is general enough for our scope: we assume that the field equations have the form proposed by \citet{Geroch_Lindblom_1991_causal},
\begin{equation}\label{fieldEquations}
M\indices{^m _A _B}\nabla_m \varphi^B = -\Xi_A \, ,
\end{equation}
where the coefficients $M\indices{^m _A _B}$ and $\Xi_A$ are algebraic functions of the fields $\varphi^A$ and of the metric $g_{ab}$ (``algebraic'' means that they do not depend on the derivatives of the fields, but only on their local value). 
We are applying Einstein's summation convention to the multi-index $B$. Since $B$ runs over the whole set of field components (that are possibly subject to constraints). It seems reasonable to restrict the dimension of the  space labeled by the multi-index $A$, or $B$, to be
equal to the number of independent observable fields in the theory. However, this restriction is not required in the following: if the number of fields $\varphi^A$ were taken to be larger than the number of independent fields, then some of the equations in \eqref{fieldEquations} can be used to implement algebraic constraints. A practical example is given in Appendix \ref{LaShear}.

%

How restrictive is equation \eqref{fieldEquations}? The fact that the system of equations is of first order does not constitute a very restrictive assumption, because higher-order systems of equations can always be written as first-order systems involving more independent fields. 

Regarding the generality of the system in \eqref{fieldEquations}, there is also a more important point to make. Given that our goal is to model also the non-hydrodynamic sector of the fluid, or at least a part of it, the derivatives in time are in general not small (by definition, a non-hydrodynamic mode is a mode whose frequency remains finite in the small $k$ limit). 
This implies that the often repeated statement that hydrodynamics arises from a derivative expansion, where to increase accuracy one increases the order, is not applicable in our case\footnote{
    One may argue that, since the gradients in space are assumed small ($k \rightarrow 0$), it should still be possible to perform a gradient expansion only in space (i.e., a small Knudsen number expansion). However, in relativity, equations whose order in space is higher than the order in time are usually pathological when boosted and do not give rise to well-posed initial value problems \citep{Kost2000,GavassinoLyapunov_2020,GavassinoFronntiers2021}. Hence the truncation in space needs to stop at the first order, leading to \eqref{fieldEquations}.} 
\citep{Grozdanov2019}.
Instead, in the present approach, each mode represents a \textit{physical} degree of freedom of the fluid, and the value of all the fields $\varphi^A$ at a given time $t$ completely defines the macro-state at $t$. Then, equation \eqref{fieldEquations} follows naturally. In principle, one could also find higher-degree terms, like $\nabla_m \varphi^A \nabla^m \varphi^B$. However, in practically all known situations one can redefine the fields and rearrange the equations in a way to recover \eqref{fieldEquations}, within a certain level of approximation. 
Furthermore, given that we will focus on linear perturbations to a homogeneous background, any term of the form $\nabla_m \varphi^A \nabla^m \varphi^B$ would anyway disappear.

The set of possible theories identified by equation \eqref{fieldEquations} is still a bit too large, hence we introduce two additional assumptions:
\begin{itemize}
\item The system \eqref{fieldEquations} is \textit{symmetric}, namely
\begin{equation}\label{symmetry}
M\indices{^m _A _B} = M\indices{^m _B _A} \, .
\end{equation}
\item The quantity $\varphi^A \Xi_A$ is strictly non-negative,
\begin{equation}
\label{dissipo}
\sigma := \varphi^A \Xi_A \geq 0 \, ,
\end{equation}
and it can be identified with the entropy production rate.
\end{itemize}
Condition \eqref{symmetry} may seem quite restrictive; however, note that we can always multiply both sides of \eqref{fieldEquations} by an arbitrary invertible matrix $\mathcal{N}\indices{^A _C}$, so that the coefficients $M\indices{^m _A _B}$ and $\Xi_A$ are redefined as follows: 
\begin{equation}
\begin{split}
 M\indices{^m _A _B} & \, \longrightarrow \, \mathcal{N}\indices{^A _C} M\indices{^m _A _B} \\
 \Xi_A & \, \longrightarrow \, \mathcal{N}\indices{^A _C} \Xi_A \, . \\
\end{split}
\end{equation}
This is implies that, in practice, many theories can be recast in a way to obey \eqref{symmetry}, at least in the linear limit (as it happens with the Israel-Stewart theory \citep{Hishcock1983}). As we shall see, equation \eqref{symmetry} allows us to easily guarantee, on general grounds, the well-posedness of the initial value problem.

Condition \eqref{dissipo} breaks the symmetry under time-reversal and gives to our system of equations a dissipative character. The identification of $\sigma$ with the entropy production rate will allow us to build a bridge with statistical mechanics.

\subsection{Hyperbolicity, causality and stability}\label{Restringo}

Now that we have identified a suitable space of theories, defined by the system in \eqref{fieldEquations}, we need to impose conditions (i,ii), as given in the introduction. The entropy production in \eqref{dissipo} takes care of condition (iii) automatically.

To take care of condition (i), we follow \citet{Geroch_Lindblom_1991_causal}: the symmetric field equations \eqref{fieldEquations} give rise to a \textit{hyperbolic} and \textit{causal} system if 
\begin{equation}\label{timelikefuture}
M\indices{^m _A _B} Z^A Z^B \, \text{ is timelike future directed}
\end{equation}
for all non-vanishing $Z^A$. Symmetric-hyperbolicity guarantees that the field equations give rise to a well-posed initial value problem, while causality guarantees that information does not propagate faster than light.

Let us move to condition (ii). By ``stability'' we mean that perturbations away from the state of global thermodynamic equilibrium can only decay and never grow. We will focus, here, on homogeneous equilibria in Minkowski space-time. If we linearise the field equations (calling $\varphi^A$ the uniform equilibrium state and $\varphi^A +\delta \varphi^A$ the perturbed state) we find
\begin{equation}\label{linearField}
M\indices{^m _A _B}\nabla_m \delta \varphi^B = -\Xi_{AB} \delta \varphi^B \quad \quad \Xi_{AB} = \dfrac{\partial \Xi_A}{\partial \varphi^B} \, .
\end{equation}
On the right-hand side, we have expanded $\Xi_A$ to the first-order in $\delta \varphi^B$ and we have used the fact that $\delta \varphi^A=0$ must be a solution of the field equations (it is the equilibrium state) to cancel the equilibrium value of $\Xi_A$. Note that the matrix $\Xi_{AB}$ is not necessarily symmetric in $A$ and $B$. Now, recalling that $M\indices{^m _A _B}$ is uniform across the spacetime, we can contract both sides of \eqref{linearField} with $\delta \varphi^A$, to obtain
\begin{equation}\label{fourdivE}
\nabla_m E^m = -\sigma
\end{equation} 
with
\begin{equation}\label{Clavicembalo}
\begin{split}
E^m & = \dfrac{1}{2} M\indices{^m _A _B} \delta \varphi^A \delta \varphi^B \\
\sigma & = \Xi_{AB} \delta \varphi^A \delta \varphi^B \geq 0 \, .\\
\end{split}
\end{equation}
Let us discuss the properties of the vector field $E^m$ defined above:
\begin{itemize}
\item From condition \eqref{timelikefuture} we know that $E^m$ is timelike future directed:
\begin{equation}\label{bramante}
E^m E_m \leq 0  \spc E^0 \geq 0 \, .
\end{equation}
\item Condition \eqref{timelikefuture} also implies that $E^m=0$ if and only if $\delta \varphi^A=0$ $\forall \, A$. 
\item From condition \eqref{dissipo}, we know that
\begin{equation}
\nabla_m E^m \leq 0 \, .
\end{equation}
\end{itemize}
All these condition together imply that we can associate to the perturbation $\delta\varphi^A$ a quadratic norm $E$ which is non-increasing in time. In fact, given an arbitrary space-like Cauchy 3D-surface $\Sigma$, the integral
\begin{equation}\label{IntroducoE}
E[\Sigma]= \int_\Sigma E^m d\Sigma_m  \quad  (\text{orientation: } d\Sigma_0 >0)
\end{equation} 
is quadratic in $\delta\varphi^A$, positive definite and vanishes only at equilibrium. 
Furthermore, given two space-like Cauchy 3D-surfaces $\Sigma_i$ (``initial'') and $\Sigma_f$ (``final''), such that $\Sigma_f$ is future to $\Sigma_i$, we have, by Gauss theorem ($\Omega$ is the volume between the two surfaces),
\begin{equation}\label{esigma}
E[\Sigma_f]-E[\Sigma_i] = \int_\Omega \nabla_m E^m d\Omega \leq 0 \, .
\end{equation}
Hence, the norm $E$ cannot increase with time, and the equilibrium state is Lyapunov stable.

In conclusion, as long as the field equations \eqref{fieldEquations} satisfy \eqref{symmetry}, \eqref{dissipo} and \eqref{timelikefuture}, and admit a homogeneous equilibrium state, the conditions (i,ii,iii) are automatically respected.

\section{Hydrodynamic and non-hydrodynamic modes}\label{sec2}

Let us imagine to randomly pick up a theory which respects all the conditions reported in subsections \ref{Assumpio} and \ref{Restringo}. What can we say about its Fourier modes? 
Thanks to the stability requirement, the imaginary part of their frequency must be non-positive,
\begin{equation}\label{stabuz}
\text{Im} \, \omega_n(k) \leq 0 \spc \forall \, k \, ,
\end{equation} 
but there is much more that we can say about $\omega_n(k)$. 

\subsection{General properties of the hydrodynamic sector: the relaxation effect}\label{takeiteasy}

Assume to prepare the fluid in an arbitrary initial state and to let it evolve for long times. If the spatial gradients are not too large, we expect that, after some time, all the non-hydrodynamic modes will have decayed and the fluid will exhibit the universal ``Navier-Stokes-type'' behaviour, in which the stress-energy tensor can be approximated using a gradient expansion around the perfect-fluid structure. If the model does not manifest such behaviour, this is a clear signal that something is wrong. Luckily, \citet{Geroch1995} and \citet{LindblomRelaxation1996} showed (in the full non-linear regime) that field equations of the form \eqref{fieldEquations}, satisfying the causality and stability requirement reported above, are always subject to a \textit{relaxation effect}, which makes them eventually indistinguishable from Navier-Stokes-like fluids as $t \rightarrow +\infty$. In other words, for this class of theories, the causality and stability requirements are enough to guarantee that the fields have the tendency to ``arrange themselves'' (after an initial transient) in a way to mimic first-order stable theories when gradients are small. The immediate consequence is that the hydrodynamic sector of these theories is always ``realistic''.

\subsection{General properties of the non-hydrodynamic sector}\label{GPOTNHS}

Let us focus on the non-hydrodynamic sector. We work in the fluid's rest frame and assume invariance under spatial translations ($\partial_j =0$), so that the field equations \eqref{linearField} reduce to
\begin{equation}\label{nonhydruz}
M\indices{^0 _A _B}\partial_t \delta \varphi^B = -\Xi_{AB} \delta \varphi^B \, .
\end{equation}
These constitute a system of $\mathfrak{D}$ ordinary differential equations for $\mathfrak{D}$ functions of time, where $\mathfrak{D}$ is the number of algebraically independent components of the fields. The non-trivial solutions of \eqref{nonhydruz} are the non-hydrodynamic modes with $k=0$.

We know from the causality condition \eqref{timelikefuture} that
\begin{equation}
M\indices{^m _A _B} Z^A Z^B (\partial_t)_m  < 0
\end{equation}
for all non-vanishing $Z^A$. It follows that the matrix $M\indices{^0 _A _B}$ is positive definite. Combining this result with the symmetry condition \eqref{symmetry}, we can conclude that there is an invertible matrix $\mathcal{N}\indices{^C _A}$ such that
\begin{equation}\label{silvestruz}
M\indices{^0 _A _B} = \delta_{CD} \, \mathcal{N}\indices{^C _A} \, \mathcal{N}\indices{^D _B} \, ,
\end{equation}
where $\delta_{CD}$ is the Kronecker symbol. This allows us to rewrite the system \eqref{nonhydruz} in the simpler form
\begin{equation}\label{semplicio}
\partial_t \delta \tilde{\varphi}^C = -\tilde{\Xi}_{CD} \delta \tilde{\varphi}^D \, ,
\end{equation}
with
\begin{equation}\label{pphhiiD}
\begin{split}
& \Xi_{AB} = \tilde{\Xi}_{CD} \, \mathcal{N}\indices{^C _A} \, \mathcal{N}\indices{^D _B}  \\
& \delta \tilde{\varphi}^D = \mathcal{N}\indices{^D _B} \delta \varphi^B \, . \\
\end{split}
\end{equation}
Note that the existence of $\tilde{\Xi}_{CD}$ is guaranteed by the fact that $\mathcal{N}\indices{^C _A}$ is invertible.

Finally, let us assume that the matrix $\tilde{\Xi}_{CD}$ is diagonalizable (the non-diagonalizable case is discussed in appendix \ref{nondiag}). Then, there is a basis of $\mathfrak{D}$ (in principle complex) eigenvectors $Y_{(n)}^D$ satisfying the eigenvalue equation
\begin{equation}\label{eigenvalue}
\tilde{\Xi}_{CD}Y_{(n)}^D  = i\omega_n \, Y_{(n)}^C \, ,
\end{equation} 
and the general solution of the field equations \eqref{semplicio} is
\begin{equation}\label{twentitree}
\delta \tilde{\varphi}^C(t) = \sum_{n=1}^{\mathfrak{D}} c_n(t) Y_{(n)}^C \, ,  
\end{equation}
with
\begin{equation}\label{twantialbero}
c_n(t) =e^{-i\omega_n t} c_n(0) \, .
\end{equation}
The complex numbers $\omega_n$ are simply the frequencies of the homogeneous modes; namely the dispersion relations $\omega_n(k)$ presented in the introduction, evaluated at $k=0$. This shows us that the mathematical structure of the non-hydrodynamic sector depends on the properties of the matrix $\tilde{\Xi}_{CD}$. For example, the dimension of the kernel of $\tilde{\Xi}_{CD}$ is the number of hydrodynamic modes, whereas the dimension of its image is the number of non-hydrodynamic modes. Note also that, if $\omega_n =0$, then $c_n$ is a conserved quantity. Assuming that the conservation of $c_n$ ``is not a coincidence'', we can expect it to reflect the existence of a more fundamental conservation law, which leads us to the intuitive rule
\begin{equation}\label{LaRegolaDoro}
\mathfrak{D} = \binom{\text{ Number of }}{\text{conservation laws}} + \binom{\text{ Number of }}{\text{non-hydro modes}} \, .
\end{equation} 
In essence, this rule tells us that every conservation law contributes with one degree of freedom and produces a hydrodynamic mode, whereas non-conserved degrees of freedom give rise to non-hydrodynamic modes \citep{GavassinoFronntiers2021}.

\subsection{The structure of the non-hydrodynamic sector}\label{LLAA}

We are now ready to prove our main result. Assume that the matrix $\tilde{\Xi}_{CD}$ (and hence also the matrix $\Xi_{AB}$) is symmetric. Then, from the spectral theorem, we know that $\tilde{\Xi}_{CD}$ is diagonalizable and $i\omega_n \in \mathbb{R}$. Hence, recalling \eqref{stabuz}, we have the following theorem:

\textit{Theorem:} given the dynamics defined by \eqref{nonhydruz}, if 
\begin{equation}\label{xiABBA}
\Xi_{AB}= \Xi_{BA} \, ,
\end{equation} 
then the frequencies of the non-hydrodynamic modes at zero $k$ have the form
\begin{equation}\label{QuandoTiRilassi}
\omega_n = -\dfrac{i}{\tau_n} \spc \tau_n >0 \, .  
\end{equation}

\begin{figure}
\begin{center}
\includegraphics[width=0.5\textwidth]{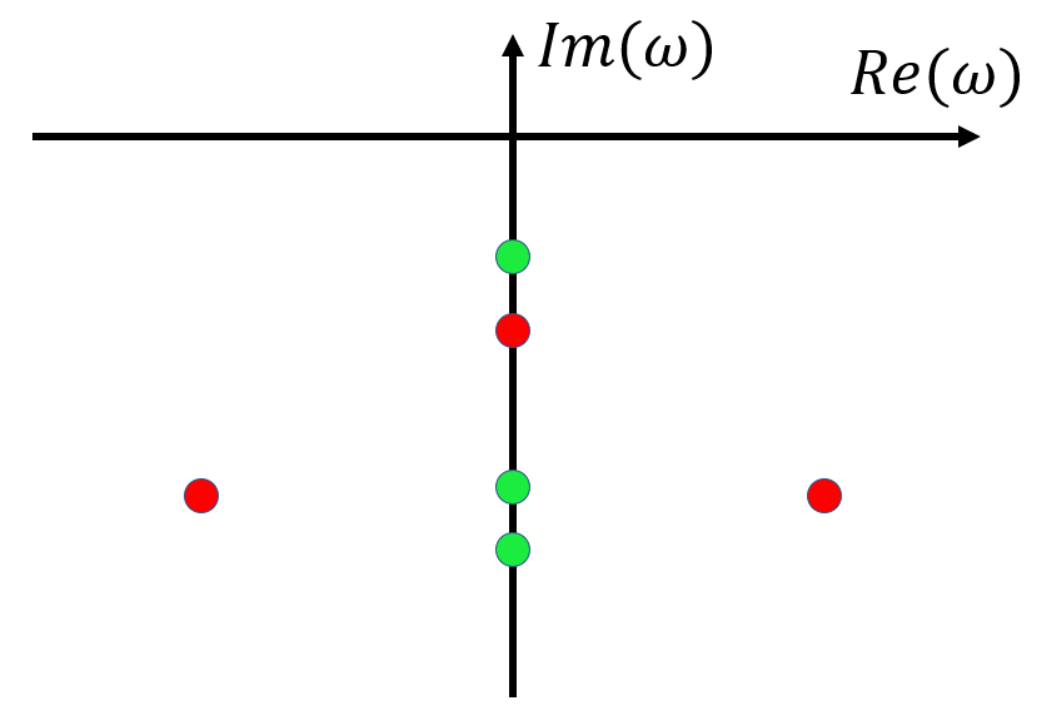}
	\caption{Geometrical disposition of the frequencies of three non-hydrodynamic modes on the complex plane. The green dots represent the modes of a fluid theory with symmetric $\Xi_{AB}$: all the frequencies sit on the imaginary axis. The red dots represent the modes of a fluid theory with $\Xi_{AB}\neq \Xi_{BA}$: the frequencies may have a real part. In both cases, the reality of $\Xi_{AB}$ guarantees perfect symmetry under inversion of the real axis. This implies, for example, that if the number of modes is odd, at least one of the frequencies must sit on the imaginary axis.}
	\label{fig:fig}
	\end{center}
\end{figure}

This theorem implies that, if we do not break the symmetry of $\Xi_{AB}$, we cannot model fluids whose non-hydrodynamic frequencies have a real part for $k=0$. Hence, if we want to go beyond Israel-Stewart-type theories, and apply the methods of Extended Irreversible Thermodynamics also to holographic plasmas, we need to include the possibility that
\begin{equation}
\Xi_{AB} \neq \Xi_{BA} \, .
\end{equation}
If we drop the assumption \eqref{xiABBA}, then $i\omega_n$ will be, in general, complex. 
Note also that, if $i\omega_n$ obeys the eigenvalue equation \eqref{eigenvalue}, then its complex conjugate, $(i\omega_n)^*$, is also eigenvalue of $\tilde{\Xi}_{CD}$, with eigenvector $(Y^D_{(n)})^*$ (to see this, just take the complex conjugate of \eqref{eigenvalue} and recall that $\tilde{\Xi}_{CD} \in \mathbb{R}$). This implies that those $\omega_n$ which are not on the imaginary axis always organise themselves into couples, whose members have the same imaginary part and opposite real part, see figure \ref{fig:fig}.

It is instructive to consider a concrete example. Assume that we want to model the internal dynamics of the shear stress $\Pi_{ab}$ of strongly coupled plasmas, under the assumption that it obeys the evolution equation \eqref{nonhydroadscft}, with $\chi \neq 1/2$. 
From equation \eqref{twofrequncies}, we know that the non-hydrodynamic sector consists of two modes for each independent component of the shear stress. Hence, by rule \eqref{LaRegolaDoro}, we can conclude that we need to double the number of viscous degrees of freedom with respect to the Israel-Stewart theory, if we want to have a double number of non-hydrodynamic modes. We are, therefore, led to postulate that
\begin{equation}
\delta \tilde{\varphi}^D = (\Pi_{ab},\Lambda_{ab}) \, ,
\end{equation}
where $\Lambda_{ab}$ is an additional field which has the same geometric properties of $\Pi_{ab}$ (it is symmetric, transverse and traceless); in this way we guarantee that they have the same number of independent components. The new variable $\Lambda_{ab}$ is just an effective field: a hydrodynamic degree of freedom which is used to parameterize the physical states of the fluid. It does not need to have any deep physical meaning, unless this is provided by microphysics. Note that, since the equilibrium state is isotropic (in the rest frame), both $\Pi_{ab}$ and $\Lambda_{ab}$ vanish at equilibrium. Hence, we dropped the symbol ``$\, \delta \,$'' for convenience.

Given that the tensor $\tilde{\Xi}_{CD}$ is evaluated at equilibrium, it must be isotropic, so that the most general field equation of the form \eqref{semplicio} is
\begin{equation}\label{chemicstab}
\partial_t \begin{pmatrix}
\Pi_{ab}    \\
\Lambda_{ab} 
\end{pmatrix} 
= -
\begin{bmatrix}
   \gamma & \gamma_I +a \\
   \gamma_I-a & \gamma'  
\end{bmatrix} 
\begin{pmatrix}
\Pi_{ab}    \\
\Lambda_{ab}  
\end{pmatrix}
\end{equation}
The $\gamma$'s represent the symmetric part of $\tilde{\Xi}_{CD}$, while $a$ is a skew-symmetric correction. The second law ($\sigma \geq 0$) produces the stability conditions
\begin{equation}
\gamma \geq 0  \quad \quad \gamma' \geq 0 \quad \quad \gamma \gamma'-\gamma_I^2 \geq 0. 
\end{equation}

With a little algebra, one can combine the first-order equations in \eqref{chemicstab} to obtain a second-order evolution equation for the stresses $\Pi_{ab}$, which has exactly the form \eqref{nonhydroadscft}, with coefficients
\begin{equation}\label{checontorto}
\begin{split}
& \tau = \dfrac{\gamma+\gamma'}{\gamma\gamma'-\gamma_I^2+a^2} \geq 0 \\
& \chi = 2 \dfrac{\gamma \gamma' -\gamma_I^2 + a^2}{(\gamma+\gamma')^2} \geq 0 \, . \\
\end{split}
\end{equation}
Now,   from equation \eqref{twofrequncies} we see that the frequencies of the non-hydrodynamic modes have a real part if we impose $\chi >1/2$. Inserting this requirement into the formula for $\chi$ we obtain the condition
\begin{equation}
4a^2 > 4 \gamma_I^2 + (\gamma-\gamma')^2  \geq 0 \, .
\end{equation}
As we can see, $\text{Re}\, \omega_\pm \neq 0$ implies $a \neq 0$.

\subsection{Do we really need all these fields?}\label{allthesefields?}

The example of the previous subsection immediately raises a question: do we really need the additional field $\Lambda_{ab}$ if we want to model a fluid subject to the field equation \eqref{nonhydroadscft}? More generally, the rule \eqref{LaRegolaDoro} tells us that, in Geroch-Lindblom theories, the more non-hydrodynamic modes we have, the more fields we need. Is all this population of fields ``physical'', or is it just a mathematical artifact that is necessary to make the theory symmetric-hyperbolic causal and stable?

To answer this question, let us consider again the field equations \eqref{chemicstab}, and assume for simplicity that $\gamma'=\gamma$, $a>0$ and $\gamma_I=0$. Then, if we focus on the evolution of the component $(1,2)$ of the stress tensor (assuming that all the other independent components are zero), it is possible to show that 
\begin{equation}\label{examplesolution}
\begin{split}
& \Pi_{12}(t) = \Pi_{12}(0) \, e^{-\gamma t} \cos (at)  \\
& \Lambda_{12}(t) = \Pi_{12}(0) \, e^{-\gamma t} \sin (at) \\
\end{split}
\end{equation}
is a solution of \eqref{chemicstab}. Moreover, both $\Pi_{12}(t)$ and $\Lambda_{12}(t)$, as given in \eqref{examplesolution}, are solutions of the damped oscillator equation \eqref{nonhydroadscft} with coefficients
\begin{equation}
\tau = \dfrac{2\gamma}{\gamma^2 + a^2}>0  \spc \chi = \dfrac{\gamma^2 + a^2}{2\gamma^2} > \dfrac{1}{2} \, .
\end{equation}

\begin{figure}
\begin{center}
\includegraphics[width=0.5\textwidth]{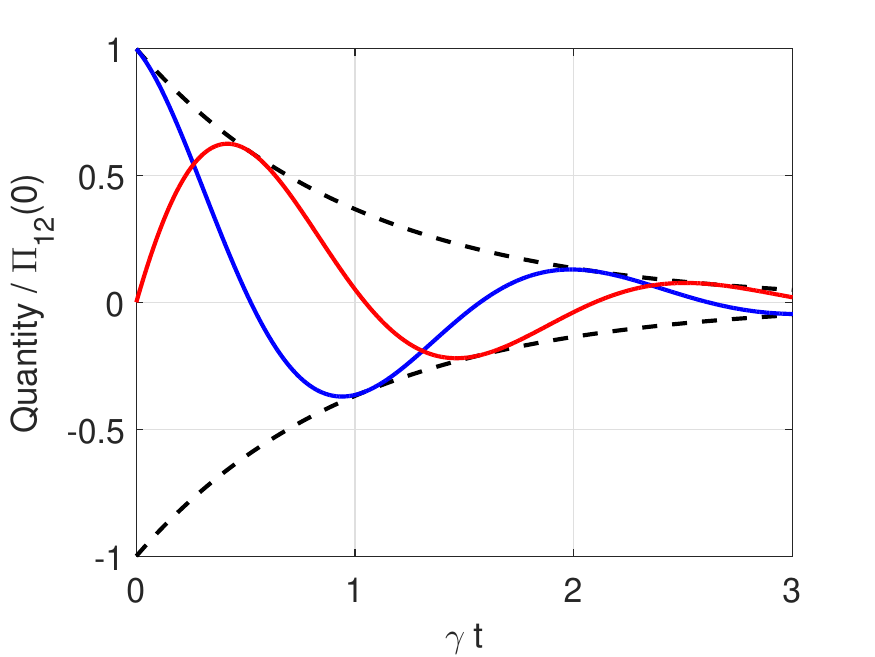}
	\caption{
	Plot of $\Pi_{12}$ (blue line) and $\Lambda_{12}$ (red line) as functions of $\gamma t$, as given by equation \eqref{examplesolution}. Both quantities are given in units of $\Pi_{12}(0)$. We imposed, for aesthetical reasons, $a=3\gamma$. The evolution of both $\Pi_{12}$ and $\Lambda_{12}$ is that of a damped harmonic oscillator, but their relative phase is such that, when $\Pi_{12}$ vanishes, $\Lambda_{12}$ is different from zero, and vice versa.}
	\label{fig:osc}
	\end{center}
\end{figure}

Let us now analyse the qualitative behaviour of the stress $\Pi_{12}$, when its evolution is given by equation \eqref{examplesolution}, see Fig. \ref{fig:osc}. At $t=0$, the stress has a comparatively high value, which then drops rapidly until, at $t=\pi/(2a)$, we have $\Pi_{12}=0$. In this precise instant of time, the shear stresses are all zero and \textit{all} the components of the stress-energy tensor $T^{ab}$ coincide with those at thermodynamic equilibrium. This implies that, if we choose to adopt the Israel-Stewart convention, according to which the state of the fluid is \textit{completely} characterised by $T^{ab}$ (we work at zero chemical potential), then we are forced to conclude that at $t=\pi/(2a)$ the fluid is in global thermodynamic equilibrium. 
The problem is that, for $t>\pi/(2a)$, the stress $\Pi_{12}$ keeps evolving, first decreasing below zero and then increasing again. This clearly shows that the state of the system at $t=\pi/(2a)$ is not the true thermodynamic equilibrium state. In other words, there must be at least one observable, besides $T^{ab}$, which
\begin{itemize}
\item is out equilibrium at $t=\pi/(2a)$ (otherwise the fluid would be in full thermodynamic equilibrium),
\item can be used to characterise the macroscopic state of the fluid at $t=\pi/(2a)$ (because evidently $T^{ab}$ is not enough),
\item is dynamically coupled to $\Pi_{12}$, so that it can be considered ``responsible'' for driving $\Pi_{12}$ out of equilibrium immediately after $t=\pi/(2a)$.
\end{itemize}
This is precisely the role of $\Lambda_{ab}$. It is a non-equilibrium thermodynamic variable \cite{landau5}, which goes out of equilibrium when $t$ approaches $\pi/(2a)$, and which (when positive) ``pushes'' $\Pi_{12}$ to assume negative values, by means of the equation
\begin{equation}
\partial_t \Pi_{12} = -\gamma \Pi_{12} - a \Lambda_{12} \, .
\end{equation}
This argument shows that $\Lambda_{ab}$ is just some thermodynamic quantity which ``resonates'' with $\Pi_{ab}$. The more non-hydrodynamic modes there are, the more resonating states are possible and, therefore, the more additional fields of the type $\Lambda_{ab}$ we need, if we want to characterise the thermodynamic state of the fluid completely. 

\subsection{Field redefinitions}\label{frammuz}

The fact that the choice of $\varphi^A$ is not uniquely prescribed (each $\varphi^A$ is just an effective field) enables us to perform changes of variables of the kind\footnote{
    The fact we are using the same notation $\tilde{\varphi}^C$ adopted in subsection \ref{GPOTNHS} is not a coincidence: the introduction of new fields $\tilde{\varphi}^C$ by means of equation \eqref{pphhiiD} was an example of a field redefinition.
    }
\begin{equation}\label{fieldsRedef}
\varphi^A = \varphi^A (\tilde{\varphi}^C) \, .
\end{equation}
The new fields $\tilde{\varphi}^C$ obey the linearised field equations
\begin{equation}\label{LinearRedef}
\tilde{M}\indices{^m _C _D}\nabla_m \delta \tilde{\varphi}^D = -\tilde{\Xi}_{CD} \delta \tilde{\varphi}^D 
\end{equation}
with
\begin{equation}
\begin{split}
 \tilde{M}\indices{^m _C _D} & = M\indices{^m _A _B} \, \dfrac{\partial \varphi^A}{\partial \tilde{\varphi}^C} \, \dfrac{\partial \varphi^B}{\partial  \tilde{\varphi}^D }\\
 \tilde{\Xi}_{CD} & = \Xi_{AB} \, \dfrac{\partial \varphi^A}{\partial \tilde{\varphi}^C} \, \dfrac{\partial \varphi^B}{\partial \tilde{\varphi}^D} \, . \\
\end{split}
\end{equation}
Note that also $ \tilde{M}\indices{^m _C _D}$ gives rise to a symmetric-hyperbolic causal system of equations, because, as long as the conditions \eqref{symmetry} and \eqref{timelikefuture} hold for $M\indices{^m _A _B}$, they are valid also for $ \tilde{M}\indices{^m _C _D}$. Furthermore,
\begin{equation}
\sigma = \Xi_{AB} \delta \varphi^A \delta \varphi^B = \tilde{\Xi}_{CD}\delta \tilde{\varphi}^C \delta \tilde{\varphi}^D \geq 0 \, . 
\end{equation}
Therefore, the field redefinition \eqref{fieldsRedef} maps (at least in the linear regime) Geroch-Lindblom theories into \textit{equivalent} Geroch-Lindblom theories, which share the same mathematical properties as the original theory. This simple result has three important consequences.

First of all, it tells us that, given a Geroch-Lindblom theory, we may always try to use field redefinitions to simplify the field equations as much as we can. An example of this kind of simplification has already been given in subsection \ref{GPOTNHS}, when we moved from \eqref{nonhydruz} to \eqref{semplicio} using the redefinition \eqref{pphhiiD} (second line).

Secondly, it shows us that, although the number of different Geroch-Lindblom theories may seem exceedingly large (apparently making the choice among them nearly impossible), they are actually less than one may think. In fact, once the conservation laws, the number of fields and their character (e.g. scalar, vector...) are given, many theories turn out to be equivalent to each other. We will show later, with some examples, how this can be used to systematically identify the ``best'' theory for a selected physical problem.

Finally, the possibility of making these field redefinitions generates the same \textit{frame ambiguities} that we see in first-order viscous theories. For example, if we have a theory such that one of the fields is the temperature $T$ and another field is the bulk-viscous stress $\Pi$, we may always define a new temperature $\tilde{T}$ by means of the equation \cite{DoreTorrieri2022}
\begin{equation}
\tilde{T}= T +c \, \Pi \, ,
\end{equation}
where $c$ is an arbitrary factor. Both $T$ and $\tilde{T}$ reduce to the equilibrium thermodynamic temperature at equilibrium, but they differ in the presence of viscous stresses. This is analogous to a change of frame of the kind described by \citet{Kovtun2019}. However, there is a fundamental mathematical difference with respect to first-order theories. In fact, as long as we are in the linear regime, the field redefinitions \eqref{fieldsRedef} reduce to the linear transformations
\begin{equation}\label{Linus}
\delta \varphi^A = \dfrac{\partial \varphi^A}{\partial \tilde{\varphi}^C} \, \delta \tilde{\varphi}^C \, .
\end{equation} 
If we take \eqref{Linus} as the mathematical relation which connects $\delta \varphi^A$ to $\delta \tilde{\varphi}^C$, we see that the system \eqref{linearField} and the system \eqref{LinearRedef} are \textit{exactly the same} system of equations, written in different variables. 
There is really no approximation in moving from one frame to the other. This implies that all the mathematical properties of the theory, such as the slope of the characteristics or the well-posedness of the initial value problem are left unchanged. We may therefore say that (in the linear regime) the \textit{frame ambiguity} of these theories is actually a \textit{frame freedom}. Which frame to use is really a matter of taste. 

This is not what happens in first-order theories, where changing from one frame to another always requires that we make some truncation of the derivative-expansion, even in the linear regime! More precisely, one always needs to neglect some term (which is present also in the linear regime) which contains a third-order derivative, such as $\partial_t^3 \delta \tilde{T}$. Such approximation changes completely the mathematical structure of the equations \cite{Grozdanov2019}; in particular, it changes the behaviour of the non-hydrodynamic sector (that is considered spurious in first-order theories), in which derivatives are not small.

\section{Gibbs stability criterion}\label{IsraeLL}


In subsection \ref{Restringo}, we have reviewed the stability-causality analysis of \citet{Geroch_Lindblom_1991_causal}. However, it has been recently shown \cite{GavassinoLyapunov_2020,GavassinoGibbs2021,GavassinoCausality2021,GavassinoFronntiers2021,GavassinoStabilityCarter2022} that, if there is a non-equilibrium entropy current $s^m$, such that $\nabla_m s^m = \sigma \geq 0$, then the conditions for linear stability (and also for linear causality!) can be derived directly from the maximum entropy principle. In this section, we provide a quick overview of such stability criterion. As we shall see in the next section, demanding the equivalence between the present stability criterion, and that of \cite{Geroch_Lindblom_1991_causal}, has far-reaching implications.

\subsection{Extremum principle}\label{asbaronz}

Assume that a fluid is in weak contact with a heat (and particle) bath. ``Weak contact'' means that, although the two systems interact with each other, the extensive quantities of the total system ``$\text{fluid}+\text{bath}$'' approximately decompose into the sum of the extensive quantities of the two parts, namely:
\begin{equation}\label{fourtizz}
\begin{split}
 S_{\text{tot}} & = S+S_H \spc \, \, \, \, \, (\text{entropy}) \\
  Q^I_{\text{tot}} & = Q^I+Q_H^I \spc (\text{conserved charges}) \\
\end{split}
\end{equation}
where the quantities without label refer to the fluid, while the quantities with the label $H$ refer to the heat bath. In equation \eqref{fourtizz}, all the quantities are computed for an assigned space-like Cauchy 3D-surface $\Sigma$, as the flux of the corresponding currents ($s^m,J^{Im}$), and one can easily show, using the Gauss theorem \cite{MTW_book}, that 
\begin{equation}\label{addictusek}
\Delta S+\Delta S_H \geq 0 \spc  \Delta Q^I = -\Delta Q_H^I ,
\end{equation} 
where $\Delta A := A[\Sigma_f]-A[\Sigma_i]$, with $\Sigma_f$ future to $\Sigma_i$.

Now, as discussed in \cite{Termo}, an ideal heat bath is defined as a body with equation of state ($I$ obeys Einstein's convention)
\begin{equation}\label{suskund}
S_H(Q^I_H)=\text{const} -\alpha_I^H Q^I_H \, , 
\end{equation}
where $\alpha_I^H$ are some fixed constant factors. Equation \eqref{suskund} expresses the fact that the bath is an effectively infinite reservoir of particles and energy, hence the second derivatives of the entropy, which scale like $[\text{Particle-Number}]^{-1}$, are effectively zero \cite{GavassinoTermometri2020}. 
Combining \eqref{addictusek} with \eqref{suskund}, we obtain (recall that $\alpha_I^H$ are constant)
\begin{equation}\label{wimcoeolc}
\Delta S_{\text{tot}} = \Delta (S+\alpha^H_I Q^I) \geq 0 \, .
\end{equation}
Thus, we have found that the quantity $\Phi =S+\alpha^H_I Q^I$ can only grow, or be constant. This implies that the state of thermodynamic equilibrium of the fluid is the state that maximizes $\Phi$ for free variations \cite{Stuekelberg1962,Israel_2009_inbook,Ottinger1997}. From this condition, one can also straightforwardly derive the expression for the equilibrium density matrix of relativistic systems \cite{GibbonsHawking1977,GavassinoGibbs2021,BecattiniBeta2016}.

\subsection{The information current}

At equilibrium, $\Phi=S+\alpha^H_I Q^I$ is maximal. Hence, if $\varphi^A$ is the equilibrium state of the fluid, and $\varphi^A + \delta \varphi^A$ is an arbitrary perturbed state, it must be true that
\begin{equation}\label{wpokdomcnimd}
\tilde{E}=-\delta \Phi > 0 \spc \forall \, \, \delta \varphi^A \neq 0 \, .
\end{equation}
By ``$\delta \Phi$'' we mean the \textit{exact} variation $\Phi[\varphi^A+\delta \varphi^A]-\Phi[\varphi^A]$. On the other hand, it follows from the definition that
\begin{equation}
\tilde{E} = \int_\Sigma \tilde{E}^m d\Sigma_m \quad (\text{orientation: } d\Sigma_0 >0) ,
\end{equation}
with
\begin{equation}\label{IlFuturo}
\tilde{E}^m =-\delta (s^m +\alpha^H_I J^{Im})= -\delta s^m -\alpha^H_I \delta J^{Im} \, .
\end{equation}
However, if the inequality \eqref{wpokdomcnimd} is respected by any variation $\delta \varphi^A$ (no constraint), and for any \textit{space-like}\footnote{$\Sigma$ must be space-like; otherwise information may propagate between different points of $\Sigma$, and $\delta \varphi^A$ would not be arbitrary across it \cite{GavassinoCausality2021}.} Cauchy 3D-surface $\Sigma$, then the four-vector $\tilde{E}^m$ is timelike future directed, namely
\begin{equation}\label{causstab}
\tilde{E}^m \tilde{E}_m \leq 0  \spc \tilde{E}^0 \geq 0 \, .
\end{equation}
The inequalities \eqref{causstab} are sufficient conditions of Lyapunov stability \cite{GavassinoGibbs2021}, and linear causality \cite{GavassinoCausality2021}, for theories that obey the second law.

The current $\tilde{E}^m$ quantifies the flow of information about the total system's microstate \cite{GavassinoCausality2021}; for this reason, we will refer to it as ``information current''. Although in equation \eqref{IlFuturo} the variations may be interpreted as finite differences, in the following we will truncate $\tilde{E}^m$ to second order in $\delta \varphi^A$. Furthermore, we see from \eqref{causstab} that the first-order contribution to $\tilde{E}^m$ must vanish identically (consider the transformation $\delta \varphi^A \rightarrow -\delta \varphi^A$), so that $\tilde{E}^m$ is a pure second-order current: $\tilde{E}^m=\mathcal{O}(\delta \varphi \delta \varphi)$. This produces the covariant Gibbs relation \cite{Israel_Stewart_1979,Israel_2009_inbook}
\begin{equation}\label{COVIS}
\delta s^m = -\alpha^H_I \delta J^{Im} +\mathcal{O}(\delta \varphi \delta \varphi) \, ,
\end{equation} 
which is, indeed, a universal property of Geroch-Lindblom theories \cite{Geroch_Lindblom_1991_causal}. 


\section{Constructing linearised Geroch-Lindblom theories}
\label{sec:costruzione}


In this section we address how to practically build a Geroch-Lindblom theory for a given physical system. Before describing the procedure we first need a useful identity relating the dynamical and statistical properties of the Geroch-Lindblom model at hand. 

\subsection{A surprising identity} 

We consider again the vector field $E^m$, defined in equation \eqref{Clavicembalo}, and we compare its properties with those of the information current $\tilde{E}^m$, defined in \eqref{IlFuturo}. Both are second-order vector fields which are confined within the future light-cone, as shown by conditions \eqref{bramante} and \eqref{causstab}. Both vanish only at equilibrium; this follows from conditions \eqref{timelikefuture} and \eqref{wpokdomcnimd}. Finally, both have four-divergence $-\sigma$, as shown by equations \eqref{fourdivE} and \eqref{IlFuturo} (use the conservation laws $\nabla_m J^{Im}=0$). On the other hand, it has been shown by \citet{GavassinoCausality2021} that a vector field that satisfies all these properties is necessarily unique, which leads us to the central identity of the paper:
\begin{equation}\label{coincidenza?}
 \tilde{E}^m= E^m \quad (\text{to second order}) \, ,
\end{equation}
which implies $\tilde{E}=E+\mathcal{O}(\delta \varphi \delta \varphi \delta \varphi)$, and is equivalent to
\begin{equation}\label{fundamentalLLL}
\delta s^m = -\alpha^H_I \delta J^{Im} - \dfrac{1}{2} M\indices{^m _{AB}} \delta \varphi^A \delta \varphi^B +\mathcal{O}(\delta \varphi \delta \varphi \delta \varphi) \, ,
 \end{equation} 
which is the second-order generalization of \eqref{COVIS}.
This formula is a bit surprising as it establishes a \textit{one-to-one} correspondence between the coefficient-matrix $M\indices{^m _A _B}$ (a dynamical property) of a generic Geroch-Lindblom theory with its information current (a statistical property). This result is particularly helpful because it tells us that we may always use some statistical arguments to estimate\footnote{
From now on, we will use the symbols $\tilde{E}^m$ and $E^m$ interchangeably. The same is true for $\tilde{E}$ and $E$.} $E^m$ using \eqref{coincidenza?}, and then extract $M\indices{^m _A _B}$ by simply performing partial derivatives:
\begin{equation}\label{esecod}
M\indices{^m _A _B} = \dfrac{\partial^2 E^m}{\partial (\delta \varphi^A) \, \partial (\delta \varphi^B)} \, .
\end{equation}
Let us see in more detail how we can use this trick to our advantage.

\subsection{Four-step procedure to construct linear~Geroch-Lindblom~models}

\label{fantastic4}

We can build a Geroch-Lindblom theory for a given fluid by following 4 steps: 
\\
\\
\textit{Step I -} We use rule \eqref{LaRegolaDoro} to ``guess'' how many fields $\varphi^A$ we need and their geometrical character (e.g. scalar, vector...). To do this, we need to have clear which conservation laws we want to explicitly include. In fact, for each conservation law, one should construct a corresponding effective field. The standard examples are
\begin{equation}\label{scheme}
\begin{split}
\text{Energy } & \longrightarrow \text{ Temperature field } T  \\
\text{Momentum } & \longrightarrow \text{ Flow field } u^a  \\
\text{Baryons } & \longrightarrow \text{ Chemical potential field } \mu \, .  \\
\end{split}
\end{equation}
The remaining fields are \textit{dissipation fields} \cite{LindblomRelaxation1996}, like the fields $\Pi_{ab}$ and $\Lambda_{ab}$ introduced in subsection \ref{LLAA}, the number of whose independent components should equal the number of non-hydrodynamic modes. Due to the possibility of making field redefinitions of the type \eqref{Linus}, it is not important how we choose to interpret the various effective fields (at this stage). It only matters their number and their geometrical character. 
\\
\\
\textit{Step II -} We build the most general expression for the information current $E^m$ and for the entropy production rate $\sigma$, given the fields $\varphi^A$. Recalling \eqref{Clavicembalo}, we see that we only need to write the most general vector field (in the case of $E^m$) and the most general scalar field (in the case of $\sigma$) which are quadratic in $\delta \varphi^A$. In principle, the number of free coefficients grows roughly like $\mathfrak{D}^2$, however, we can always use
\begin{itemize}
\item symmetries,
\item field redefinitions,
\item insights from statistical mechanics,
\end{itemize} 
to simplify the structure of the theory as much as possible and restrict our attention to formulas for $E^m$ and $\sigma$ that are well-motivated (from a statistical-mechanical viewpoint).
\\
\\
\textit{Step III -} If we compare \eqref{linearField} with \eqref{Clavicembalo}, we see that the linearised field equations are given by
\begin{equation}\label{themethoddizzzuz}
\nabla_m \, \dfrac{\partial E^m}{\partial (\delta \varphi^A)} = -\dfrac{1}{2} \dfrac{\partial \sigma}{\partial (\delta \varphi^A)} - \Xi_{[AB]} \, \delta \varphi^B \, .
\end{equation}
The only ingredient that the study of $E^m$ and $\sigma$ cannot provide is $\Xi_{[AB]}$ (the antisymmetric part of $\Xi_{AB}$). This needs to be adjusted to correctly reproduce the non-hydrodynamic sector of the fluid. In the particular case in which we know that all the non-hydrodynamic frequencies $\omega_n$ sit on the imaginary axis, we can just set $\Xi_{[AB]}=0$. If, instead, some frequencies have a non-vanishing real part, $\Xi_{[AB]}$ \textit{must} be non-vanishing.
\\
\\
\textit{Step IV -} To make sure that the field equations are hyperbolic, causal and stable, we only need to require that $E^m$ is timelike future directed and $\sigma$ is non-negative, for any non-vanishing value of $\delta \varphi^A$. Recalling \eqref{Clavicembalo}, and considering that the background is isotropic, we only need to require the positive definiteness of two quadratic forms: $E^0-E^1$ and $\sigma$. Therefore, there is no need to perform the direct Fourier analysis.
\\
\\
In order to show how this whole four-step procedure works in practice, we apply it to the case of heat diffusion in section~\ref{sec_heat}. In particular, the concrete example outlined in section~\ref{atvf} will allow us to derive a universal model for diffusion in strongly coupled plasmas and holographic fluids, outlined in section~\ref{diffusione_holo}.   

\subsection{The bridge~between~information~and~action~principle:  hydrodynamics~as~a~field~theory}
\label{azione}

Equation \eqref{themethoddizzzuz} reminds an Euler-Lagrange equation. Furthermore, the four steps above are similar to the standard procedure for formulating effective field theories: \textit{Step I} = ``choose the fields'', \textit{Step II} = ``build the most general Lagrangian density $\mathcal{L \,}$'', \textit{Step III} = ``compute the field equations from $\mathcal{L \,}$'', \textit{Step IV} = ``rule out unphysical equations''. 
In addition, the whole idea of using $E$ to derive the field equations is not so new: we are expanding the information $E$ (analogue to the energy, $\mathcal{H}$) to second order around its absolute minimum (equilibrium state, analogue of vacuum), with the goal of obtaining linear field equations. Finally, the dynamics defined by these equations is such that $E$ does not increase (somehow analogous to the conservation of  $\mathcal{H}$, that is exactly conserved). 
In view of this set of analogies, equation \eqref{coincidenza?} is not surprising: like the Hamiltonian determines the equations of motion, the information $E$ determines the matrix $M\indices{^m _A _B}$.


The interesting thing is that the above analogy between our hydrodynamic framework and classical Lagrangian field theory can be pushed even further: we can we convert the intuitive correspondence
 \begin{equation}\label{burbur}
 (\, E^m \,, \, \sigma \,, \,\Xi_{[AB]} \,) 
 \quad  \leftrightarrow \quad
 \mathcal{L} 
 \end{equation}
 into a mathematically precise equivalence. In fact, the field equations \eqref{linearField} can be obtained from an action principle, with Lagrangian density
\begin{equation}\label{lagrangian}
\mathcal{L}=\mathcal{L}_E + \mathcal{L}_\sigma + \mathcal{L}_{ND} \, ,
\end{equation}
where\footnote{
    We adopt the notation $\phi\overset{\leftrightarrow}{\nabla}\psi= \phi (\nabla \psi)- (\nabla \phi)\psi $.
}
\begin{equation}\label{lagrangian33}
\begin{split}
& \mathcal{L}_E =  M\indices{^m_A _B} \delta \varphi_1^A \overset{\leftrightarrow}{\nabla}_m \delta \varphi_2^B \\
& \mathcal{L}_\sigma = \Xi_{(AB)}(\delta \varphi_1^A \delta \varphi_1^B-\delta \varphi_2^A \delta \varphi_2^B)  \\
& \mathcal{L}_{ND}=  2\, \Xi_{[AB]} \delta \varphi_1^A \delta \varphi_2^B \, . \\
\end{split}
\end{equation}
Here, $\delta \varphi_1^A$ and $\delta \varphi_2^A$ are formally doubled fields \cite{Galley2013,Galley2014},
\begin{equation}\label{doubling}
\delta \varphi^A \rightarrow (\delta \varphi_1^A,\delta \varphi_2^A),
\end{equation}
which should be set equal to each other ($\delta \varphi_1^A=\delta \varphi_2^A=\delta \varphi^A$) in the physical limit, \textit{after} the variation has been taken.

Specifying any of the three physical quantities on the left-hand side of \eqref{burbur} is equivalent to fixing one of the three terms in the Lagrangian density \eqref{lagrangian}:
\begin{itemize}
    \item prescribing the information current allows one to write the kinetic term $\mathcal{L}_E$ via \eqref{esecod};
    \item prescribing the entropy production rate $\sigma$, which depends only on $\Xi_{(AB)}$ and not on $\Xi_{[AB]}$, fixes the dissipative term $\mathcal{L}_\sigma = \sigma_1 - \sigma_2$, see \eqref{Clavicembalo};
    \item prescribing $\Xi_{[AB]}$ fixes the non-dissipative term $\mathcal{L}_{ND}$. 
\end{itemize}  
We also note that $\Xi_{(AB)}$ and $\Xi_{[AB]}$ are explicitly separated in the Lagrangian $\mathcal{L}$, in the sense that they are contracted with different couples $\delta \varphi^A_{1,2} \delta \varphi^B_{1,2} \,$. This makes clear why they play a different role in the dynamics of the system, as also explicitly verified in section \ref{LLAA}. 
To understand the role of $\Xi_{(AB)}$ and $\Xi_{[AB]}$, we can compute the Euler-Lagrange equations for the total Lagrangian in \eqref{lagrangian}:
\begin{equation}
\label{piccione}
\begin{split}
& M\indices{^m_A _B} \nabla_m \delta \varphi_2^B = - \Xi_{(AB)}\delta \varphi_1^B - \Xi_{[AB]} \delta \varphi_2^B \\
& M\indices{^m_A _B} \nabla_m \delta \varphi_1^B = - \Xi_{(AB)}\delta \varphi_2^B - \Xi_{[AB]} \delta \varphi_1^B \, .
\end{split}
\end{equation}
By performing the change of variables \cite{Galley2014}
\begin{equation}
\delta \varphi_+^A = \dfrac{\delta \varphi_1^A + \delta \varphi_2^A}{2} \spc \delta \varphi_-^A = \delta \varphi_1^A-\delta \varphi_2^A \, ,
\end{equation}
the equations of motion \eqref{piccione} read
\begin{equation}
 M\indices{^m_A _B} \nabla_m \delta \varphi_\pm^B = -\big(\pm \Xi_{(AB)} + \Xi_{[AB]} \big) \delta \varphi_\pm^B \, .  
\end{equation}
As can be seen, the fields $\delta \varphi_+^A$, which in the physical limit reduce to $\delta \varphi^A$, obey the dissipative field equation \eqref{linearField}, while the additional fields $\delta \varphi_-^A$, which vanish the physical limit, obey an antidissipative field equation, with the same $\Xi_{[AB]}$ but opposite $\Xi_{(AB)}$. This confirms the idea (presented in subsection \ref{allthesefields?}) that $\Xi_{(AB)}$ is responsible for the damping of the perturbations and $\Xi_{[AB]}$ is responsible for the dynamical oscillation of the gapped modes seen in figure \ref{fig:osc}, which exists independently from the second law of thermodynamics.

Finally, we stress that the action principle described above is formulated using the Eulerian specification of the flow field, as opposed to the Lagrangian description based on keeping track of the fluid elements' worldlines. This is consistent with the UEIT framework described in \citep{GavassinoFronntiers2021}, as discussed in section 2.2 therein: contrarily to what is done in most of the literature, e.g.  \citep{andersson2007review,Endlich2011,TorrieriIS2016,Termo,GavassinoRadiazione}, here the fields $\varphi^A$ on which the action principle builds are not the comoving coordinates of the fluid elements. They are, instead, local non-equilibrium thermodynamic variables (in the ``extended'' sense of UEIT), like the effective temperature $T$ or the viscous stress $\Pi^{ab}$.

\section{Israel-Stewart hydrodynamics as a Geroch-Lindblom theory}
\label{IsEck}

Before considering some examples of theories built directly by using the methodology outlined in subsection \ref{fantastic4}, let us prove that the linearised Israel-Stewart theory is a Geroch-Lindblom theory. In some sense, this is already known, because the divergence-type theory (which is a Geroch-Lindblom theory \cite{Geroch_Lindblom_1991_causal}) is equivalent to the Israel-Stewart theory, in the linear regime \cite{Liu1986}. However, we will verify explicitly that the Israel-Stewart field equations can be obtained directly from equation \eqref{themethoddizzzuz}.

We will also employ our formalism to move from the Eckart frame to the Landau frame, showing the equivalence of the two approaches.

\subsection{Eckart frame}

The fields of the Israel-Stewart theory in the Eckart frame can be chosen to be\footnote{Following the scheme \eqref{scheme}, one could take $(T,\mu)$ instead of $(\mathfrak{s},P)$ as degree of freedom of the theory. However, $E^m$ is more conveniently written in terms of $\mathfrak{s}$ and $P$, thus we have used \eqref{Linus} to make the change of variables $(T,\mu) \rightarrow (\mathfrak{s},P)$.}
\begin{equation}\label{FieldsIS}
(\varphi^A)= (\mathfrak{s},P,u^a,\Pi,q^a,\Pi^{ab})\, ,
\end{equation}
being the entropy per particle, the non-viscous pressure, the velocity field, the bulk-viscous stress, the heat flux and the shear-viscous stress, respectively. 
In the equilibrium state, $P$ and $\mathfrak{s}$ are uniform, $u^a=\delta\indices{^a_t}$ (recall that we work in the equilibrium global rest frame of the fluid) and $\Pi=q^a=\Pi^{ab}=0$. At the linear level, the fields \eqref{FieldsIS} obey the geometrical constraints
\begin{equation}\label{construzzo}
\delta u^0 = \delta q^0 = \delta \Pi^{0a}  = \delta \Pi^{[ab]} = \delta \Pi\indices{^j _j}=0 \, .
\end{equation}
The information current $E^m$ and the entropy production rate $\sigma$ are given by \cite{Hishcock1983,GavassinoGibbs2021} (notation: $j,k\in \{1,2,3\}$)
\begin{equation}\label{E0MIS}
\begin{split}
E^0 =  \dfrac{1}{2T} \bigg[ & \dfrac{nT}{c_p} (\delta \mathfrak{s})^2 + \dfrac{1}{c_s^2} \dfrac{(\delta P)^2}{\rho+P}   \\ & +(\rho+P)\delta u^j \delta u_j +  2\delta u^j \delta q_j  \\
& + \beta_0 (\delta \Pi)^2 + \beta_1 \delta q^j \delta q_j +\beta_2 \delta \Pi^{jk} \delta \Pi_{jk}   \bigg] \\
\end{split}
\end{equation}
\begin{equation}\label{EjMIS}
\begin{split}
E^j =  \dfrac{1}{T} \bigg[ & \delta P \delta u^j + \delta \Pi \delta u^j + \delta \Pi^{jk} \delta u_k  \\
& + \dfrac{\delta T \delta q^j}{T} -\alpha_0 \delta \Pi \delta q^j - \alpha_1 \delta \Pi^{jk} \delta q_k   \bigg] \\
\end{split}
\end{equation}
\begin{equation}\label{SigmaMis}
\sigma = \dfrac{(\delta \Pi)^2}{\zeta T} + \dfrac{\delta q^j \delta q_j}{\kappa T^2} + \dfrac{\delta \Pi^{jk} \delta \Pi_{jk}}{2\eta T} \, ,
\end{equation}
where $n$, $\rho$, $c_s^2$, $c_p$, $\zeta$, $\kappa$ and $\eta$ are baryon density, energy density, sound-speed squared, specific heat at constant pressure, bulk viscosity, heat conductivity and shear viscosity. The factors $\alpha_i$ [not to be confused with $\alpha^H_I$, introduced in \eqref{suskund}] and $\beta_i$ are second-order expansion coefficients of the non-equilibrium entropy current. The first-order temperature perturbation $\delta T$ is related to $\delta P$ and $\delta \mathfrak{s}$ by the equation
\begin{equation}\label{temeprature}
\delta T = \dfrac{T}{c_p} \bigg( \delta \mathfrak{s} + \dfrac{\kappa_p}{n} \delta P  \bigg)\, ,
\end{equation}  
where $\kappa_p$ is the isobaric thermal expansivity (a.k.a. expansion coefficient). 

The thermodynamic quantities presented above can be computed from an equation of state for the enthalpy per particle,
\begin{equation}
\dfrac{\rho +P}{n}  = \tilde{h}(\mathfrak{s},P) \, ,
\end{equation}
by means of the thermodynamic relations
\begin{equation}
\begin{split}
& d\tilde{h} = T d \mathfrak{s} + \dfrac{dP}{n} \, ,\\
& H_{\tilde{h}} =
\begin{bmatrix}
   \dfrac{T}{c_p} & & \dfrac{T \kappa_p}{n c_p}  \\
   \\
   \dfrac{T \kappa_p}{n c_p} & & -\dfrac{1}{n^2 \tilde{h} c_s^2}  \\
\end{bmatrix} \, ,
 \\
\end{split}
\end{equation}
where $H_{\tilde{h}}$ is the Hessian of $\tilde{h}$ in the variables $( \mathfrak{s},P )$. 

We can, now, derive the Israel-Stewart field equations using \eqref{themethoddizzzuz}. Given that all the non-hydrodynamic modes of the theory have imaginary gap, we set $\Xi_{[AB]}=0$. To keep trace of the fact that a given field equation has been obtained from \eqref{themethoddizzzuz} for a given choice of $\delta \varphi^A$, we write the respective field $\varphi^A$ inside a box before the equation. This said, the 14 field equations of the Israel-Stewart theory are (we have multiplied all the equations by $T$)
\begin{flalign}\label{settatntuno}
\boxed{\mathfrak{s}} \quad \dfrac{nT}{c_p} \partial_t \delta \mathfrak{s} + \dfrac{\partial_j \delta q^j}{c_p}=0 &&
\end{flalign}
\begin{flalign}\label{settantadue}
\boxed{P} \quad \dfrac{\partial_t \delta P}{(\rho+P)c_s^2} + \partial_j \delta u^j + \dfrac{\kappa_p \partial_j \delta q^j}{nc_p}=0 &&
\end{flalign}
\begin{flalign}\label{settantatre}
\boxed{u^k} \quad \! \partial_t[ (\rho \! + \! \! P)  \delta u_k \! \! + \!  \delta q_k] \! + \! \partial_k (\delta P \! \! + \! \delta \Pi) \! + \! \partial_j \delta \Pi^j_k \! \! = \! 0  &&
\end{flalign}
\begin{flalign}\label{settantquattro}
\boxed{\Pi} \quad \beta_0 \partial_t \delta \Pi  + \partial_j (\delta u^j -\alpha_0 \delta q^j) = -\dfrac{\delta \Pi}{\zeta} &&
\end{flalign}
\begin{flalign}\label{settantcinque}
\boxed{q^k} \quad \! \partial_t(\beta_1 \! \delta q_k \! + \! \delta u_k) \! + \! \partial_k \! \bigg[ \! \dfrac{\delta T}{T} \! - \! \alpha_0 \delta\Pi  \bigg] \! \! - \! \alpha_1 \! \partial_j \delta \Pi^j_k \! \! = \! -\dfrac{\delta q_k}{\kappa T} &&
\end{flalign}
\begin{flalign}\label{settantasei}
\boxed{\Pi^{kl}} \quad \beta_2 \partial_t \delta \Pi_{kl} + \braket{\partial_k \delta u_l -\alpha_1 \partial_k \delta q_l }= -\dfrac{\delta \Pi_{kl}}{2\eta} \, , &&
\end{flalign}
where $\braket{A_{kl}}$ denotes the symmetric traceless part of $A_{kl}$. Equations \eqref{settatntuno} and \eqref{settantadue} are simply $\nabla_a \delta s^a=\sigma$ and $\nabla_a \delta (nu^a)=0$, truncated to the first order (and rescaled by some constant). We also recognise equation \eqref{settantatre} as the conservation of linear momentum. Equations \eqref{settantquattro}, \eqref{settantcinque} and \eqref{settantasei} are the telegraph-type equations for the dissipation fields. They coincide with equations (37), (38) and (39) of \citet{Hishcock1983} (under the assumption of a homogeneous background). 

As always happens with variational methods, the existence of the constraints \eqref{construzzo} can generate some complications when we perform the derivatives with respect to $\delta \varphi^A$, because we need to make sure that the final field equations are compatible with such constraints. The transversality conditions ($\delta u^0 = \delta q^0 = \delta \Pi^{0a} =\delta \Pi^{a0}=0$) are automatically taken care of by our choice of degrees of freedom (e.g. we take as degree of freedom $\delta u^k$ rather than $\delta u^a$), but the same is not true for the conditions of symmetry and tracelessness of the shear-stress tensor ($\delta \Pi^{[jk]}=\delta \Pi\indices{^j _j}=0$). In appendix \ref{LaShear} we explain how to overcome this obstacle and obtain the correct form of \eqref{settantasei}.

As expected, the system of equations \eqref{settatntuno}-\eqref{settantasei} is manifestly symmetric (it is well-known that the linearised Israel-Stewart theory admits a symmetric formulation \cite{Hishcock1983}). For example, if we compare \eqref{settatntuno} with \eqref{settantcinque}, using \eqref{temeprature}, we immediately see that (recall that all the equations have been multiplied by $T$)
\begin{equation}
TM^j_{\mathfrak{s},q^k} = \dfrac{1}{c_p} \delta^j_k =TM^j_{q^k,\mathfrak{s}} \, . 
\end{equation} 
As another example, we can compare \eqref{settantatre} with \eqref{settantasei} and find
\begin{equation}
TM^j_{u^k,\Pi^{hl}} =  \dfrac{\delta^j_h g_{kl} + \delta^j_l g_{kh}}{2} -\dfrac{\delta^j_k g_{hl}}{3}   = TM^j_{\Pi^{hl}, u^k} \, . 
\end{equation}
These formulas could also be deduced directly from $E^j$ using \eqref{esecod}.

If we impose hyperbolicity, causality and stability using the technique of Step IV, we end up repeating the same stability analysis of \citet{Hishcock1983}. Causality follows automatically \cite{GavassinoCausality2021,GavassinoSuperlum2021}.

\subsection{A first example of change of frame}

In subsection \ref{frammuz}, we showed that every linearised Geroch-Lindblom theory is invariant under changes of variables, as given in \eqref{Linus}. As a first application of our formalism, we use this invariance to prove that (in the linear regime) the Israel-Stewart theory in the Eckart frame \cite{Hishcock1983} is mathematically equivalent to the Israel-Stewart theory in the Landau frame \cite{OlsonLifsh1990}.

In order to move from the fields \eqref{FieldsIS} of the Israel-Stewart theory in the Eckart frame, to the fields
\begin{equation}\label{FieldsISLanda}
(\tilde{\varphi}^C)= (\mathfrak{s},P,\tilde{u}^a,\Pi,\nu^a,\Pi^{ab}) 
\end{equation}
of the Israel-Stewart theory in the Landau frame, we only need to make the change of variables \cite{Israel_Stewart_1979,Salazar2020}
\begin{equation}\label{sbafuz}
\delta u^a = \delta \tilde{u}^a + \dfrac{\delta \nu^a}{n} \spc \delta q^a = -\dfrac{\rho + P}{n} \delta \nu^a \, .
\end{equation}
The transverse vector field $\nu^a$ is the particle-diffusion current, while $\tilde{u}^a$ is the Landau-frame flow velocity. Plugging \eqref{sbafuz} into \eqref{E0MIS}-\eqref{SigmaMis} we obtain
\begin{equation}\label{E0MISLL}
\begin{split}
E^0 =  \dfrac{1}{2T} \bigg[ & \dfrac{nT}{c_p} (\delta \mathfrak{s})^2 + \dfrac{1}{c_s^2} \dfrac{(\delta P)^2}{\rho+P}  + (\rho+P)\delta \tilde{u}^j \delta \tilde{u}_j   \\
& + \beta_0 (\delta \Pi)^2 + \tilde{\beta}_1 \delta \nu^j \delta \nu_j +\beta_2 \delta \Pi^{jk} \delta \Pi_{jk}   \bigg] \\
\end{split}
\end{equation}
\begin{equation}\label{EjMISLL}
\begin{split}
E^j =  \dfrac{1}{T} \bigg[ & \delta P \delta \tilde{u}^j + \delta \Pi \delta \tilde{u}^j + \delta \Pi^{jk} \delta \tilde{u}_k  \\
& + T \delta \bigg( \dfrac{\mu}{T} \bigg) \delta \nu^j -\tilde{\alpha}_0 \delta \Pi \delta \nu^j - \tilde{\alpha}_1 \delta \Pi^{jk} \delta \nu_k   \bigg] \\
\end{split}
\end{equation}
\begin{equation}\label{SigmaMisLL}
\sigma = \dfrac{(\delta \Pi)^2}{\zeta T} + \dfrac{\delta \nu^j \delta \nu_j}{\tilde{\kappa} T^2} + \dfrac{\delta \Pi^{jk} \delta \Pi_{jk}}{2\eta T} \, ,
\end{equation}
where $\mu$ is the chemical potential, satisfying the thermodynamic relation
\begin{equation}\label{chemuzzo}
nT\delta \bigg( \dfrac{\mu}{T} \bigg) = \delta P - \dfrac{\rho + P}{T} \delta T  \, ,
\end{equation}
$\tilde{\kappa}$ is the charge-diffusivity, given by \cite{Kovtun2019}
\begin{equation}
\tilde{\kappa} = \dfrac{n^2 \kappa}{(\rho +P)^2} \, ,
\end{equation}
and we have introduced the coefficients
\begin{equation}\label{GUEITTTT}
\begin{split}
& \tilde{\beta}_1 = \dfrac{(\rho+P)^2}{n^2}\bigg( \beta_1 - \dfrac{1}{\rho +P} \bigg) \\ & \tilde{\alpha}_i = - \dfrac{\rho +P}{n} \bigg( \alpha_i + \dfrac{1}{\rho + P}  \bigg)\, .\\
\end{split}
\end{equation}
Equations \eqref{E0MISLL}-\eqref{SigmaMisLL} coincide with equations (39) and (42) of \citet{OlsonLifsh1990}, confirming that $E^m$ and $\sigma$ are indeed the information current and the entropy production rate of the Israel-Stewart theory in the Landau frame. Again, we can use \eqref{themethoddizzzuz} to compute the field equations:

\begin{flalign}\label{LLsettatntuno}
\boxed{\mathfrak{s}} \quad \dfrac{nT}{c_p} \partial_t \delta \mathfrak{s} -\dfrac{\rho +P}{n c_p} \, \partial_j \delta \nu^j=0 &&
\end{flalign}
\begin{flalign}\label{LLsettantadue}
\boxed{P} \quad \dfrac{\partial_t \delta P}{(\rho+P)c_s^2} \! + \! \partial_j \delta \tilde{u}^j \! + \! T\dfrac{\partial}{\partial P} \bigg( \dfrac{\mu}{T} \bigg)\bigg|_{\mathfrak{s}} \! \!\partial_j \delta \nu^j \! = \! 0 &&
\end{flalign}
\begin{flalign}\label{LLsettantatre}
\boxed{\tilde{u}^k} \quad (\rho+P) \partial_t\delta \tilde{u}_k  + \partial_k (\delta P + \delta \Pi) +\partial_j \delta \Pi^j_k=0  &&
\end{flalign}
\begin{flalign}\label{LLsettantquattro}
\boxed{\Pi} \quad \beta_0 \partial_t \delta \Pi + \partial_j (\delta \tilde{u}^j -\tilde{\alpha}_0 \delta \nu^j) = -\dfrac{\delta \Pi}{\zeta} &&
\end{flalign}
\begin{flalign}\label{LLsettantcinque}
\boxed{\nu^k} \quad \! \tilde{\beta}_1  \partial_t \delta \nu_k \! \! + \! \partial_k \! \bigg[  T  \delta \bigg(  \! \dfrac{\mu}{T}  \! \bigg) \! - \! \tilde{\alpha}_0 \delta\Pi  \bigg] \! \! - \! \tilde{\alpha}_1  \partial_j \delta \Pi^j_k \! \! = \! -  \dfrac{\delta \nu_k}{\tilde{\kappa} T} &&
\end{flalign}
\begin{flalign}\label{LLsettantasei}
\boxed{\Pi^{kl}} \quad \beta_2 \partial_t \delta \Pi_{kl} + \braket{\partial_k \delta \tilde{u}_l -\tilde{\alpha}_1 \partial_k \delta \nu_l }= -\dfrac{\delta \Pi_{kl}}{2\eta} \, . &&
\end{flalign}
Given that these equations arise from the same action principle (see subsection \ref{azione}) as those of the Israel-Stewart theory in the Eckart frame, it is clear that the system \eqref{settatntuno}-\eqref{settantasei} is equivalent to \eqref{LLsettatntuno}-\eqref{LLsettantasei}. This can be easily verified explicitly, with the aid of the thermodynamic identity 
\begin{equation}
nT \dfrac{\partial}{\partial P} \bigg( \dfrac{\mu}{T} \bigg)\bigg|_\mathfrak{s} = 1- \dfrac{(\rho+P)k_p}{nc_p} \, ,
\end{equation}
which follows from \eqref{temeprature} and \eqref{chemuzzo}. Furthermore, it is straightforward to verify that \eqref{LLsettatntuno}-\eqref{LLsettantasei} are, indeed, the field equations of the Israel-Stewart theory in the Landau frame [compare with equations (32)-(37) of \citet{OlsonLifsh1990}, for homogeneous backgrounds].

In conclusion, we have proved that, for linear perturbations around homogeneous equilibria, the Israel-Stewart theory in the Eckart frame and the Israel-Stewart theory in the Landau frame are \textit{exactly} the same theory. This implies that the stability-causality conditions found by \citet{OlsonLifsh1990}  (in the Landau frame) are just a rewrite of those found by \citet{Hishcock1983}  (in the Eckart frame). This can also be checked explicitly. For example, the positivity conditions for $\Omega_4$ and $\Omega_7$ of \citet{OlsonLifsh1990} read
\begin{equation}
\begin{split}
&   \dfrac{2(\rho \! + \! P)^2 \beta_2  \! + \! n^2 \tilde{\beta}_1 \! - \! 2n(\rho \! + \! P)\tilde{\alpha}_1 \! - \! \rho \! - \! P}{2(n^2 \tilde{\beta}_1 +\rho +P)\beta_2 -(n\tilde{\alpha}_1+1)^2} \! < \! \rho \! + \! P \\
& \dfrac{1}{(\rho \! + \! P)^2} \bigg[ n^2 \tilde{\beta}_1 +\rho \! + \! P - \dfrac{(n\tilde{\alpha}_1+1)^2}{2\beta_2} \bigg]>0 \, . \\
\end{split}
\end{equation}
Using \eqref{GUEITTTT}, these conditions can be rewritten as follows:
\begin{equation}
\begin{split}
&   \dfrac{2\beta_2 + \beta_1 + 2\alpha_1}{2\beta_1 \beta_2  - \alpha_1^2}  <  \rho \! + \! P \\
& \beta_1 - \dfrac{\alpha_1^2}{2\beta_2}>0 \, . \\
\end{split}
\end{equation}
These are, indeed, the positivity conditions for $\Omega_4$ and $\Omega_7$ of \citet{Hishcock1983}.

\subsection{Other frames}

The argument above, for the equivalence between the Eckart and the Landau frame, is valid for any field redefinition (see subsection \ref{frammuz}), and, hence, for any change of frame. For example, if one makes the field redefinition
\begin{equation}
\tilde{P}= P +\Pi \, ,
\end{equation}
they will end up with an Israel-Stewart theory in which, instead of there being an ``equilibrium pressure'', plus a viscous stress, there is just the total pressure $\tilde{P}$. This, however, comes at the expense of having a coupling $\delta \tilde{P} \, \delta \Pi$ in $E^0$ (replace $P$ in equation \eqref{E0MIS} with $\tilde{P}-\Pi$), which converts $\Pi$ into a viscous correction to energy and particle densities. The equations of the resulting theory will, then, look very different from \eqref{settatntuno}-\eqref{settantasei}, but their physical content is the same (in the linear regime). 

However, it is important to note that our argument for the equivalence between different hydrodynamic frames does not extend to the so-called ``Israel-Stewart theory in a general frame'', proposed by \citet{NoronhaGeneralFrame2021}. The reason is that such theory has more degrees of freedom than the Israel-Stewart theory and, therefore, more non-hydrodynamic modes.

\section{Models for heat conduction beyond Cattaneo}
\label{sec_heat}

We give some examples of linearised theories constructed through the ``four-step procedure'' outlined in subsection \ref{fantastic4}. For simplicity, we focus on models for pure heat conduction, namely models in which the only relevant conservation law is the energy: $ Q^I  = U $. Following the logical scheme of \eqref{scheme}, we assume that among the effective fields of the theory there is a field $T$, which in local thermodynamic equilibrium becomes the temperature field. All other fields ``mediate'' dissipation \cite{LindblomRelaxation1996}.

\subsection{A warm-up case: Cattaneo's model}

We begin with a pedagogical example, in $1+1$ dimensions. Let's assume to have only two effective fields:
\begin{equation}\label{Cattuz}
(\varphi^A)= (T,q)\, ,
\end{equation}
where $q$ may be interpreted as the heat flux in the positive direction. By rule \eqref{LaRegolaDoro}, we know that the theory will have only one non-hydrodynamic mode. Let us write down the most general expressions for $E^m$ and $\sigma$:
\begin{equation}\label{Cattiz}
\begin{split}
 E^0  & = m_1 \, (\delta T)^2 + m_2 \, \delta T \delta q + m_3  \, (\delta q)^2 \\
 E^1 & = m_4  \, (\delta T)^2 + m_5  \, \delta T \delta q + m_6  \, (\delta q)^2 \\
 \sigma & = m_7  \, (\delta T)^2 + m_8  \, \delta T \delta q + m_9  \, (\delta q)^2 \, , \\
\end{split}
\end{equation}
where the coefficients $m_n$ are some background constants. Following \textit{Step II} (see subsection \ref{fantastic4}), we can use symmetries, field redefinitions and insights from statistical mechanics to simplify the formulas of $E^m$ and $\sigma$. Let us see what we can argue from first principles:
\begin{itemize}
\item \textit{Symmetries}: we can assume that the background state is symmetric under inversion of the $x^1-$axis. This implies that, under the transformation $(\delta T, \delta q)\rightarrow (\delta T , -\delta q)$, we should observe a change $(E^0,E^1,\sigma)\rightarrow (E^0,-E^1,\sigma)$, which implies
\begin{equation}
m_2 = m_4 = m_6 = m_8 =0\, .
\end{equation} 
\item \textit{Field redefinitions}: we are always free to rescale the field $q$ as we wish (making the transformation $\delta q \rightarrow a\, \delta q$). We want to set the scale of $\delta q$ in a way for it to quantify the flux of energy in the direction $x^1$. To do it, we can compare the second line of \eqref{Cattiz} with the corresponding Israel-Stewart formula, equation \eqref{EjMIS}, and we conclude that, for $q$ to represent the actual energy flux (and not just a quantity proportional to it), we must have
\begin{equation}
m_5 = 1/T^2 \, .
\end{equation}
Comparison with Israel-Stewart also suggest renaming $m_3$ and $m_9$ as $m_3 =\beta_1/(2T)$ and $m_9=1/(\kappa T^2)$.
\item \textit{Statistical mechanics}: In local thermodynamic equilibrium, the fluid's state must become that of a perfect fluid. Therefore, we can impose that, for $\delta q=0$, $(E^0,E^1,\sigma)$ are indistinguishable from those of a perfect fluid in hydrostatic equilibrium\footnote{Equation \eqref{fundamentalLLL} is necessary for making this step, because it tells us that $E^m$ has a statistical interpretation, which makes it ``theory-independent''. Without equation \eqref{fundamentalLLL}, the same physical state, modelled with two different hydrodynamic theories, could have different $E^m$. }, which can be computed directly from the Israel-Stewart formulas \eqref{E0MIS}-\eqref{SigmaMis} imposing $\delta P=\delta u^k=0$ (hydrostatic equilibrium) and $\delta \Pi =\delta q^k=\delta \Pi^{jk}=0$ (local thermodynamic equilibrium). From this we obtain, recalling \eqref{temeprature},
\begin{equation}
m_1 = \dfrac{nc_p}{2T^2} \spc m_7 =0 \, .
\end{equation}
\end{itemize}

Putting all these results together, equation \eqref{Cattiz} reduces to
\begin{equation}
\begin{split}
& E^0 =  \dfrac{nc_p}{2T^2} \, (\delta T)^2 +  \dfrac{\beta_1}{2T}  \, (\delta q)^2 \\
& E^1 = \dfrac{\delta T \delta q}{T^2} \spc \sigma =\dfrac{(\delta q)^2}{\kappa T^2} \, . \\
\end{split}
\end{equation}
Now we can use \eqref{themethoddizzzuz} to derive the field equations. Since the non-hydrodynamic frequency is only one, it must sit of the imaginary axis, and we can impose $\Xi_{[AB]}=0$. The result is (we multiply both equations by $T^2$)
\begin{flalign}\label{settantasei22}
\boxed{T} \quad nc_p \partial_t \delta T + \partial_x \delta q =0 &&
\end{flalign}
\begin{flalign}\label{settantasei33}
\boxed{q} \quad T\beta_1 \partial_t \delta q + \partial_x \delta T = -\dfrac{\delta q}{\kappa} \, . &&
\end{flalign}
These are the equations of Cattaneo's model for heat conduction \cite{cattaneo1958}. Note that the correct thermodynamic coefficient in equation \eqref{settantasei22} is indeed $c_p$, and not $c_v$. In fact, while the pressure must be uniform in a fluid at rest (we can set $\delta P=0$), individual fluid elements expand  and contract, in response to a change in temperature \cite{landau6}. Hence, the density is not constant. 

To give an idea of how the action of these theories looks like, we also write explicitly the Lagrangian density of Cattaneo's model, as given in \eqref{lagrangian}-\eqref{lagrangian33}:
\begin{equation}
\begin{split}
T^2 \mathcal{L} = & nc_p \, \delta T_1 \overset{\leftrightarrow}{\partial}_t \delta T_2 + T\beta_1 \, \delta q_1 \overset{\leftrightarrow}{\partial}_t \delta q_2 \, +  \\
& \delta T_1 \overset{\leftrightarrow}{\partial}_x \delta q_2 + \delta q_1 \overset{\leftrightarrow}{\partial}_x \delta T_2 + \dfrac{\delta q_1^2 - \delta q_2^2}{\kappa} \, .\\
\end{split}
\end{equation} 
One can easily verify that the field equations \eqref{settantasei22} and \eqref{settantasei33} can be recovered as the physical limit ($\delta T_1 = \delta T_2=\delta T$ and $\delta q_1 = \delta q_2 = \delta q$) of the Euler-Lagrange equations computed from the Lagrangian density above.

In conclusion, we have shown that Cattaneo's model is the \textit{only possible} Geroch-Lindblom theory for heat conduction one can build (in the linear regime) with the choice of fields \eqref{Cattuz}. The frequency of the single non-hydrodynamic mode has the standard form \eqref{QuandoTiRilassi} (for $k=0$), with relaxation time $\tau=\kappa T\beta_1$. The stability and causality conditions are computed directly from the information current (\textit{Step IV}) and have been extensively discussed in \cite{GavassinoCausality2021}.

\subsection{Adding a scalar field: duality between bulk viscosity and chemical reactions}\label{addingascalar}

In order go beyond Cattaneo's model, we need to add new fields to \eqref{Cattuz}. Let us work in $3+1$ dimensions and consider a theory with three fields,
\begin{equation}\label{fielduz}
(\varphi^A)=(T,\mathbb{A},q^j) \, ,
\end{equation}
where $\mathbb{A}$ is an additional non-conserved scalar degree of freedom, which vanishes at equilibrium. The physical interpretation of $\mathbb{A}$ may change from system to system. In view of \eqref{scheme}, one may interpret $\mathbb{A}$ to be the chemical potential of a non-conserved current, which is non-zero only out of equilibrium (e.g. the chemical potential of photons for a fluid coupled to radiation \cite{GavassinoRadiazione}, or the one of phonons in a neutral superfluid \cite{GavassinoKhalatnikov2022}). The presence of $\mathbb{A}$ has the effect of adding a non-hydrodynamic mode to the theory. Furthermore, the dynamics of $\mathbb{A}$ can couple with that of $T$ and $q^j$, effectively changing the way in which heat propagates across the system.   

It is always possible to perform a change of hydrodynamic frame (namely, a redefinition of the temperature) of the form
\begin{equation}\label{gabbani}
T = \tilde{T}+c \, \mathbb{A} \, ,
\end{equation}
where $c$ is an arbitrary constant. We can use this freedom to set to zero the term proportional to $\delta T  \delta \mathbb{A}$ in $E^0$. Following the same procedure as in the previous subsection, the most general expression for $E^m$ and $\sigma$ in the aforementioned frame is
\begin{equation}
\begin{split}
E^0 & = \dfrac{1}{2T} \bigg[ \dfrac{nc_p}{T} (\delta T)^2 + \beta_0 (\delta \mathbb{A})^2 + \beta_1 \delta q^j \delta q_j \bigg]\\
E^j & = \dfrac{1}{T}\bigg[ \dfrac{\delta T \delta q^j}{T} - \alpha_0 \delta \mathbb{A} \delta q^j \bigg]  \\
\sigma & =\dfrac{1}{T} \bigg[ \dfrac{(\delta \mathbb{A})^2}{\xi } + \dfrac{\delta q^j \delta q_j}{\kappa T} \bigg] \, ,  \\
\end{split}
\end{equation}
where the term proportional to $(\delta T)^2$ in $\sigma$ must be zero because in local thermodynamic equilibrium (i.e. for $\delta \mathbb{A}=\delta q^j=0$)  no entropy should be produced. Therefore, also the term proportional to $\delta T\delta \mathbb{A}$ must be zero, otherwise $\sigma$ would not be non-negative definite, violating the second law of thermodynamics.

The field equations, computed using \eqref{themethoddizzzuz}, are (again we multiply them by $T^2$)
\begin{flalign}\label{settantasei222}
\boxed{T} \quad nc_p \partial_t \delta T + \partial_j \delta q^j = 0 &&
\end{flalign}
\begin{flalign}\label{settantasei444}
\boxed{\mathbb{A}} \quad T(\beta_0 \partial_t \delta \mathbb{A} -  \alpha_0 \partial_j \delta q^j) =  - T\delta \mathbb{A}/\xi  &&
\end{flalign}
\begin{flalign}\label{settantasei333}
\boxed{q^k} \quad T\beta_1 \partial_t \delta q_k + \partial_k (\delta T-T\alpha_0 \delta \mathbb{A}) = -\delta q_k/\kappa \, . &&
\end{flalign}
We could not add any skew-symmetric part to $\Xi_{AB}$. In fact, by isotropy (recall that $\Xi_{AB}$ are background quantities),
\begin{equation}
\Xi_{q^k,T} = \Xi_{T, q^k} = \Xi_{q^k,\mathbb{A}} = \Xi_{\mathbb{A},q^k} =0 \, ,
\end{equation}
hence the only non-zero term of $\Xi_{[AB]}$ could be $\Xi_{[T\mathbb{A}]}$. However, given that the model needs to obey the conservation of energy, there should be at least one equation having the form of a conservation law 
\begin{equation}\label{consurvo}
    \partial_t (c_1 \delta T + c_2 \delta \mathbb{A}) + \partial_j (c_3 \delta q^j)=0 \, .
\end{equation}
The only way for this to be possible is to require that $\Xi_{T\mathbb{A}}=0$, so that equation \eqref{settantasei222} takes the form \eqref{consurvo}. 

What is the physical content of this theory? In the limit of small $\omega$ and $k$ (hydrodynamic sector), we see from \eqref{settantasei333} that $q$ is of first order in the gradients, $q=\mathcal{O}(\nabla)$. Inserting this estimate into \eqref{settantasei444}, we find that $\mathbb{A}=\mathcal{O}(\nabla^2)$. Hence, to first order in gradients, $\mathbb{A}$ disappears, and we recover Fick's law  $q \propto -\nabla T$. This is a manifestation of the relaxation effect \cite{LindblomRelaxation1996} discussed in subsection \ref{takeiteasy}. 
The non-hydrodynamic sector, on the other hand, presents four modes. Three of them are the standard ``Cattaneo-type'' heat relaxation modes (one for each spatial dimension), with relaxation time $\tau_q=\kappa T\beta_1$. There is, however, an additional relaxation mode, with relaxation time $\tau_\mathbb{A}= \xi \beta_0$. If we interpret $\mathbb{A}$ as a chemical potential, this additional mode models a chemical relaxation towards chemical equilibrium, that is attained at $\mathbb{A}=0$ when the direct and inverse chemical reactions balance.

We also note that the field equations  \eqref{settantasei222}-\eqref{settantasei333} can be equivalently obtained imposing $\delta P=\delta u^k=\delta \Pi^{jk}=0$ in the Israel-Stewart theory\footnote{
    This automatically implies that the stability-causality conditions of a theory with the additional field $\mathbb{A}$ are the same as those of its Israel-Stewart dual.
}
\eqref{settatntuno}-\eqref{settantasei}, and making the identification $\mathbb{A}= \Pi$. 
This shows that the bulk-viscous stress $\Pi$ is dynamically indistinguishable from the effective chemical potential $\mathbb{A}$ of a non-conserved chemical species, showing that the mathematical duality  
\begin{equation}
(\text{Bulk viscosity}) \longleftrightarrow (\text{Effective chemistry})
\end{equation}
for isotropic fluids \citep{BulkGavassino,camelio2022arXiv,Camelio2022IIarXiv}, survives also in the presence of heat conduction (that destroys isotropicity). This is consistent with the recent treatment of bulk viscosity and heat conduction in relativistic superfluids \citep{GavassinoKhalatnikov2022}, see Fig.~1 therein.

\subsection{Adding a transverse vector field}
\label{atvf}

In the previous subsection we have seen that adding a scalar field to \eqref{Cattuz} endows the Cattaneo's dynamics with a gapped mode that can be interpreted as arising from chemical relaxation: such a theory is equivalent to the one of Israel and Stewart. 
To see if we can produce a model that is qualitatively different from the Israel-Stewart one, it is interesting to upgrade \eqref{Cattuz} to a theory based on three fields,
\begin{equation}
(\varphi^A)=(T,q^j,p^j)\, ,
\end{equation}
where $p^j$ is a transverse vector field, geometrically analogous to $q^j$, which vanishes at equilibrium. Using the invariance of the theory under field redefinitions of the kind
\begin{equation}
\delta q^j = c_1 \delta \tilde{q}^j + c_2 \delta \tilde{p}^j \spc \delta p^j = c_3 \delta \tilde{p}^j + c_4 \delta \tilde{q}^j \, ,
\end{equation}
we can always choose our fields in such a way that the information current and the entropy production rate take the form\footnote{
    We are assuming, for simplicity, invariance under parity: we are neglecting possible contributions to $E^j$ proportional to $\varepsilon^{jkl}\delta q_k \delta p_l$, where $\epsilon^{jkl}$ is the 3D Levi-Civita symbol.}
\begin{equation}
\begin{split}
E^0 & =   \dfrac{nc_p}{2T^2} (\delta T)^2 +  \dfrac{\beta_1}{2T} ( \delta q^j \delta q_j + \delta p^j \delta p_j) \\
E^j & = \dfrac{\delta T \delta q^j}{T^2}  \\
\sigma & =\dfrac{1}{T^2} (\xi_1 \, \delta q^j \delta q_j +2 \xi_2 \, \delta q^j \delta p_j + \xi_3 \, \delta p^j \delta p_j ) \, .  \\
\end{split}
\end{equation}
The field equations, computed using \eqref{themethoddizzzuz}, are
\begin{flalign}\label{nanni1}
\boxed{T} \quad nc_p \partial_t \delta T + \partial_j \delta q^j =0 &&
\end{flalign}
\begin{flalign}\label{nanni2}
\boxed{q^k} \quad T\beta_1 \partial_t  \delta q_k + \partial_k \delta T = -\xi_1 \delta q_k - (\xi_2+b) \delta p_k  &&
\end{flalign}
\begin{flalign}\label{nanni3}
\boxed{p^k} \quad T\beta_1 \partial_t \delta p_k  = - (\xi_2-b) \delta q_k -\xi_3 \delta p_k \, , &&
\end{flalign}
where $b$ is the skew-symmetric part of $\Xi_{AB}$. 
This is the first complete model that we meet which cannot be recovered as a limit of the Israel-Stewart theory. Hence, it is worth exploring its properties in greater detail.

\textit{Reliability criteria -} The stability-causality conditions for the present theory are summarised below:
\begin{itemize}
\item Rest-frame Gibbs criterion ($E^0>0$): 
\begin{equation}\label{ubf}
c_p>0 \spc \beta_1>0 \, ,
\end{equation}
\item Causality condition ($E^0 E^0 \geq E^j E_j$): 
\begin{equation}\label{ubff}
nc_p T\beta_1 \geq 1 \, ,
\end{equation}
\item Second law of thermodynamics ($\sigma \geq 0$):
\begin{equation}\label{ubfff}
\xi_1\geq 0 \spc \xi_3 \geq 0 \spc \xi_1 \xi_3 \geq \xi_2^2 \, .
\end{equation}
\end{itemize}
As a consistency check, in appendix \ref{modebymode} we verify explicitly that, if conditions \eqref{ubf}-\eqref{ubfff} are obeyed, the standard stability-causality criteria discussed in  \cite{Krotscheck1978,Kovtun2019}
\begin{equation}\label{crittuz}
\text{Im }\omega \leq 0  \spc \lim_{k \rightarrow \infty} \bigg| \dfrac{\text{Re }\omega}{k} \bigg| \leq 1
\end{equation}
are, indeed, respected.

\textit{Non-hydrodynamic sector -}  In the homogeneous limit, equations \eqref{ubff} and \eqref{ubfff} may be compactly written in the form
\begin{equation}\label{qpippo}
T \beta_1 \partial_t \begin{pmatrix}
\delta q_k   \\
\delta p_k \\
\end{pmatrix} 
= -
\begin{bmatrix}
   \xi_1 & \xi_2 +b \\
   \xi_2-b & \xi_3  \\
\end{bmatrix} 
\begin{pmatrix}
\delta q_k   \\
\delta p_k \\
\end{pmatrix} \, ,
\end{equation}
which is analogous to \eqref{chemicstab}. This kind of equation always appears when there are two dissipation fields with the same geometric character. Again, we can combine the two equations to obtain a second-order differential equation for $\delta q_k$:
\begin{equation}\label{ewmodkc}
  \dfrac{1}{2} \chi  \tau^2 \, \partial_t^2 \delta q_k + \tau \, \partial_t \delta q_k + \delta q_k=0 \, ,
\end{equation}
with
\begin{equation}\label{ermofk}
\tau = \dfrac{T \beta_1 (\xi_1 + \xi_3)}{\xi_1 \xi_3 - \xi_2^2+b^2}  \spc \chi = 2\dfrac{\xi_1 \xi_3 - \xi_2^2 + b^2}{(\xi_1 + \xi_3)^2} \, .
\end{equation}
Equations \eqref{ewmodkc} and \eqref{ermofk} are analogous to \eqref{nonhydroadscft} and \eqref{checontorto}. Again, if $b$ is sufficiently large, it is possible to have $\chi>1/2$, which produces oscillations of the same kind as those shown in figure \ref{fig:osc}.

\textit{Hydrodynamic sector -} As anticipated in subsection \ref{GPOTNHS}, Geroch-Lindblom theories relax to Navier-Stokes states. This means, following \citet{LindblomRelaxation1996}, that, as $t \rightarrow +\infty$ (and for small $k$), the relaxation terms $T\beta_1 \partial_t  \delta q_k$ and $T\beta_1 \partial_t  \delta p_k$ in \eqref{nanni2} and \eqref{nanni3} become negligible, converting the dynamical equations \eqref{nanni2} and \eqref{nanni3} into constraints:
\begin{equation}
 \begin{pmatrix}
\partial_k \delta T   \\
0 \\
\end{pmatrix} 
\approx -
\begin{bmatrix}
   \xi_1 & \xi_2 +b \\
   \xi_2-b & \xi_3  \\
\end{bmatrix} 
\begin{pmatrix}
\delta q_k   \\
\delta p_k \\
\end{pmatrix} \, .
\end{equation}
These constraints can be used (inverting the matrix on the right-hand side) to write the dissipation fields in terms of the gradient of the temperature:
\begin{equation}\label{cunz}
\delta q_k \approx \dfrac{-\xi_3 \partial_k \delta T}{\xi_1 \xi_3 -\xi_2^2 +b^2} \quad \quad \delta p_k \approx \dfrac{(\xi_2 -b) \partial_k \delta T}{\xi_1 \xi_3 -\xi_2^2 +b^2} \, .
\end{equation}
The first equation is simply Fourier's law, with heat conductivity coefficient
\begin{equation}\label{kappuccio}
\kappa = \dfrac{\xi_3}{\xi_1 \xi_3 -\xi_2^2 +b^2} \geq 0 \, .
\end{equation}
Plugging \eqref{cunz} into \eqref{nanni1}, we obtain the heat equation:
\begin{equation}\label{diffondo}
\partial_t \delta T \approx \dfrac{\kappa}{nc_p} \partial_j \partial^j \delta T \, ,
\end{equation}
proving that the universal diffusive (Fick-type) behaviour is, indeed, recovered. Note that, because of the stability-causality conditions \eqref{ubf}-\eqref{ubfff}, one always has
\begin{equation}\label{inequol}
\tau \geq T\beta_1 \kappa \geq \dfrac{\kappa}{nc_p} \, .
\end{equation}

In appendix \ref{smallk} we show that the formulas \eqref{ermofk} and \eqref{kappuccio} for the transport coefficients $\tau$, $\chi$ and $\kappa$ could be computed directly from the study of the dispersion relations of the theory, $\omega_n(k)$, in the limit of small wave-vectors.

\subsection{Adding a skew-symmetric transverse tensor field}

In subsection \ref{addingascalar} we have shown that, if we include in our model for heat conduction a single scalar field $\mathbb{A}$, the resulting dynamics is just that of Israel-Stewart, where $\mathbb{A}$ plays the role of the bulk stress $\Pi$. It is easy to verify that something similar happens if we insert a transverse symmetric traceless tensor field $\mathbb{A}^{jk}$: the resulting dynamics is just that of Israel-Stewart, where $\mathbb{A}^{jk}$ plays the role of the shear stress $\Pi^{jk}$. 

Instead, we obtain something really new if we consider a theory built on the fields
\begin{equation}\label{wenomdl}
(\varphi^A)=(T,q^j,\Omega^{jk}) \, ,
\end{equation}
where $\Omega^{jk}$ is a transverse \textit{skew-symmetric} tensor field:
\begin{equation}\label{skew}
\Omega^{jk}=-\Omega^{kj} \, .
\end{equation}
Note that, by Hodge duality, one might change degree of freedom and work with the vector field 
\begin{equation}\label{pom}
p^j=\dfrac{1}{2}\varepsilon^{jkl} \Omega_{kl} \, ,
\end{equation}
reducing the present case to that of the previous subsection. However, since we are focusing, for simplicity, on theories that are invariant under parity, the theory built on $\Omega^{jk}$ and the one built on $p^j$ are different. If parity is broken, one may just build $E^m$ and $\sigma$ using all the terms of both the present subsection and the previous one, imposing \eqref{pom}.

The most general information current and entropy production rate, built from the fields given in \eqref{wenomdl}, and compatible with the symmetries of the problem, are
\begin{equation}
\begin{split}
E^0 & =  \dfrac{1}{2T} \bigg[   \dfrac{nc_p}{T} (\delta T)^2 +  \beta_1 \delta q^j \delta q_j  + \gamma_1 \delta \Omega^{jk} \delta\Omega_{jk}  \bigg] \\
E^j & = \dfrac{\delta T \delta q^j}{T^2} + \gamma_2 \dfrac{ \delta\Omega^{jk} \delta q_k}{T^2} \\
\sigma  & = \dfrac{\delta q^j \delta q_j}{\kappa T^2} + \dfrac{\delta \Omega^{jk} \delta \Omega_{jk}}{2\xi T}  \, . \\
\end{split}  
\end{equation}
Note that, in the expressions above, the order of the indices of $\delta \Omega^{jk}$ matters. For example, in the formula for $E^j$, one could replace $\Omega^{jk} \delta q_k$ with $\Omega^{kj} \delta q_k$, at the price of changing the sign of $\gamma_2$. In $E^0$ and $\sigma$, the indices have been contracted in such a way that the coefficients $\gamma_1$ and $\xi$ must be positive, to ensure stability and thermodynamic consistency.

Similarly to what happened with the shear stress, we need to be careful with applying \eqref{themethoddizzzuz} for computing the field equations, because the constraint \eqref{skew} must be respected. This problem is easily solved adapting to $\Omega^{jk}$ the technique outlined in appendix \ref{LaShear} for $\Pi^{jk}$. The result are the equations given below:
\begin{flalign}\label{nanni11}
\boxed{T} \quad nc_p \partial_t \delta T + \partial_j \delta q^j =0 &&
\end{flalign}
\begin{flalign}\label{nanni22}
\boxed{q^k} \quad T\beta_1 \partial_t  \delta q_k + \partial_k  \delta T  +  \gamma_2 \, \partial_j \delta \Omega\indices{^j _k} = -\delta q_k /\kappa &&
\end{flalign}
\begin{flalign}\label{nanni33}
\boxed{\Omega^{kl}} \quad T \gamma_1 \partial_t \delta \Omega_{kl} + \gamma_2 \, \partial_{[k} \delta q_{l]} = -T \delta \Omega_{kl}/2\xi \, . &&
\end{flalign}
We can use \eqref{pom} (and its inverse: $\Omega^{kl}=\varepsilon^{klj}p_j$) to rewrite \eqref{nanni22} and \eqref{nanni33} in the alternative form
\begin{equation}
\begin{split}
&  T\beta_1 \partial_t  \delta q_k + \partial_k  \delta T  -\gamma_2 (\nabla \times \delta p)_k = -\delta q_k /\kappa \\
& 2T \gamma_1 \partial_t \delta p_k + \gamma_2 (\nabla \times \delta q)_k =-T\delta p_k/\xi \, , \\
\end{split}
\end{equation}
where $(\nabla \times  p)^j = \varepsilon^{jkl} \partial_k p_l$ is the standard 3D-curl of $p$. We see that, when the additional field is a transverse skew-symmetric tensor field, the resulting theory is that of two vector fields, which are, however, dynamically coupled only through curls (as long as parity is not broken). Hence, our method may also be used to model the non-hydrodynamic sector of MHD and spin-hydrodynamics.

\section{Onsager-Casimir relations}

We have shown that equation \eqref{fundamentalLLL} can be used as a starting point for the systematic construction of causal and stable fluid theories. On the other hand, it also enables us to connect our newly-born theories with non-equilibrium thermodynamics. We can use this bridge to derive some general Onsager-Casimir relations. In this section we explain how to do it.

\subsection{The symmetry principle}\label{Casimiro}

We focus on homogeneous perturbations which conserve the value of the total integrals of motion (i.e. we focus on purely non-hydrodynamic modes). Then,
\begin{equation}
\delta J^{I0} = 0 \, ,
\end{equation}
which, using \eqref{fundamentalLLL}, implies
\begin{equation}\label{wewe}
E^0 = -\delta s^0 \, .
\end{equation}
Let us consider again the Lyapunov functional $E$, defined in \eqref{IntroducoE}, taking $\Sigma$ to be a surface at constant time (recall that we are working in the fluid's equilibrium rest-frame). Homogeneity implies
\begin{equation}\label{wewe2}
E = E^0 \, V \, ,
\end{equation}
where $V$ is the total volume occupied by the fluid. Combining \eqref{wewe} with \eqref{wewe2}, and recalling \eqref{Clavicembalo}, we have
\begin{equation}\label{sxsx}
\delta S = - \dfrac{V}{2} \, M\indices{^0_{AB}} \delta \varphi^A \delta \varphi^B \, .
\end{equation}
Our goal, now, is to establish a direct connection with the notation of \citet{Onsager_Casimir}. Let us assume that we can group the fields $\varphi^A$ into two categories, depending on their behaviour under time-reversal\footnote{Instead of just time-reversal, one may consider $CPT$ \cite{Glorioso2018}.}: the even fields (for which we use the multi-indices $P,Q,...$) and the odd fields (for which we use the multi-indices $X,Y,...$), so that, adopting the notation of \cite{Onsager_Casimir}, we can write
\begin{equation}
\alpha^P = \delta \varphi^P  \spc \beta^X = \delta \varphi^X \, .
\end{equation} 
Since the entropy is even in time, we know from \eqref{sxsx} that \cite{Onsager_Casimir}
\begin{equation}\label{mxp}
M\indices{^0 _P _X} = M\indices{^0 _X _P} =0 \, ,
\end{equation} 
so that the conjugate variables to respectively $\alpha^P$ and $\beta^X$ are
\begin{equation}
\gamma_P = V M\indices{^0 _P _Q} \delta \varphi^Q \spc \gamma_X = V M\indices{^0 _X _Y} \delta \varphi^Y \, .
\end{equation}
Then, the field equations \eqref{nonhydruz} can be rewritten in Casimir's form
\begin{equation}
\begin{split}
& \dot{\alpha}^P = p^{PQ}\gamma_Q + p^{PY} \gamma_Y \\
& \dot{\beta}^X = p^{XQ}\gamma_Q + p^{XY} \gamma_Y \, , \\
\end{split}
\end{equation}
provided that we make the identification
\begin{equation}\label{mxp2}
\Xi_{AB}=-V p^{CD} M\indices{^0 _A _C} M\indices{^0 _B _D} \, .
\end{equation}
The coefficients $p^{CD}$ obey the Onsager-Casimir relations
\begin{equation}\label{maxp3}
p^{PQ}=p^{QP} \quad \quad p^{PX}=-p^{XP} \quad \quad p^{XY}=p^{YX} \, .
\end{equation}
If we combine \eqref{mxp}, \eqref{mxp2} and \eqref{maxp3}, we obtain
\begin{equation}\label{max4p}
\Xi_{PQ}=\Xi_{QP} \quad \quad \Xi_{PX}=-\Xi_{XP} \quad \quad \Xi_{XY}=\Xi_{YX} \, .
\end{equation}
Equations \eqref{mxp} and \eqref{max4p} are powerful constraints arising from microscopic reversibility, which can be used to further simplify the equations of a fluid model. We can summarise them as follows:
\begin{itemize}
\item If $\varphi^A$ and $\varphi^B$ have equal behaviour under time reversal (even-even or odd-odd case), then
\begin{equation}
\Xi_{[AB]}=0;
\end{equation}
\item If $\varphi^A$ and $\varphi^B$ have opposite behaviour under time reversal (even-odd case), then
\begin{equation}\label{xaxaxux}
M\indices{^0 _A _B}=\Xi_{(AB)}=0 \, .
\end{equation}
\end{itemize}

\subsection{Application: Onsager and holography}

We can, now, employ the Onsager-Casimir symmetry principle to draw interesting conclusions about the hydrodynamics (and non-equilibrium thermodynamics) of strongly coupled plasmas. 

In the introduction, we mentioned that the non-hydrodynamic sector of strongly coupled holographic theories differs from that of Israel-Stewart-type theories because the frequencies are no longer purely imaginary. In subsection \ref{LLAA}, we showed that the only way for the frequencies to have a real part is that $\Xi_{[AB]}\neq 0$. Finally, in subsection \ref{Casimiro} we verified that, if the hydrodynamic theory is consistent with the Onsager-Casimir principle, then $\Xi_{[AB]}$ must vanish whenever $\varphi^A$ and $\varphi^B$ acquire the same phase under time reversal (even-even and odd-odd cases). In conclusion, the \textit{only way} that we have to reproduce the non-hydrodynamic sector of holographic plasmas is to impose that each dissipative flux (like the heat flux $q^a$ and the viscous stresses $\Pi$ and $\Pi^{ab}$) is dynamically coupled (through $\Xi_{AB}$) to a ``thermodynamic partner'', with the same geometric character but \textit{opposite} behaviour under time reversal. In other words, the non-equilibrium degrees of freedom must appear in couples,
\begin{equation}\label{propriopartner}
(q^a,p^a) \quad \quad (\Pi,\Lambda) \quad \quad (\Pi^{ab},\Lambda^{ab}) \, ,
\end{equation}
whose members have exactly the same geometrical properties (e.g. both $\Pi^{ab}$ and $\Lambda^{ab}$ are transverse, symmetric and traceless) but acquire a different phase under time reversal, see the table below:
\begin{center}
\begin{tabular}{ |l| c| l| c|}
\hline
Flux & Phase of flux & Partner & Phase of partner \\
\hline
 $\, q^a$ & $-1$ & $\, p^a$ & $+1$ \\ 
 $\, \Pi$ & $+1$ & $\, \Lambda$ & $-1$ \\  
 $\, \Pi^{ab}$ & $+1$ & $\, \Lambda^{ab}$ & $-1$ \\
 \hline  
\end{tabular}
\end{center}
In this way, we can couple each flux with its partner in the same way as we did in equations \eqref{chemicstab} and \eqref{qpippo}, with the difference that now we can use the Onsager-Casimir principle to require the anti-symmetry of the coupling term, e.g.
\begin{equation}
 \Xi_{\Pi,\Lambda}= -\Xi_{\Lambda,\Pi} \, ,
\end{equation} 
for the case of bulk viscosity.

\subsection{The intuitive interpretation of the partners}\label{partnerizzodibrutto}

If, at first, the introduction of the partners may seem to be an ad-hoc assumption, it is actually a direct consequence of equation \eqref{nonhydroadscft}. In fact, equation \eqref{nonhydroadscft} models a damped armonic oscillator \cite{Denicol_Relaxation_2011}, which is the simplest example of a couple of thermodynamic degrees of freedom having opposite behaviour under time reversal. Let us explore this analogy in more detail; the setting is similar to that proposed by \citet{Heimburg2017}.

We consider a harmonic oscillator with mass $M$ and Hooke's constant $\kappa$, which is weakly interacting with a heat bath with entropy $S_H$, energy $U_H$ and (constant) temperature $T_H$. Then, the energy and entropy of the total isolated system ``$\text{oscillator}+\text{bath}$'' are
\begin{equation}
\begin{split}
& S_{\text{tot}} \approx \text{const} + \dfrac{U_H}{T_H} \\
& U_{\text{tot}} = \dfrac{1}{2} \kappa x^2 + \dfrac{1}{2} M v^2 + U_H \, , \\
\end{split}
\end{equation}
where $x$ and $v$ are the position and velocity of the mass $M$. In deriving the equation for the entropy, we assumed that the oscillator has no additional internal degrees of freedom, so that $S_{\text{tot}} = S_H$. 

It is evident that the equilibrium state is $x=v=0$. Hence, if a spontaneous fluctuation  $(\delta x, \delta v)$ from equilibrium occurs, the resulting change of entropy is (recall that the fluctuation must conserve $U_{\text{tot}}$)
\begin{equation}\label{oscillaturzonzonz}
\delta S_{\text{tot}} = -\dfrac{1}{2} \bigg[ \dfrac{\kappa}{T_H} (\delta x)^2 + \dfrac{M}{T_H} (\delta v)^2  \bigg] \, .
\end{equation}
We can use equation \eqref{oscillaturzonzonz} to compute the conjugate ``$\gamma$-variables'' \cite{Onsager_Casimir} explicitly, and use them to write the dynamical equations for $x$ and $v$ in the Onsager-Casimir form:
\begin{equation}
\begin{split}
& \delta \dot{x}= p^{xx} \, \, \dfrac{\kappa \, \delta x}{ T_H} + p^{xv} \, \, \dfrac{M \, \delta v}{T_H} \\
& \delta \dot{v}= p^{vx} \, \, \dfrac{\kappa \, \delta x}{ T_H} + p^{vv} \, \, \dfrac{M \, \delta v}{T_H} \, . \\
\end{split}
\end{equation}
Now we can apply the symmetry principle, and require that
\begin{equation}
p^{xv}=-p^{vx} \, ,
\end{equation}
where the minus comes from the fact that $x$ is even in time, while $v$ is odd in time. In addition, we recall that by definition $v=\dot{x}$, so that
\begin{equation}
p^{xx}=0  \spc p^{xv} = T_H/M \, ,
\end{equation}
and, rewriting $p^{vv}$ as $-\gamma T_H/M$, we obtain
\begin{equation}
\delta \dot{x} = \delta v \spc \delta \dot{v}= - \dfrac{\kappa}{M} \delta x - \gamma \delta v \, .
\end{equation}
Combining these two equations we arrive at
\begin{equation}
\ddot{x} + \gamma \dot{x} + \dfrac{\kappa}{M} x =0 \, ,
\end{equation}
which is the equation of the damped harmonic oscillator.

In conclusion, we have just shown that the equation of the damped harmonic oscillator is the by-product of applying the Onsager-Casimir principle to a system in which both $x$ and its time-derivative $v=\dot{x}$ are regarded as independent non-equilibrium degrees of freedom. Their skew-symmetric coupling is what gives rise to the oscillatory part of the evolution. An analogous mechanism occurs in holographic fluids, where the partners of the fluxes are just proportional to the time-derivatives of the fluxes themselves (we will show this explicitly in the next section).

\section{Diffusion processes in holographic plasmas}
\label{diffusione_holo}

We finally have all the ingredients we need to model diffusion processes in strongly coupled plasmas. In this final section, we will use our formalism to derive a universal (albeit approximate) thermodynamically consistent analogue of Cattaneo's model for diffusion, valid for $\mathcal{N}=4$ SYM holographic fluids. 

\subsection{A universal equation for diffusive phenomena}

Let us consider again the scalar-vector-vector theory, developed in subsection \ref{atvf}. It was meant to describe the propagation of heat in a fluid at finite chemical potential, but it may equivalently be used to describe the diffusion of a generic conserved charge (including transverse momentum, as in the shear channel), provided that one assigns a different interpretation to all the quantities involved. 

If we take $p^j$ to be the partner of $q^j$, as given in \eqref{propriopartner}, we can apply the Onsager-Casimir principle and set
\begin{equation}
\xi_2 =0 \, .
\end{equation} 
Furthermore, coherently with the quantitative analysis of \citet{KovtunHolography2005}, we can impose
\begin{equation}\label{impongoperbene}
\tau = \dfrac{\kappa}{nc_p}= \dfrac{1}{2\pi m T} \spc \quad \chi=2 \, ,
\end{equation}
where the numerical factor $m$ depends on the quantity that is diffusing (e.g. $m=1$ for R-charge  and $m=2$ for transverse momentum). The value of $\kappa/(nc_p)$ is regulated to match the diffusion coefficients computed in \cite{KovtunHolography2005}, while $\tau$ and $\chi$ have been chosen in such a way that the two non-hydrodynamic frequencies \eqref{twofrequncies} take the simple form
\begin{equation}
\omega_\pm = 2\pi m T \bigg( -\dfrac{i}{2} \pm \dfrac{\sqrt{3}}{2}  \bigg)\, .
\end{equation}
This is not the exact value of the poles computed in \cite{KovtunHolography2005}, but it is a reasonable approximation, valid for both R-charge density ($m=1$) and transverse momentum ($m=2$). With the choice of transport coefficients above, the inequalities \eqref{inequol} are saturated, producing the conditions
\begin{equation}
nc_p T \beta_1 =1  \spc \quad \xi_1 =0 \, .
\end{equation}
The first equation implies that in these fluids information propagates at the speed of light. This means that the front (but not necessarily the whole shape) of a localised perturbation drifts with speed $1$. Finally, if we compare \eqref{impongoperbene} with \eqref{ermofk}, we can fix the value of the remaining transport coefficients:
\begin{equation}
\xi_3 = b = \dfrac{2\pi m T}{n c_p} \, .
\end{equation}
Plugging all these results into the system of field equations \eqref{nanni1}-\eqref{nanni3}, introducing the new fields
\begin{equation}
\tilde{q}^j := \dfrac{q^j}{nc_p} \spc \tilde{p}^j := \dfrac{p^j}{nc_p} \, , 
\end{equation}
and setting the scale of the space-time coordinates $x^a$ in such a way that $2\pi mT=1$, we find that all the coefficients disappear, leaving a universal system:
\begin{equation}\label{centosettanatacinqe}
\begin{split}
&  \partial_t \delta T + \partial_j \delta \tilde{q}^j =0 \\
&   \partial_t  \delta \tilde{q}_k + \partial_k \delta T = -  \delta \tilde{p}_k\\
&   \partial_t \delta \tilde{p}_k  =  \delta \tilde{q}_k - \delta \tilde{p}_k \, . \\
\end{split}
\end{equation}
In the homogeneous limit, i.e. in the limit in which the analysis of subsection \ref{Casimiro} is valid, the second equation of \eqref{centosettanatacinqe} becomes $p_k = -\dot{q}_k$. This establishes the thermodynamic analogy with the damped harmonic oscillator, as outlined in subsection \ref{partnerizzodibrutto}.

Combining all the equations of \eqref{centosettanatacinqe}, we obtain a closed equation for the effective dynamics of $T$:
\begin{equation}
\partial_t \, \delta T = (1+\partial_t) \, \partial_a \partial^a \, \delta T \, ,
\end{equation}
which, restoring the constants $2\pi m T$, $c$, $k_B$ and $\hbar$ becomes
\begin{equation}
\dfrac{\partial  }{\partial t} \, \delta T =  \bigg( z + z^2 \dfrac{\partial}{\partial t} \bigg) \bigg( c^2 \dfrac{\partial^2}{\partial x_j \partial x^j}  - \dfrac{\partial^2}{\partial t^2}  \bigg)\, \delta  T \, ,
\end{equation}
with
\begin{equation}
z = \dfrac{\hbar}{2\pi m k_B T} \, .
\end{equation}
This is the equation we were looking for.

\subsection{Model comparison}


In order to have an intuitive idea of how different models behave, it is interesting to  solve numerically their dynamical equations, 
\begin{equation}
\label{pizzaiolo}
\begin{split}
& \partial_t \, \delta T =  \partial_j \partial^j \, \delta T \spc \spc \, \, \, \, (\text{Fick}), \\
& \partial_t \, \delta T =  \partial_a \partial^a \, \delta T \spc \spc \, \, \, (\text{Cattaneo}), \\
& \partial_t \, \delta T = (1+\partial_t) \, \partial_a \partial^a \, \delta T \spc (\text{Holography}), \\
\end{split}
\end{equation}
for a temperature perturbation $ \delta T$. For simplicity, we consider a flat 1+1 spacetime, namely $\partial_j \partial^j=\partial_{xx}$ and $\partial_a \partial^a=\partial_{xx}-\partial_{tt}$. 

In the Fick case we just have to set a single initial condition $\delta T_0(x)$ on the slice $t=0$. The Cattaneo and Holography cases are higher order in time: the initial condition for $\partial_t \delta T$ is set from the heat equation, namely $\partial_t\delta T(0,x)=\partial_{xx}\delta T_0(x)$, for both the Cattaneo and the Holography scenarios. 
Similarly, the  initial condition for $\partial_{tt} \delta T$ is $\partial_{tt}\delta T(0,x)=\partial_{xxxx}\delta T_0(x)$ for the Holography case.

\begin{figure*}
    \centering
    \includegraphics[width=0.98\textwidth]{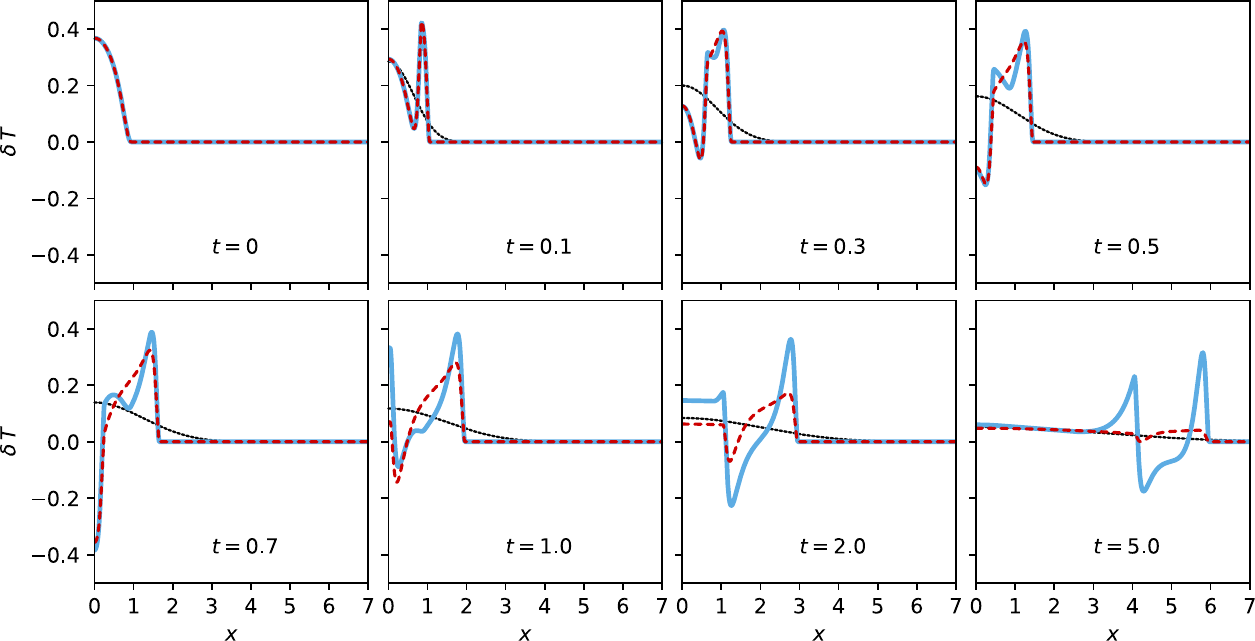}
    \caption{Evolution of the perturbation $\delta T$ with compact support for the three models in \eqref{pizzaiolo}: ``Fick'' (thin black line), ``Cattaneo'' (red, dashed) and ``Holography'' (thick blue line).
    Only the $x>0$ region is shown as the evolution is symmetric. At $t=0,1,3,5$ the front of the Cattaneo and Holography models is respectively located at $x=1,2,4,6$, meaning that the signal propagates at the speed of light. The relaxation effect is visible in the last frame: the three models tend to be more and more indistinguishable close to the origin, as the amplitude of the front travels and decreases in amplitude. }
    \label{fig:comparison}
\end{figure*}

Numerical comparison of the three models is given in Fig \ref{fig:comparison}, where several snapshots of the evolution are shown. To check that signals are indeed subliminal, we take the usual smooth bump function as our initial condition,  
\begin{equation}
    \delta T_0(x) = \Theta(1-x^2) \, \exp{\bigg( \dfrac{1}{x^2-1} \bigg)} \, , 
\end{equation}
since it has compact support ($\Theta$ is the Heaviside step function). In Fig \eqref{fig:comparison} we can see that the boundary of the support of $\delta T(x,t)$ propagates at the speed of light for the Cattaneo and Holography models: despite the fact that  from a numerical point of view it's not immediate to identify the exact location of the support's boundary, we see that the large spatial gradients of $ \delta T_0$ close to $x=\pm 1$ quickly give rise to a propagating front that is easy to follow. While the amplitude of such front quickly goes to zero for the Cattaneo case, we find it to be more long-lived for the Holography model. Eventually, all the three models have to relax to the trivial solution $\delta T=0$. 

\begin{figure}
    \centering
    \includegraphics[width=0.47\textwidth]{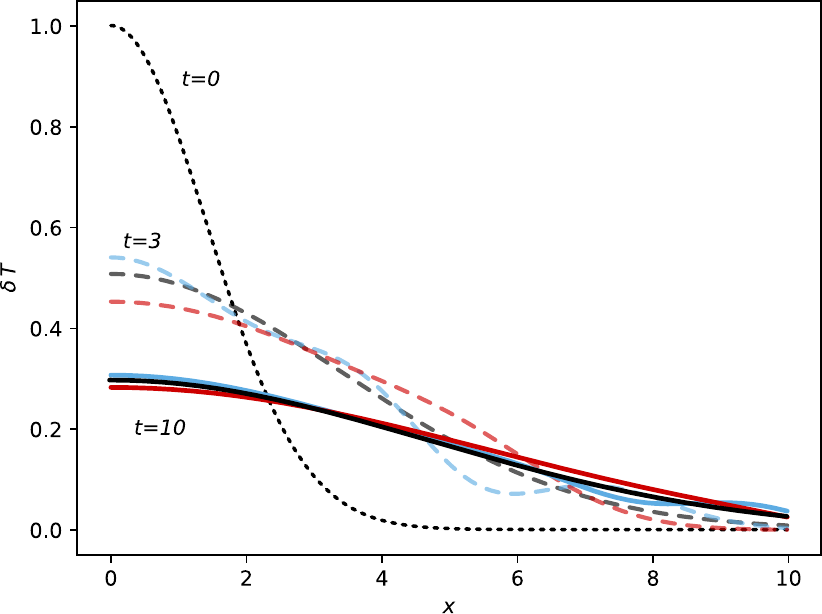}
    \caption{
    Relaxation of $\delta T$ for the three models in \eqref{pizzaiolo}: ``Fick'' (black), ``Cattaneo'' (red) and ``Holography'' (blue).
    The initial condition (black, dotted) at $t=0$ is the same for all the three models and is a Gaussian of variance 2 (only the $x>0$ region is shown). 
    The dashed curves correspond to $t=3$ and the solid ones to $t=10$: as time advances the models become indistinguishable. }
    \label{fig:relax}
\end{figure}

To test the relaxation property of the system, it is more convenient  to choose an initial condition with smaller gradients, so that the relaxation will be already evident at early times. In Fig \eqref{fig:relax} we take $\delta T_0$ to be  a Gaussian of variance 2. 
In fact, with this numerical test we are not interested in checking the causality property, so that now the support of the initial data can be non-compact. Since we now have no regions with large gradients in the initial condition, we see no clear propagating fronts as in the previous numerical experiment. However, it is evident that the three models relax to the usual Gaussian solution of the non-relativistic heat equation, the differences being smaller than the  $10\%$ already for $t \sim 10 $.

\section{Conclusions}
\label{concussioni}

We have developed a classical field theory for modelling dissipative fluids close to equilibrium that is ``quasi-hydrodynamic'' \citep{Grozdanov2019}, in the sense that the model includes the usual degrees of freedom of a perfect fluid plus a finite number of quasi-conserved local degrees of freedom. 
These quasi-conserved internal quantities are treated as genuine hydrodynamic fields: within this approach, the non-hydrodynamic sector is physical and describes the local relaxation/oscillation of quasi-conserved quantities \cite{GavassinoFronntiers2021}. 

Our quasi-hydrodynamic formalism arises from an action principle, constructed using the method of the doubling of variables \cite{Galley2013,Galley2014}. The associated Euler-Lagrange equations have the appealing mathematical structure envisaged by \citet{Geroch_Lindblom_1991_causal}, which implies that the theories derived using our approach are symmetric-hyperbolic, causal, and Lyapunov stable by construction. 
\\
\\
Thanks to this construction, the contact with non-equilibrium thermodynamics is straightforward and the resulting quasi-hydrodynamic models are automatically consistent with:
\begin{enumerate}
    \item[-] the second law of thermodynamics,
    \item[-] the Gibbs stability criterion,
    \item[-] the Onsager-Casimir symmetry principle.
\end{enumerate}
This is valid for both the hydrodynamic and the non-hydrodynamic sectors of the quasi-hydrodynamic model.

Starting from our construction, we have seen that the Cattaneo model for heat conduction \cite{cattaneo1958} and the Israel-Stewart theory for dissipation \cite{Israel_Stewart_1979} are recovered for some particular choices of the effective fields. 
Furthermore, using the invariance of the field equations under field redefinitions, we could use the formalism to prove that the linearised Israel-Stewart theory in the Eckart frame \cite{Hishcock1983} and the linearised Israel-Stewart theory in the Landau frame \cite{OlsonLifsh1990} are the same theory. 
We proved this \textit{exact} mathematical equivalence explicitly, through a change of variables. As a corollary, we could show that the stability/causality conditions of \citet{OlsonLifsh1990} are just a rewrite of those of \citet{Hishcock1983}. We verified this equivalence also explicitly. 
The most important novelty of our approach is the possibility of producing theories with arbitrary non-hydrodynamic sector, including that of strongly coupled plasmas (e.g. the holographic dual of the $\mathcal{N}=4$ super-symmetric Yang-Mills theory \cite{Heller2014}). 
By applying the Onsager-Casimir symmetry principle, we could show that the non-hydrodynamic degrees of freedom of such fluids are \textit{thermodynamically} (and not only dynamically) equivalent to those of a damped harmonic oscillator. This means that, in a fluid of this kind, the time-derivative of the viscous stress, namely $\braket{u^m \nabla_m \Pi_{ab}}$, is a quasi-conserved quantity (just like $\Pi_{ab}$ itself \cite{Grozdanov2019}), so that $\Pi_{ab}$ acquires an effective ``inertia'' \cite{Heimburg2017}, which forces the viscous stress to oscillate around zero.

Finally, we stress that some physical systems may display non-hydrodynamic modes that come in the form of branch cuts, e.g. \citep{moore2018cuts,Romatschke2019PhRvL,Perna_21_linear_cuts}.
This poses a difficulty that probably cannot be addressed within our framework. In fact, 
\citet{Denicol_Relaxation_2011} have shown that it is always possible to take a Green function with an arbitrary number of frequency poles at $k=0$, and extract from it a differential equation in time that reproduces such frequency poles. In the presence of cuts, this procedure may need some modification. Ultimately, it is not clear if a branch cut can be reproduced with a differential equation involving a finite number of time-derivatives. From the perspective of our approach, infinite time-derivatives would correspond to infinite algebraic degrees of freedom, and hence to an infinite number of effective fields. Therefore, the problem of reproducing branch cuts within our framework is still open, and it constitutes an interesting subject for future investigations.

\section*{Acknowledgements}

We acknowledge support from the Polish National Science Centre grants OPUS 2129/33/B/ST9/00942. Partial support comes from PHAROS, COST Action CA16214. We thank J. Noronha for introducing us to the problem of reconciling EIT with holography. LG is also grateful to D. Tsang and G. Torrieri for introducing him to the action principle of dissipative systems.

\appendix

\section{Non-diagonalisable case}\label{nondiag}

Let $\delta\tilde{\boldsymbol{\varphi}}$ be the $\mathfrak{D}$-dimensional array whose components are $\delta \tilde{\varphi}^D$ and $\tilde{\Xi}$ the $\mathfrak{D}\times \mathfrak{D}$ matrix whose elements are $\tilde{\Xi}_{CD}$. Then, the system of field equation \eqref{semplicio} can be rewritten in the compact form
\begin{equation}
\partial_t \, \delta\tilde{\boldsymbol{\varphi}} = -\tilde{\Xi} \, \delta\tilde{\boldsymbol{\varphi}} \, .
\end{equation}
The general solution of this equation
\begin{equation}\label{generalsolutionnonhydro}
 \delta\tilde{\boldsymbol{\varphi}} (t) = e^{-\tilde{\Xi} \, t} \delta\tilde{\boldsymbol{\varphi}} (0) \, .
\end{equation}
Now, let us focus on the case in which $\tilde{\Xi}$ is non-diagonalisable. Then, the Jordan-Chevalley decomposition guarantees that we can write $\tilde{\Xi}$ as 
\begin{equation}
\tilde{\Xi} = \hat{D} + \hat{N} \, ,
\end{equation}
where $\hat{D}$ and $\hat{N}$ are two (possibly complex) $\mathfrak{D}\times \mathfrak{D}$ matrices such that
\begin{itemize}
\item $\hat{D}$ is diagonalizable: there is a basis of (possibly complex) arrays $\textbf{Y}_{(n)}$ such that $\hat{D} \textbf{Y}_{(n)} = i\omega_n \textbf{Y}_{(n)}$,
\item $\hat{N}$ is nilpotent: there is a finite positive integer $Q$ such that $\hat{N}^Q=0$,
\item $\hat{D}$ and $\hat{N}$ commute: $\big[ \hat{D} , \hat{N} \big]=0$.
\end{itemize}
Given these properties, the general solution of \eqref{generalsolutionnonhydro} can be decomposed as:
\begin{equation}
\delta\tilde{\boldsymbol{\varphi}} (t) = \sum_{q=0}^{Q-1} \, \sum_{n=1}^{\mathfrak{D}}\, (-1)^q\, \dfrac{c_n}{q!} \, t^q \,e^{-i\omega_n t} \, \hat{N}^q \, \textbf{Y}_{(n)} \, .
\end{equation}
This solution is structurally different from \eqref{twentitree}-\eqref{twantialbero}, because of the presence of the factors $t^q$. Such factors will appear also in the retarded linear-response Green's functions of the non-hydrodynamic sector, which will take the generic form (in the homogeneous limit)
\begin{equation}
\mathcal{G}_R (\tau) =\Theta(\tau) \sum_{nq} \dfrac{a_{nq}}{q!} \tau^q e^{-i\omega_n \tau} \, .
\end{equation}
In the frequency space, it becomes
\begin{equation}
\mathcal{G}_R (\omega) = \sum_{nq} \dfrac{a_{nq}}{[-i(\omega-\omega_n)]^{q+1}} \, .
\end{equation}
Thus, the direct signature of the non-diagonalizability of the matrix $\tilde{\Xi}_{CD}$ is the presence of higher-order poles in the retarded Green's functions.   

\section{Handling the constraints of the shear stress}
\label{LaShear}

Let us focus on a simplified problem. We consider only the fields $(u^k,\Pi^{jk})$ and assume that 
\begin{equation}\label{puffo}
E^0 =  \dfrac{1}{2T} \bigg[  (\rho+P)\delta u^j \delta u_j +\beta_2 \delta \Pi^{jk} \delta \Pi_{jk}   \bigg] 
\end{equation}
\begin{equation}\label{puffi}
E^j =  \dfrac{\delta \Pi^{jk} \delta u_k}{T} \spc \sigma =  \dfrac{\delta \Pi^{jk} \delta \Pi_{jk}}{2\eta T} \, .  
\end{equation}
We need to compute the field equations using \eqref{themethoddizzzuz}, but we know that there are the constraints $\delta \Pi^{[jk]}=\delta \Pi\indices{^j _j}=0$. This means that, while performing the derivatives, we cannot treat all the components of $\delta \Pi^{jk}$ as independent. How do we account for such constraints?

The trick is to introduce 5 independent unconstrained functions $Z^A$ (5 is the number of independent components of the stress tensor), which characterise the state of the stress tensor completely, namely $\Pi^{jk}=\Pi^{jk}(Z^A)$. For example, one may have
\begin{equation}\label{akj}
\Pi^{jk} =
\begin{bmatrix}
   Z^1+Z^2 & Z^3 & Z^4  \\
  Z^3 & Z^1-Z^2 & Z^5\\
  Z^4 & Z^5 & -2Z^1  \\
\end{bmatrix} \, .
\end{equation}
Then, we can make the change of variables
\begin{equation}\label{daPiaZ}
\delta \Pi^{jk} = \dfrac{\partial \Pi^{jk}}{\partial Z^A} \delta Z^A \, .
\end{equation}
Given that the constraints $\delta \Pi^{[jk]}=\delta \Pi\indices{^j _j}=0$ must be verified for any possible value of $\delta Z^A$, it follows that 
\begin{equation}
\dfrac{\partial \Pi^{[jk]}}{\partial Z^A} =\dfrac{\partial \Pi^j_j}{\partial Z^A}=0 \, ,
\end{equation}
as we can see in the examples below, computed from \eqref{akj}:
\begin{equation}\label{exampio}
\dfrac{\partial\Pi^{jk}}{\partial Z^2} =
\begin{bmatrix}
   1 & 0 & 0  \\
  0 & -1 & 0\\
  0 & 0 & 0  \\
\end{bmatrix} \spc
\dfrac{\partial\Pi^{jk}}{\partial Z^3} =
\begin{bmatrix}
   0 & 1 & 0  \\
  1 & 0 & 0\\
  0 & 0 & 0  \\
\end{bmatrix} \, .
\end{equation}
Using the change of variables \eqref{daPiaZ}, equations \eqref{puffo} and \eqref{puffi} become
\begin{equation}\label{puffo2}
E^0 =  \dfrac{1}{2T} \bigg[  (\rho+P)\delta u^j \delta u_j +\beta_2 \dfrac{\partial \Pi^{jk}}{\partial Z^A} \dfrac{\partial \Pi_{jk}}{\partial Z^B} \delta Z^A \delta Z^B \bigg] 
\end{equation}
\begin{equation}\label{puffi2}
E^j = \dfrac{\partial \Pi^{jk}}{\partial Z^A} \dfrac{\delta Z^A\delta u_k}{T} \quad \, \, \,  \sigma = \dfrac{\partial \Pi^{jk}}{\partial Z^A} \dfrac{\partial \Pi_{jk}}{\partial Z^B}  \dfrac{ \delta Z^A \delta Z^B}{2\eta T} \, .  
\end{equation}
Now our degrees of freedom are just $(u^k,Z^A)$, which are unconstrained. Therefore, we are free to use \eqref{themethoddizzzuz} to compute the field equations:
\begin{flalign}\label{settantasei2}
\boxed{u^{k}} \quad (\rho+P)\partial_t \delta u_k +  \dfrac{\partial \Pi^j_k}{\partial Z^A}\partial_j \delta Z^A =0 &&
\end{flalign}
\begin{flalign}\label{settantasei3}
\boxed{Z^A} \quad \dfrac{\partial \Pi^{jk}}{\partial Z^A} \bigg[ \dfrac{\partial \Pi_{jk}}{\partial Z^B} \bigg(\beta_2 \partial_t  +\dfrac{1}{2\eta } \bigg) \delta Z^B +  \partial_j \delta u_k  \bigg]=0 \, . &&
\end{flalign}
We may just stop here. However, typically one wants the equations to be written directly in terms of the shear stresses. We can reabsorb the variables $\delta Z^A$ into $\delta \Pi^{jk}$ using \eqref{daPiaZ} (recall that the matrix of partial derivatives is a background constant). Furthermore, we see from the examples \eqref{exampio} that the only role of the matrix $\partial \Pi^{jk}/\partial Z^A$ in equation \eqref{settantasei3} is to extract the symmetric traceless part of the term in the square brackets, so that our field equations can be equivalently rewritten as follows:
\begin{equation}
\begin{split}
& (\rho+P)\partial_t \delta u_k +  \partial_j \delta \Pi^j_k =0 \\
& \beta_2 \partial_t \delta \Pi_{jk} + \braket{\partial_j \delta u_k} = -\dfrac{\delta \Pi_{jk}}{2\eta} \\
\end{split}
\end{equation}
The reader can verify that if we computed the field equations directly from \eqref{puffo}-\eqref{puffi}, using \eqref{themethoddizzzuz} and treating $\Pi^{jk}$ as an unconstrained variable, we would obtain exactly the same equations above, with the only difference that, in the second equation, $\braket{\partial_j \delta u_k} $ would be replaced by $\partial_j \delta u_k$. Such equation would be clearly incompatible with the constraint of symmetry and tracelessness of the shear stresses. Our trick of using the degrees of freedom $Z^A$ enforces the constraints $\delta \Pi^{[jk]}=\delta \Pi\indices{^j _j}=0$ by construction.

\section{Fourier analysis of the scalar-vector-vector theory for heat conduction}

Working in the Fourier space, we study the dynamical properties of the model for heat conduction presented in subsection~\ref{atvf}. For simplicity, we work in 1+1 dimensions.

\subsection{Stability and causality}\label{modebymode}

We consider equations \eqref{nanni1}-\eqref{nanni3} and assume a space-time dependence of the kind
\begin{equation}\label{algebrizzo}
\delta \varphi^A(t,x) = \delta \varphi^A(0,0) \, e^{\Gamma t +ikx} \, ,
\end{equation}
where $\Gamma = -i\omega \in \mathbb{C}$ and $k \in \mathbb{R}$. The stability requirement is $\text{Re} \, \Gamma \leq 0$, see equation \eqref{crittuz}. 

Using \eqref{algebrizzo}, the field equations become algebraic. The dispersion relations $\Gamma=\Gamma(k)$, associated with the field equations \eqref{nanni1}-\eqref{nanni3}, are given by the condition \cite{Hishcock1988}
\begin{equation}\label{dettoxic}
\text{det}
\begin{bmatrix}
   nc_p \Gamma & ik & 0  \\
  ik & T\beta_1 \Gamma+\xi_1 & \xi_2+b \\
  0 & \xi_2 -b & T\beta_1 \Gamma + \xi_3  \\
\end{bmatrix} =0 \, .
\end{equation}
The determinant above is a third-order polynomial in $\Gamma$, so that we are left with the following root-finding problem:
\begin{equation}\label{polluz}
A_3 \Gamma^3 + A_2 \Gamma^2 + A_1 \Gamma + A_0 =0
\end{equation}
with
\begin{equation}
\begin{split}
A_0 &=\xi_3 k^2 \\
A_1 &=nc_p (\xi_1 \xi_3-\xi_2^2 +b^2) + T\beta_1 k^2 \\
A_2 &=nTc_p \beta_1 (\xi_1 + \xi_3) \\
A_3 &=nT^2 c_p \beta_1^2 \, . \\
\end{split}
\end{equation}
From conditions \eqref{ubf}-\eqref{ubfff}, we see that all the coefficients $A_i$ are non-negative. Hence, there are no real positive roots of \eqref{polluz}. 
If we decompose $\Gamma$ into its real and imaginary parts, $\Gamma = \Gamma_R - i \omega_R$ (with $\Gamma_R , \omega_R \in \mathbb{R}$), and we assume that $\omega_R \neq 0$, then  $\Gamma_R$ must be solution of \citep{Hishcock1988}
\begin{equation}\label{gagaga}
8 A_3^2 \Gamma_R^3+8A_2 A_3 \Gamma_R^2+2(A_1 A_3+A_2^2)\Gamma_R+A_1 A_2 - A_0 A_3 =0.
\end{equation}
One can verify directly that $A_1 A_2 - A_0 A_3 >0$. Hence, the coefficients of each power of $\Gamma_R$ in \eqref{gagaga} are all positive, meaning that there is no positive root for $\Gamma_R$. In conclusion, all the modes are damped and the system is stable in the fluid's rest frame.

Let us move to the causality condition. We need to take the limit $k \rightarrow \infty$, in which \eqref{dettoxic} becomes
\begin{equation}
\text{det}
\begin{bmatrix}
   nc_p \Gamma & ik & 0  \\
  ik & T\beta_1 \Gamma & 0 \\
  0 & 0 & T\beta_1 \Gamma  \\
\end{bmatrix} \approx 0 \, .
\end{equation}
The only non-trivial solutions are plane waves with dispersion relation (use $\omega = i \Gamma$)
\begin{equation}
\omega = \pm \dfrac{k}{\sqrt{nTc_p \beta_1}} \, .
\end{equation}
The causality condition \eqref{ubff} is, therefore, equivalent to the second condition of \eqref{crittuz}, which is the standard mode-based causality condition.

Finally, from theorem 2 of \citet{GavassinoSuperlum2021}, we know that, since the theory is stable in the fluid's rest frame, and it is causal, it is also stable in any boosted frame, completing our analysis.

\subsection{Small wave-vector limit}\label{smallk}

Let us work in the limit $k \rightarrow 0$. We can expand the dispersion relations $\Gamma=\Gamma(k)$ to the second order in $k$,
\begin{equation}
\Gamma = \Gamma_0 + \Gamma_2 k^2,
\end{equation}
where the first-order term vanishes by symmetry. Inserting this expansion into \eqref{polluz}, and truncating to the second order in $k$, we obtain an equation of the form $B_0+B_2 k^2=0$, where $B_0$ and $B_2$ are some polynomials in $\Gamma_0$ and $\Gamma_2$. 
Given that this equation must be valid for any small $k$, we must impose $B_0 =B_2 =0$. 
This allows us to determine both $\Gamma_0$ and $\Gamma_2$. The solutions, keeping only the leading order in $k$, are (recall that $\omega=i\Gamma$)
\begin{equation}
\begin{split}
& \omega_\pm = \dfrac{-i(\xi_1+ \xi_3) \pm \sqrt{4(b^2-\xi_2^2)-(\xi_1-\xi_3)^2}}{2T\beta_1} \\
& \omega_3 = \dfrac{-i\xi_3 k^2}{nc_p (\xi_1 \xi_3 -\xi_2^2+b^2)}  
\end{split}
\end{equation}
The two frequencies $\omega_+$ and $\omega_-$ belong to the non-hydrodynamic sector. 
The relative modes are governed by a differential equation of the form \eqref{ewmodkc} only if $\omega_\pm$ are related to $\tau$ and $\chi$ by means of \eqref{twofrequncies}. This allows us to compute $\chi$ and $\tau$ directly: the result is equation \eqref{ermofk}. Analogously, the  frequency $\omega_3$ lives in the hydrodynamic sector and is consistent with \eqref{diffondo} only if \eqref{kappuccio} holds.

\bibliography{Biblio}

\begin{thebibliography}{63}%
\makeatletter
\providecommand \@ifxundefined [1]{%
 \@ifx{#1\undefined}
}%
\providecommand \@ifnum [1]{%
 \ifnum #1\expandafter \@firstoftwo
 \else \expandafter \@secondoftwo
 \fi
}%
\providecommand \@ifx [1]{%
 \ifx #1\expandafter \@firstoftwo
 \else \expandafter \@secondoftwo
 \fi
}%
\providecommand \natexlab [1]{#1}%
\providecommand \enquote  [1]{``#1''}%
\providecommand \bibnamefont  [1]{#1}%
\providecommand \bibfnamefont [1]{#1}%
\providecommand \citenamefont [1]{#1}%
\providecommand \href@noop [0]{\@secondoftwo}%
\providecommand \href [0]{\begingroup \@sanitize@url \@href}%
\providecommand \@href[1]{\@@startlink{#1}\@@href}%
\providecommand \@@href[1]{\endgroup#1\@@endlink}%
\providecommand \@sanitize@url [0]{\catcode `\\12\catcode `\$12\catcode
  `\&12\catcode `\#12\catcode `\^12\catcode `\_12\catcode `\%12\relax}%
\providecommand \@@startlink[1]{}%
\providecommand \@@endlink[0]{}%
\providecommand \url  [0]{\begingroup\@sanitize@url \@url }%
\providecommand \@url [1]{\endgroup\@href {#1}{\urlprefix }}%
\providecommand \urlprefix  [0]{URL }%
\providecommand \Eprint [0]{\href }%
\providecommand \doibase [0]{http://dx.doi.org/}%
\providecommand \selectlanguage [0]{\@gobble}%
\providecommand \bibinfo  [0]{\@secondoftwo}%
\providecommand \bibfield  [0]{\@secondoftwo}%
\providecommand \translation [1]{[#1]}%
\providecommand \BibitemOpen [0]{}%
\providecommand \bibitemStop [0]{}%
\providecommand \bibitemNoStop [0]{.\EOS\space}%
\providecommand \EOS [0]{\spacefactor3000\relax}%
\providecommand \BibitemShut  [1]{\csname bibitem#1\endcsname}%
\let\auto@bib@innerbib\@empty
\bibitem [{\citenamefont {{Kadanoff}}\ and\ \citenamefont
  {{Martin}}(1963)}]{Kadanoff1963}%
  \BibitemOpen
  \bibfield  {author} {\bibinfo {author} {\bibfnamefont {L.~P.}\ \bibnamefont
  {{Kadanoff}}}\ and\ \bibinfo {author} {\bibfnamefont {P.~C.}\ \bibnamefont
  {{Martin}}},\ }\href {\doibase 10.1016/0003-4916(63)90078-2} {\bibfield
  {journal} {\bibinfo  {journal} {Annals of Physics}\ }\textbf {\bibinfo
  {volume} {24}},\ \bibinfo {pages} {419} (\bibinfo {year} {1963})}\BibitemShut
  {NoStop}%
\bibitem [{\citenamefont {{Peliti}}(2011)}]{peliti_book}%
  \BibitemOpen
  \bibfield  {author} {\bibinfo {author} {\bibfnamefont {L.}~\bibnamefont
  {{Peliti}}},\ }\href@noop {} {\emph {\bibinfo {title} {{Statistical Mechanics
  in a Nutshell}}}},\ In a nutshell\ (\bibinfo  {publisher} {Princeton
  University Press},\ \bibinfo {year} {2011})\BibitemShut {NoStop}%
\bibitem [{\citenamefont {{Czajka}}\ and\ \citenamefont
  {{Jeon}}(2017)}]{Kubos2017}%
  \BibitemOpen
  \bibfield  {author} {\bibinfo {author} {\bibfnamefont {A.}~\bibnamefont
  {{Czajka}}}\ and\ \bibinfo {author} {\bibfnamefont {S.}~\bibnamefont
  {{Jeon}}},\ }\href {\doibase 10.1103/PhysRevC.95.064906} {\bibfield
  {journal} {\bibinfo  {journal} {\prc}\ }\textbf {\bibinfo {volume} {95}},\
  \bibinfo {eid} {064906} (\bibinfo {year} {2017})},\ \Eprint
  {http://arxiv.org/abs/1701.07580} {arXiv:1701.07580 [nucl-th]} \BibitemShut
  {NoStop}%
\bibitem [{\citenamefont {{Geroch}}(1995)}]{Geroch1995}%
  \BibitemOpen
  \bibfield  {author} {\bibinfo {author} {\bibfnamefont {R.}~\bibnamefont
  {{Geroch}}},\ }\href {\doibase 10.1063/1.530958} {\bibfield  {journal}
  {\bibinfo  {journal} {Journal of Mathematical Physics}\ }\textbf {\bibinfo
  {volume} {36}},\ \bibinfo {pages} {4226} (\bibinfo {year}
  {1995})}\BibitemShut {NoStop}%
\bibitem [{\citenamefont {{Lindblom}}(1996)}]{LindblomRelaxation1996}%
  \BibitemOpen
  \bibfield  {author} {\bibinfo {author} {\bibfnamefont {L.}~\bibnamefont
  {{Lindblom}}},\ }\href {\doibase 10.1006/aphy.1996.0036} {\bibfield
  {journal} {\bibinfo  {journal} {Annals of Physics}\ }\textbf {\bibinfo
  {volume} {247}},\ \bibinfo {pages} {1} (\bibinfo {year} {1996})},\ \Eprint
  {http://arxiv.org/abs/gr-qc/9508058} {arXiv:gr-qc/9508058 [gr-qc]}
  \BibitemShut {NoStop}%
\bibitem [{\citenamefont {{Policastro}}\ \emph {et~al.}(2002)\citenamefont
  {{Policastro}}, \citenamefont {{Son}},\ and\ \citenamefont
  {{Starinets}}}]{Policastro2002}%
  \BibitemOpen
  \bibfield  {author} {\bibinfo {author} {\bibfnamefont {G.}~\bibnamefont
  {{Policastro}}}, \bibinfo {author} {\bibfnamefont {D.~T.}\ \bibnamefont
  {{Son}}}, \ and\ \bibinfo {author} {\bibfnamefont {A.~O.}\ \bibnamefont
  {{Starinets}}},\ }\href {\doibase 10.1088/1126-6708/2002/09/043} {\bibfield
  {journal} {\bibinfo  {journal} {Journal of High Energy Physics}\ }\textbf
  {\bibinfo {volume} {2002}},\ \bibinfo {eid} {043} (\bibinfo {year} {2002})},\
  \Eprint {http://arxiv.org/abs/hep-th/0205052} {arXiv:hep-th/0205052 [hep-th]}
  \BibitemShut {NoStop}%
\bibitem [{\citenamefont {{Kovtun}}(2012)}]{kovtun_lectures_2012}%
  \BibitemOpen
  \bibfield  {author} {\bibinfo {author} {\bibfnamefont {P.}~\bibnamefont
  {{Kovtun}}},\ }\href {\doibase 10.1088/1751-8113/45/47/473001} {\bibfield
  {journal} {\bibinfo  {journal} {Journal of Physics A Mathematical General}\
  }\textbf {\bibinfo {volume} {45}},\ \bibinfo {eid} {473001} (\bibinfo {year}
  {2012})},\ \Eprint {http://arxiv.org/abs/1205.5040} {arXiv:1205.5040
  [hep-th]} \BibitemShut {NoStop}%
\bibitem [{\citenamefont {{Glorioso}}\ and\ \citenamefont
  {{Liu}}(2018)}]{Glorioso2018}%
  \BibitemOpen
  \bibfield  {author} {\bibinfo {author} {\bibfnamefont {P.}~\bibnamefont
  {{Glorioso}}}\ and\ \bibinfo {author} {\bibfnamefont {H.}~\bibnamefont
  {{Liu}}},\ }\href@noop {} {\bibfield  {journal} {\bibinfo  {journal} {arXiv
  e-prints}\ ,\ \bibinfo {eid} {arXiv:1805.09331}} (\bibinfo {year} {2018})},\
  \Eprint {http://arxiv.org/abs/1805.09331} {arXiv:1805.09331 [hep-th]}
  \BibitemShut {NoStop}%
\bibitem [{\citenamefont {Denicol}\ \emph {et~al.}(2011)\citenamefont
  {Denicol}, \citenamefont {Noronha}, \citenamefont {Niemi},\ and\
  \citenamefont {Rischke}}]{Denicol_Relaxation_2011}%
  \BibitemOpen
  \bibfield  {author} {\bibinfo {author} {\bibfnamefont {G.~S.}\ \bibnamefont
  {Denicol}}, \bibinfo {author} {\bibfnamefont {J.}~\bibnamefont {Noronha}},
  \bibinfo {author} {\bibfnamefont {H.}~\bibnamefont {Niemi}}, \ and\ \bibinfo
  {author} {\bibfnamefont {D.~H.}\ \bibnamefont {Rischke}},\ }\href {\doibase
  10.1103/PhysRevD.83.074019} {\bibfield  {journal} {\bibinfo  {journal} {Phys.
  Rev. D}\ }\textbf {\bibinfo {volume} {83}},\ \bibinfo {pages} {074019}
  (\bibinfo {year} {2011})}\BibitemShut {NoStop}%
\bibitem [{\citenamefont {Gavassino}\ \emph {et~al.}(2021)\citenamefont
  {Gavassino}, \citenamefont {Antonelli},\ and\ \citenamefont
  {Haskell}}]{BulkGavassino}%
  \BibitemOpen
  \bibfield  {author} {\bibinfo {author} {\bibfnamefont {L.}~\bibnamefont
  {Gavassino}}, \bibinfo {author} {\bibfnamefont {M.}~\bibnamefont
  {Antonelli}}, \ and\ \bibinfo {author} {\bibfnamefont {B.}~\bibnamefont
  {Haskell}},\ }\href {\doibase 10.1088/1361-6382/abe588} {\bibfield  {journal}
  {\bibinfo  {journal} {Classical and Quantum Gravity}\ }\textbf {\bibinfo
  {volume} {38}},\ \bibinfo {pages} {075001} (\bibinfo {year}
  {2021})}\BibitemShut {NoStop}%
\bibitem [{\citenamefont {Gavassino}\ and\ \citenamefont
  {Antonelli}(2021)}]{GavassinoFronntiers2021}%
  \BibitemOpen
  \bibfield  {author} {\bibinfo {author} {\bibfnamefont {L.}~\bibnamefont
  {Gavassino}}\ and\ \bibinfo {author} {\bibfnamefont {M.}~\bibnamefont
  {Antonelli}},\ }\href {\doibase 10.3389/fspas.2021.686344} {\bibfield
  {journal} {\bibinfo  {journal} {Front. Astron. Space Sci.}\ }\textbf
  {\bibinfo {volume} {8}},\ \bibinfo {pages} {686344} (\bibinfo {year}
  {2021})},\ \Eprint {http://arxiv.org/abs/2105.15184} {arXiv:2105.15184
  [gr-qc]} \BibitemShut {NoStop}%
\bibitem [{\citenamefont {Hiscock}\ and\ \citenamefont
  {Lindblom}(1983)}]{Hishcock1983}%
  \BibitemOpen
  \bibfield  {author} {\bibinfo {author} {\bibfnamefont {W.~A.}\ \bibnamefont
  {Hiscock}}\ and\ \bibinfo {author} {\bibfnamefont {L.}~\bibnamefont
  {Lindblom}},\ }\href {\doibase https://doi.org/10.1016/0003-4916(83)90288-9}
  {\bibfield  {journal} {\bibinfo  {journal} {Annals of Physics}\ }\textbf
  {\bibinfo {volume} {151}},\ \bibinfo {pages} {466 } (\bibinfo {year}
  {1983})}\BibitemShut {NoStop}%
\bibitem [{\citenamefont {Israel}\ and\ \citenamefont
  {Stewart}(1979)}]{Israel_Stewart_1979}%
  \BibitemOpen
  \bibfield  {author} {\bibinfo {author} {\bibfnamefont {W.}~\bibnamefont
  {Israel}}\ and\ \bibinfo {author} {\bibfnamefont {J.}~\bibnamefont
  {Stewart}},\ }\href {\doibase https://doi.org/10.1016/0003-4916(79)90130-1}
  {\bibfield  {journal} {\bibinfo  {journal} {Annals of Physics}\ }\textbf
  {\bibinfo {volume} {118}},\ \bibinfo {pages} {341 } (\bibinfo {year}
  {1979})}\BibitemShut {NoStop}%
\bibitem [{\citenamefont {{Denicol}}\ \emph {et~al.}(2012)\citenamefont
  {{Denicol}}, \citenamefont {{Moln{\'a}r}}, \citenamefont {{Niemi}},\ and\
  \citenamefont {{Rischke}}}]{DMNR2012}%
  \BibitemOpen
  \bibfield  {author} {\bibinfo {author} {\bibfnamefont {G.~S.}\ \bibnamefont
  {{Denicol}}}, \bibinfo {author} {\bibfnamefont {E.}~\bibnamefont
  {{Moln{\'a}r}}}, \bibinfo {author} {\bibfnamefont {H.}~\bibnamefont
  {{Niemi}}}, \ and\ \bibinfo {author} {\bibfnamefont {D.~H.}\ \bibnamefont
  {{Rischke}}},\ }\href {\doibase 10.1140/epja/i2012-12170-x} {\bibfield
  {journal} {\bibinfo  {journal} {European Physical Journal A}\ }\textbf
  {\bibinfo {volume} {48}},\ \bibinfo {eid} {170} (\bibinfo {year} {2012})},\
  \Eprint {http://arxiv.org/abs/1206.1554} {arXiv:1206.1554 [nucl-th]}
  \BibitemShut {NoStop}%
\bibitem [{\citenamefont {{Chabanov}}\ \emph {et~al.}(2021)\citenamefont
  {{Chabanov}}, \citenamefont {{Rezzolla}},\ and\ \citenamefont
  {{Rischke}}}]{Chab2021}%
  \BibitemOpen
  \bibfield  {author} {\bibinfo {author} {\bibfnamefont {M.}~\bibnamefont
  {{Chabanov}}}, \bibinfo {author} {\bibfnamefont {L.}~\bibnamefont
  {{Rezzolla}}}, \ and\ \bibinfo {author} {\bibfnamefont {D.~H.}\ \bibnamefont
  {{Rischke}}},\ }\href {\doibase 10.1093/mnras/stab1384} {\bibfield  {journal}
  {\bibinfo  {journal} {\mnras}\ }\textbf {\bibinfo {volume} {505}},\ \bibinfo
  {pages} {5910} (\bibinfo {year} {2021})},\ \Eprint
  {http://arxiv.org/abs/2102.10419} {arXiv:2102.10419 [gr-qc]} \BibitemShut
  {NoStop}%
\bibitem [{\citenamefont {Cattaneo}(1958)}]{cattaneo1958}%
  \BibitemOpen
  \bibfield  {author} {\bibinfo {author} {\bibfnamefont {C.}~\bibnamefont
  {Cattaneo}},\ }\href {https://books.google.pl/books?id=mHGeQwAACAAJ} {\emph
  {\bibinfo {title} {Sur une forme de l'{\'e}quation de la chaleur
  {\'e}liminant le paradoxe d'une propagation instantan{\'e}e}}},\ Comptes
  rendus hebdomadaires des s{\'e}ances de l'Acad{\'e}mie des sciences\
  (\bibinfo  {publisher} {Gauthier-Villars},\ \bibinfo {year}
  {1958})\BibitemShut {NoStop}%
\bibitem [{\citenamefont {Heller}\ \emph {et~al.}(2014)\citenamefont {Heller},
  \citenamefont {Janik}, \citenamefont {Spali\ifmmode~\acute{n}\else
  \'{n}\fi{}ski},\ and\ \citenamefont {Witaszczyk}}]{Heller2014}%
  \BibitemOpen
  \bibfield  {author} {\bibinfo {author} {\bibfnamefont {M.~P.}\ \bibnamefont
  {Heller}}, \bibinfo {author} {\bibfnamefont {R.~A.}\ \bibnamefont {Janik}},
  \bibinfo {author} {\bibfnamefont {M.}~\bibnamefont
  {Spali\ifmmode~\acute{n}\else \'{n}\fi{}ski}}, \ and\ \bibinfo {author}
  {\bibfnamefont {P.}~\bibnamefont {Witaszczyk}},\ }\href {\doibase
  10.1103/PhysRevLett.113.261601} {\bibfield  {journal} {\bibinfo  {journal}
  {Phys. Rev. Lett.}\ }\textbf {\bibinfo {volume} {113}},\ \bibinfo {pages}
  {261601} (\bibinfo {year} {2014})}\BibitemShut {NoStop}%
\bibitem [{\citenamefont {Kovtun}\ and\ \citenamefont
  {Starinets}(2005)}]{KovtunHolography2005}%
  \BibitemOpen
  \bibfield  {author} {\bibinfo {author} {\bibfnamefont {P.~K.}\ \bibnamefont
  {Kovtun}}\ and\ \bibinfo {author} {\bibfnamefont {A.~O.}\ \bibnamefont
  {Starinets}},\ }\href {\doibase 10.1103/PhysRevD.72.086009} {\bibfield
  {journal} {\bibinfo  {journal} {Phys. Rev. D}\ }\textbf {\bibinfo {volume}
  {72}},\ \bibinfo {pages} {086009} (\bibinfo {year} {2005})}\BibitemShut
  {NoStop}%
\bibitem [{\citenamefont {Onsager}(1931)}]{Onsager_1931}%
  \BibitemOpen
  \bibfield  {author} {\bibinfo {author} {\bibfnamefont {L.}~\bibnamefont
  {Onsager}},\ }\href {\doibase 10.1103/PhysRev.37.405} {\bibfield  {journal}
  {\bibinfo  {journal} {Phys. Rev.}\ }\textbf {\bibinfo {volume} {37}},\
  \bibinfo {pages} {405} (\bibinfo {year} {1931})}\BibitemShut {NoStop}%
\bibitem [{\citenamefont {Casimir}(1945)}]{Onsager_Casimir}%
  \BibitemOpen
  \bibfield  {author} {\bibinfo {author} {\bibfnamefont {H.~B.~G.}\
  \bibnamefont {Casimir}},\ }\href {\doibase 10.1103/RevModPhys.17.343}
  {\bibfield  {journal} {\bibinfo  {journal} {Rev. Mod. Phys.}\ }\textbf
  {\bibinfo {volume} {17}},\ \bibinfo {pages} {343} (\bibinfo {year}
  {1945})}\BibitemShut {NoStop}%
\bibitem [{\citenamefont
  {{Gavassino}}(2021{\natexlab{a}})}]{GavassinoGibbs2021}%
  \BibitemOpen
  \bibfield  {author} {\bibinfo {author} {\bibfnamefont {L.}~\bibnamefont
  {{Gavassino}}},\ }\href {\doibase 10.1088/1361-6382/ac2b0e} {\bibfield
  {journal} {\bibinfo  {journal} {Classical and Quantum Gravity}\ }\textbf
  {\bibinfo {volume} {38}},\ \bibinfo {eid} {21LT02} (\bibinfo {year}
  {2021}{\natexlab{a}})},\ \Eprint {http://arxiv.org/abs/2104.09142}
  {arXiv:2104.09142 [gr-qc]} \BibitemShut {NoStop}%
\bibitem [{\citenamefont {Geroch}\ and\ \citenamefont
  {Lindblom}(1991)}]{Geroch_Lindblom_1991_causal}%
  \BibitemOpen
  \bibfield  {author} {\bibinfo {author} {\bibfnamefont {R.}~\bibnamefont
  {Geroch}}\ and\ \bibinfo {author} {\bibfnamefont {L.}~\bibnamefont
  {Lindblom}},\ }\href {\doibase https://doi.org/10.1016/0003-4916(91)90063-E}
  {\bibfield  {journal} {\bibinfo  {journal} {Annals of Physics}\ }\textbf
  {\bibinfo {volume} {207}},\ \bibinfo {pages} {394 } (\bibinfo {year}
  {1991})}\BibitemShut {NoStop}%
\bibitem [{\citenamefont {Jou}\ \emph {et~al.}(1999)\citenamefont {Jou},
  \citenamefont {Casas-Vázquez},\ and\ \citenamefont {Lebon}}]{Jou_Extended}%
  \BibitemOpen
  \bibfield  {author} {\bibinfo {author} {\bibfnamefont {D.}~\bibnamefont
  {Jou}}, \bibinfo {author} {\bibfnamefont {J.}~\bibnamefont {Casas-Vázquez}},
  \ and\ \bibinfo {author} {\bibfnamefont {G.}~\bibnamefont {Lebon}},\ }\href
  {\doibase 10.1088/0034-4885/51/8/002} {\bibfield  {journal} {\bibinfo
  {journal} {Reports on Progress in Physics}\ }\textbf {\bibinfo {volume}
  {51}},\ \bibinfo {pages} {1105} (\bibinfo {year} {1999})}\BibitemShut
  {NoStop}%
\bibitem [{\citenamefont {{Grozdanov}}\ \emph {et~al.}(2019)\citenamefont
  {{Grozdanov}}, \citenamefont {{Lucas}},\ and\ \citenamefont
  {{Poovuttikul}}}]{Grozdanov2019}%
  \BibitemOpen
  \bibfield  {author} {\bibinfo {author} {\bibfnamefont {S.}~\bibnamefont
  {{Grozdanov}}}, \bibinfo {author} {\bibfnamefont {A.}~\bibnamefont
  {{Lucas}}}, \ and\ \bibinfo {author} {\bibfnamefont {N.}~\bibnamefont
  {{Poovuttikul}}},\ }\href {\doibase 10.1103/PhysRevD.99.086012} {\bibfield
  {journal} {\bibinfo  {journal} {\prd}\ }\textbf {\bibinfo {volume} {99}},\
  \bibinfo {eid} {086012} (\bibinfo {year} {2019})},\ \Eprint
  {http://arxiv.org/abs/1810.10016} {arXiv:1810.10016 [hep-th]} \BibitemShut
  {NoStop}%
\bibitem [{\citenamefont {{Camelio}}\ \emph
  {et~al.}(2022{\natexlab{a}})\citenamefont {{Camelio}}, \citenamefont
  {{Gavassino}}, \citenamefont {{Antonelli}}, \citenamefont {{Bernuzzi}},\ and\
  \citenamefont {{Haskell}}}]{camelio2022arXiv}%
  \BibitemOpen
  \bibfield  {author} {\bibinfo {author} {\bibfnamefont {G.}~\bibnamefont
  {{Camelio}}}, \bibinfo {author} {\bibfnamefont {L.}~\bibnamefont
  {{Gavassino}}}, \bibinfo {author} {\bibfnamefont {M.}~\bibnamefont
  {{Antonelli}}}, \bibinfo {author} {\bibfnamefont {S.}~\bibnamefont
  {{Bernuzzi}}}, \ and\ \bibinfo {author} {\bibfnamefont {B.}~\bibnamefont
  {{Haskell}}},\ }\href@noop {} {\bibfield  {journal} {\bibinfo  {journal}
  {arXiv e-prints}\ ,\ \bibinfo {eid} {arXiv:2204.11809}} (\bibinfo {year}
  {2022}{\natexlab{a}})},\ \Eprint {http://arxiv.org/abs/2204.11809}
  {arXiv:2204.11809 [gr-qc]} \BibitemShut {NoStop}%
\bibitem [{\citenamefont {Ernst}\ \emph {et~al.}(1970)\citenamefont {Ernst},
  \citenamefont {Hauge},\ and\ \citenamefont {van Leeuwen}}]{ErnstLetters1970}%
  \BibitemOpen
  \bibfield  {author} {\bibinfo {author} {\bibfnamefont {M.~H.}\ \bibnamefont
  {Ernst}}, \bibinfo {author} {\bibfnamefont {E.~H.}\ \bibnamefont {Hauge}}, \
  and\ \bibinfo {author} {\bibfnamefont {J.~M.~J.}\ \bibnamefont {van
  Leeuwen}},\ }\href {\doibase 10.1103/PhysRevLett.25.1254} {\bibfield
  {journal} {\bibinfo  {journal} {Phys. Rev. Lett.}\ }\textbf {\bibinfo
  {volume} {25}},\ \bibinfo {pages} {1254} (\bibinfo {year}
  {1970})}\BibitemShut {NoStop}%
\bibitem [{\citenamefont {{De Schepper}}\ \emph {et~al.}(1974)\citenamefont
  {{De Schepper}}, \citenamefont {{Van Beyeren}},\ and\ \citenamefont
  {Ernst}}]{Deschepper1974}%
  \BibitemOpen
  \bibfield  {author} {\bibinfo {author} {\bibfnamefont {I.}~\bibnamefont {{De
  Schepper}}}, \bibinfo {author} {\bibfnamefont {H.}~\bibnamefont {{Van
  Beyeren}}}, \ and\ \bibinfo {author} {\bibfnamefont {M.}~\bibnamefont
  {Ernst}},\ }\href {\doibase https://doi.org/10.1016/0031-8914(74)90290-0}
  {\bibfield  {journal} {\bibinfo  {journal} {Physica}\ }\textbf {\bibinfo
  {volume} {75}},\ \bibinfo {pages} {1} (\bibinfo {year} {1974})}\BibitemShut
  {NoStop}%
\bibitem [{\citenamefont {{Kovtun}}\ \emph {et~al.}(2011)\citenamefont
  {{Kovtun}}, \citenamefont {{Moore}},\ and\ \citenamefont
  {{Romatschke}}}]{KovtunStickiness2011}%
  \BibitemOpen
  \bibfield  {author} {\bibinfo {author} {\bibfnamefont {P.}~\bibnamefont
  {{Kovtun}}}, \bibinfo {author} {\bibfnamefont {G.~D.}\ \bibnamefont
  {{Moore}}}, \ and\ \bibinfo {author} {\bibfnamefont {P.}~\bibnamefont
  {{Romatschke}}},\ }\href {\doibase 10.1103/PhysRevD.84.025006} {\bibfield
  {journal} {\bibinfo  {journal} {\prd}\ }\textbf {\bibinfo {volume} {84}},\
  \bibinfo {eid} {025006} (\bibinfo {year} {2011})},\ \Eprint
  {http://arxiv.org/abs/1104.1586} {arXiv:1104.1586 [hep-ph]} \BibitemShut
  {NoStop}%
\bibitem [{\citenamefont {{Kost{\"a}dt}}\ and\ \citenamefont
  {{Liu}}(2000)}]{Kost2000}%
  \BibitemOpen
  \bibfield  {author} {\bibinfo {author} {\bibfnamefont {P.}~\bibnamefont
  {{Kost{\"a}dt}}}\ and\ \bibinfo {author} {\bibfnamefont {M.}~\bibnamefont
  {{Liu}}},\ }\href {\doibase 10.1103/PhysRevD.62.023003} {\bibfield  {journal}
  {\bibinfo  {journal} {\prd}\ }\textbf {\bibinfo {volume} {62}},\ \bibinfo
  {eid} {023003} (\bibinfo {year} {2000})},\ \Eprint
  {http://arxiv.org/abs/cond-mat/0010276} {arXiv:cond-mat/0010276
  [cond-mat.stat-mech]} \BibitemShut {NoStop}%
\bibitem [{\citenamefont {Gavassino}\ \emph
  {et~al.}(2020{\natexlab{a}})\citenamefont {Gavassino}, \citenamefont
  {Antonelli},\ and\ \citenamefont {Haskell}}]{GavassinoLyapunov_2020}%
  \BibitemOpen
  \bibfield  {author} {\bibinfo {author} {\bibfnamefont {L.}~\bibnamefont
  {Gavassino}}, \bibinfo {author} {\bibfnamefont {M.}~\bibnamefont
  {Antonelli}}, \ and\ \bibinfo {author} {\bibfnamefont {B.}~\bibnamefont
  {Haskell}},\ }\href {\doibase 10.1103/physrevd.102.043018} {\bibfield
  {journal} {\bibinfo  {journal} {Physical Review D}\ }\textbf {\bibinfo
  {volume} {102}} (\bibinfo {year} {2020}{\natexlab{a}}),\
  10.1103/physrevd.102.043018}\BibitemShut {NoStop}%
\bibitem [{\citenamefont {Landau}\ and\ \citenamefont
  {Lifshitz}(2013{\natexlab{a}})}]{landau5}%
  \BibitemOpen
  \bibfield  {author} {\bibinfo {author} {\bibfnamefont {L.}~\bibnamefont
  {Landau}}\ and\ \bibinfo {author} {\bibfnamefont {E.}~\bibnamefont
  {Lifshitz}},\ }\href {https://books.google.pl/books?id=VzgJN-XPTRsC} {\emph
  {\bibinfo {title} {Statistical Physics}}},\ \bibinfo {number} {v. 5}\
  (\bibinfo  {publisher} {Elsevier Science},\ \bibinfo {year}
  {2013})\BibitemShut {NoStop}%
\bibitem [{\citenamefont {{Dore}}\ \emph {et~al.}(2022)\citenamefont {{Dore}},
  \citenamefont {{Gavassino}}, \citenamefont {{Montenegro}}, \citenamefont
  {{Shokri}},\ and\ \citenamefont {{Torrieri}}}]{DoreTorrieri2022}%
  \BibitemOpen
  \bibfield  {author} {\bibinfo {author} {\bibfnamefont {T.}~\bibnamefont
  {{Dore}}}, \bibinfo {author} {\bibfnamefont {L.}~\bibnamefont {{Gavassino}}},
  \bibinfo {author} {\bibfnamefont {D.}~\bibnamefont {{Montenegro}}}, \bibinfo
  {author} {\bibfnamefont {M.}~\bibnamefont {{Shokri}}}, \ and\ \bibinfo
  {author} {\bibfnamefont {G.}~\bibnamefont {{Torrieri}}},\ }\href {\doibase
  10.1016/j.aop.2022.168902} {\bibfield  {journal} {\bibinfo  {journal} {Annals
  of Physics}\ }\textbf {\bibinfo {volume} {442}},\ \bibinfo {eid} {168902}
  (\bibinfo {year} {2022})},\ \Eprint {http://arxiv.org/abs/2109.06389}
  {arXiv:2109.06389 [hep-th]} \BibitemShut {NoStop}%
\bibitem [{\citenamefont {{Kovtun}}(2019)}]{Kovtun2019}%
  \BibitemOpen
  \bibfield  {author} {\bibinfo {author} {\bibfnamefont {P.}~\bibnamefont
  {{Kovtun}}},\ }\href {\doibase 10.1007/JHEP10(2019)034} {\bibfield  {journal}
  {\bibinfo  {journal} {Journal of High Energy Physics}\ }\textbf {\bibinfo
  {volume} {2019}},\ \bibinfo {eid} {34} (\bibinfo {year} {2019})},\ \Eprint
  {http://arxiv.org/abs/1907.08191} {arXiv:1907.08191 [hep-th]} \BibitemShut
  {NoStop}%
\bibitem [{\citenamefont {{Gavassino}}\ \emph {et~al.}(2022)\citenamefont
  {{Gavassino}}, \citenamefont {{Antonelli}},\ and\ \citenamefont
  {{Haskell}}}]{GavassinoCausality2021}%
  \BibitemOpen
  \bibfield  {author} {\bibinfo {author} {\bibfnamefont {L.}~\bibnamefont
  {{Gavassino}}}, \bibinfo {author} {\bibfnamefont {M.}~\bibnamefont
  {{Antonelli}}}, \ and\ \bibinfo {author} {\bibfnamefont {B.}~\bibnamefont
  {{Haskell}}},\ }\href {\doibase 10.1103/PhysRevLett.128.010606} {\bibfield
  {journal} {\bibinfo  {journal} {\prl}\ }\textbf {\bibinfo {volume} {128}},\
  \bibinfo {eid} {010606} (\bibinfo {year} {2022})},\ \Eprint
  {http://arxiv.org/abs/2105.14621} {arXiv:2105.14621 [gr-qc]} \BibitemShut
  {NoStop}%
\bibitem [{\citenamefont {{Gavassino}}(2022)}]{GavassinoStabilityCarter2022}%
  \BibitemOpen
  \bibfield  {author} {\bibinfo {author} {\bibfnamefont {L.}~\bibnamefont
  {{Gavassino}}},\ }\href@noop {} {\bibfield  {journal} {\bibinfo  {journal}
  {arXiv e-prints}\ ,\ \bibinfo {eid} {arXiv:2202.06760}} (\bibinfo {year}
  {2022})},\ \Eprint {http://arxiv.org/abs/2202.06760} {arXiv:2202.06760
  [gr-qc]} \BibitemShut {NoStop}%
\bibitem [{\citenamefont {{Misner}}\ \emph {et~al.}(1973)\citenamefont
  {{Misner}}, \citenamefont {{Thorne}},\ and\ \citenamefont
  {{Wheeler}}}]{MTW_book}%
  \BibitemOpen
  \bibfield  {author} {\bibinfo {author} {\bibfnamefont {C.~W.}\ \bibnamefont
  {{Misner}}}, \bibinfo {author} {\bibfnamefont {K.~S.}\ \bibnamefont
  {{Thorne}}}, \ and\ \bibinfo {author} {\bibfnamefont {J.~A.}\ \bibnamefont
  {{Wheeler}}},\ }\href@noop {} {\emph {\bibinfo {title} {{Gravitation}}}}\
  (\bibinfo  {publisher} {San Francisco: W.H.~Freeman and Co.},\ \bibinfo
  {year} {1973})\BibitemShut {NoStop}%
\bibitem [{\citenamefont {{Gavassino}}\ and\ \citenamefont
  {{Antonelli}}(2020)}]{Termo}%
  \BibitemOpen
  \bibfield  {author} {\bibinfo {author} {\bibfnamefont {L.}~\bibnamefont
  {{Gavassino}}}\ and\ \bibinfo {author} {\bibfnamefont {M.}~\bibnamefont
  {{Antonelli}}},\ }\href {\doibase 10.1088/1361-6382/ab5f23} {\bibfield
  {journal} {\bibinfo  {journal} {Classical and Quantum Gravity}\ }\textbf
  {\bibinfo {volume} {37}},\ \bibinfo {eid} {025014} (\bibinfo {year}
  {2020})},\ \Eprint {http://arxiv.org/abs/1906.03140} {arXiv:1906.03140
  [gr-qc]} \BibitemShut {NoStop}%
\bibitem [{\citenamefont {{Gavassino}}(2020)}]{GavassinoTermometri2020}%
  \BibitemOpen
  \bibfield  {author} {\bibinfo {author} {\bibfnamefont {L.}~\bibnamefont
  {{Gavassino}}},\ }\href {\doibase 10.1007/s10701-020-00393-x} {\bibfield
  {journal} {\bibinfo  {journal} {Foundations of Physics}\ } (\bibinfo {year}
  {2020}),\ 10.1007/s10701-020-00393-x},\ \Eprint
  {http://arxiv.org/abs/2005.06396} {arXiv:2005.06396 [gr-qc]} \BibitemShut
  {NoStop}%
\bibitem [{\citenamefont {Stueckelberg}(1962)}]{Stuekelberg1962}%
  \BibitemOpen
  \bibfield  {author} {\bibinfo {author} {\bibfnamefont {E.}~\bibnamefont
  {Stueckelberg}},\ }\href@noop {} {\bibfield  {journal} {\bibinfo  {journal}
  {Helvetica Physica Acta}\ }\textbf {\bibinfo {volume} {35}} (\bibinfo {year}
  {1962})}\BibitemShut {NoStop}%
\bibitem [{\citenamefont {Israel}(2009)}]{Israel_2009_inbook}%
  \BibitemOpen
  \bibfield  {author} {\bibinfo {author} {\bibfnamefont {W.}~\bibnamefont
  {Israel}},\ }\enquote {\bibinfo {title} {Relativistic thermodynamics},}\ in\
  \href {\doibase 10.1007/978-3-7643-8878-2_8} {\emph {\bibinfo {booktitle}
  {E.C.G. Stueckelberg, An Unconventional Figure of Twentieth Century Physics:
  Selected Scientific Papers with Commentaries}}},\ \bibinfo {editor} {edited
  by\ \bibinfo {editor} {\bibfnamefont {J.}~\bibnamefont {Lacki}}, \bibinfo
  {editor} {\bibfnamefont {H.}~\bibnamefont {Ruegg}}, \ and\ \bibinfo {editor}
  {\bibfnamefont {G.}~\bibnamefont {Wanders}}}\ (\bibinfo  {publisher}
  {Birkh{\"a}user Basel},\ \bibinfo {address} {Basel},\ \bibinfo {year}
  {2009})\ pp.\ \bibinfo {pages} {101--113}\BibitemShut {NoStop}%
\bibitem [{\citenamefont {Grmela}\ and\ \citenamefont
  {\"Ottinger}(1997)}]{Ottinger1997}%
  \BibitemOpen
  \bibfield  {author} {\bibinfo {author} {\bibfnamefont {M.}~\bibnamefont
  {Grmela}}\ and\ \bibinfo {author} {\bibfnamefont {H.~C.}\ \bibnamefont
  {\"Ottinger}},\ }\href {\doibase 10.1103/PhysRevE.56.6620} {\bibfield
  {journal} {\bibinfo  {journal} {Phys. Rev. E}\ }\textbf {\bibinfo {volume}
  {56}},\ \bibinfo {pages} {6620} (\bibinfo {year} {1997})}\BibitemShut
  {NoStop}%
\bibitem [{\citenamefont {Gibbons}\ and\ \citenamefont
  {Hawking}(1977)}]{GibbonsHawking1977}%
  \BibitemOpen
  \bibfield  {author} {\bibinfo {author} {\bibfnamefont {G.~W.}\ \bibnamefont
  {Gibbons}}\ and\ \bibinfo {author} {\bibfnamefont {S.~W.}\ \bibnamefont
  {Hawking}},\ }\href {\doibase 10.1103/PhysRevD.15.2752} {\bibfield  {journal}
  {\bibinfo  {journal} {Phys. Rev. D}\ }\textbf {\bibinfo {volume} {15}},\
  \bibinfo {pages} {2752} (\bibinfo {year} {1977})}\BibitemShut {NoStop}%
\bibitem [{\citenamefont {{Becattini}}(2016)}]{BecattiniBeta2016}%
  \BibitemOpen
  \bibfield  {author} {\bibinfo {author} {\bibfnamefont {F.}~\bibnamefont
  {{Becattini}}},\ }\href {\doibase 10.5506/APhysPolB.47.1819} {\bibfield
  {journal} {\bibinfo  {journal} {Acta Physica Polonica B}\ }\textbf {\bibinfo
  {volume} {47}},\ \bibinfo {pages} {1819} (\bibinfo {year} {2016})},\ \Eprint
  {http://arxiv.org/abs/1606.06605} {arXiv:1606.06605 [gr-qc]} \BibitemShut
  {NoStop}%
\bibitem [{\citenamefont {Galley}(2013)}]{Galley2013}%
  \BibitemOpen
  \bibfield  {author} {\bibinfo {author} {\bibfnamefont {C.~R.}\ \bibnamefont
  {Galley}},\ }\href {\doibase 10.1103/PhysRevLett.110.174301} {\bibfield
  {journal} {\bibinfo  {journal} {Phys. Rev. Lett.}\ }\textbf {\bibinfo
  {volume} {110}},\ \bibinfo {pages} {174301} (\bibinfo {year}
  {2013})}\BibitemShut {NoStop}%
\bibitem [{\citenamefont {{Galley}}\ \emph {et~al.}(2014)\citenamefont
  {{Galley}}, \citenamefont {{Tsang}},\ and\ \citenamefont
  {{Stein}}}]{Galley2014}%
  \BibitemOpen
  \bibfield  {author} {\bibinfo {author} {\bibfnamefont {C.~R.}\ \bibnamefont
  {{Galley}}}, \bibinfo {author} {\bibfnamefont {D.}~\bibnamefont {{Tsang}}}, \
  and\ \bibinfo {author} {\bibfnamefont {L.~C.}\ \bibnamefont {{Stein}}},\
  }\href@noop {} {\bibfield  {journal} {\bibinfo  {journal} {arXiv e-prints}\
  ,\ \bibinfo {eid} {arXiv:1412.3082}} (\bibinfo {year} {2014})},\ \Eprint
  {http://arxiv.org/abs/1412.3082} {arXiv:1412.3082 [math-ph]} \BibitemShut
  {NoStop}%
\bibitem [{\citenamefont {{Andersson}}\ and\ \citenamefont
  {{Comer}}(2007)}]{andersson2007review}%
  \BibitemOpen
  \bibfield  {author} {\bibinfo {author} {\bibfnamefont {N.}~\bibnamefont
  {{Andersson}}}\ and\ \bibinfo {author} {\bibfnamefont {G.~L.}\ \bibnamefont
  {{Comer}}},\ }\href {\doibase 10.12942/lrr-2007-1} {\bibfield  {journal}
  {\bibinfo  {journal} {Living Reviews in Relativity}\ }\textbf {\bibinfo
  {volume} {10}},\ \bibinfo {eid} {1} (\bibinfo {year} {2007})},\ \Eprint
  {http://arxiv.org/abs/gr-qc/0605010} {gr-qc/0605010} \BibitemShut {NoStop}%
\bibitem [{\citenamefont {{Endlich}}\ \emph {et~al.}(2011)\citenamefont
  {{Endlich}}, \citenamefont {{Nicolis}}, \citenamefont {{Rattazzi}},\ and\
  \citenamefont {{Wang}}}]{Endlich2011}%
  \BibitemOpen
  \bibfield  {author} {\bibinfo {author} {\bibfnamefont {S.}~\bibnamefont
  {{Endlich}}}, \bibinfo {author} {\bibfnamefont {A.}~\bibnamefont
  {{Nicolis}}}, \bibinfo {author} {\bibfnamefont {R.}~\bibnamefont
  {{Rattazzi}}}, \ and\ \bibinfo {author} {\bibfnamefont {J.}~\bibnamefont
  {{Wang}}},\ }\href {\doibase 10.1007/JHEP04(2011)102} {\bibfield  {journal}
  {\bibinfo  {journal} {Journal of High Energy Physics}\ }\textbf {\bibinfo
  {volume} {2011}},\ \bibinfo {eid} {102} (\bibinfo {year} {2011})},\ \Eprint
  {http://arxiv.org/abs/1011.6396} {arXiv:1011.6396 [hep-th]} \BibitemShut
  {NoStop}%
\bibitem [{\citenamefont {{Montenegro}}\ and\ \citenamefont
  {{Torrieri}}(2016)}]{TorrieriIS2016}%
  \BibitemOpen
  \bibfield  {author} {\bibinfo {author} {\bibfnamefont {D.}~\bibnamefont
  {{Montenegro}}}\ and\ \bibinfo {author} {\bibfnamefont {G.}~\bibnamefont
  {{Torrieri}}},\ }\href {\doibase 10.1103/PhysRevD.94.065042} {\bibfield
  {journal} {\bibinfo  {journal} {\prd}\ }\textbf {\bibinfo {volume} {94}},\
  \bibinfo {eid} {065042} (\bibinfo {year} {2016})},\ \Eprint
  {http://arxiv.org/abs/1604.05291} {arXiv:1604.05291 [hep-th]} \BibitemShut
  {NoStop}%
\bibitem [{\citenamefont {Gavassino}\ \emph
  {et~al.}(2020{\natexlab{b}})\citenamefont {Gavassino}, \citenamefont
  {Antonelli},\ and\ \citenamefont {Haskell}}]{GavassinoRadiazione}%
  \BibitemOpen
  \bibfield  {author} {\bibinfo {author} {\bibfnamefont {L.}~\bibnamefont
  {Gavassino}}, \bibinfo {author} {\bibfnamefont {M.}~\bibnamefont
  {Antonelli}}, \ and\ \bibinfo {author} {\bibfnamefont {B.}~\bibnamefont
  {Haskell}},\ }\href {\doibase 10.3390/sym12091543} {\bibfield  {journal}
  {\bibinfo  {journal} {Symmetry}\ }\textbf {\bibinfo {volume} {12}},\ \bibinfo
  {pages} {1543} (\bibinfo {year} {2020}{\natexlab{b}})}\BibitemShut {NoStop}%
\bibitem [{\citenamefont {{Liu}}\ \emph {et~al.}(1986)\citenamefont {{Liu}},
  \citenamefont {{M{\"u}ller}},\ and\ \citenamefont {{Ruggeri}}}]{Liu1986}%
  \BibitemOpen
  \bibfield  {author} {\bibinfo {author} {\bibfnamefont {I.~S.}\ \bibnamefont
  {{Liu}}}, \bibinfo {author} {\bibfnamefont {I.}~\bibnamefont {{M{\"u}ller}}},
  \ and\ \bibinfo {author} {\bibfnamefont {T.}~\bibnamefont {{Ruggeri}}},\
  }\href {\doibase 10.1016/0003-4916(86)90164-8} {\bibfield  {journal}
  {\bibinfo  {journal} {Annals of Physics}\ }\textbf {\bibinfo {volume}
  {169}},\ \bibinfo {pages} {191} (\bibinfo {year} {1986})}\BibitemShut
  {NoStop}%
\bibitem [{\citenamefont
  {{Gavassino}}(2021{\natexlab{b}})}]{GavassinoSuperlum2021}%
  \BibitemOpen
  \bibfield  {author} {\bibinfo {author} {\bibfnamefont {L.}~\bibnamefont
  {{Gavassino}}},\ }\href@noop {} {\bibfield  {journal} {\bibinfo  {journal}
  {arXiv e-prints}\ ,\ \bibinfo {eid} {arXiv:2111.05254}} (\bibinfo {year}
  {2021}{\natexlab{b}})},\ \Eprint {http://arxiv.org/abs/2111.05254}
  {arXiv:2111.05254 [gr-qc]} \BibitemShut {NoStop}%
\bibitem [{\citenamefont {{Olson}}(1990)}]{OlsonLifsh1990}%
  \BibitemOpen
  \bibfield  {author} {\bibinfo {author} {\bibfnamefont {T.~S.}\ \bibnamefont
  {{Olson}}},\ }\href {\doibase 10.1016/0003-4916(90)90366-V} {\bibfield
  {journal} {\bibinfo  {journal} {Annals of Physics}\ }\textbf {\bibinfo
  {volume} {199}},\ \bibinfo {pages} {18} (\bibinfo {year} {1990})}\BibitemShut
  {NoStop}%
\bibitem [{\citenamefont {{Salazar}}\ and\ \citenamefont
  {{Zannias}}(2020)}]{Salazar2020}%
  \BibitemOpen
  \bibfield  {author} {\bibinfo {author} {\bibfnamefont {J.~F.}\ \bibnamefont
  {{Salazar}}}\ and\ \bibinfo {author} {\bibfnamefont {T.}~\bibnamefont
  {{Zannias}}},\ }\href {\doibase 10.1142/S0218271820300104} {\bibfield
  {journal} {\bibinfo  {journal} {International Journal of Modern Physics D}\
  }\textbf {\bibinfo {volume} {29}},\ \bibinfo {eid} {2030010} (\bibinfo {year}
  {2020})},\ \Eprint {http://arxiv.org/abs/1904.04368} {arXiv:1904.04368
  [gr-qc]} \BibitemShut {NoStop}%
\bibitem [{\citenamefont {{Noronha}}\ \emph {et~al.}(2021)\citenamefont
  {{Noronha}}, \citenamefont {{Spali{\'n}ski}},\ and\ \citenamefont
  {{Speranza}}}]{NoronhaGeneralFrame2021}%
  \BibitemOpen
  \bibfield  {author} {\bibinfo {author} {\bibfnamefont {J.}~\bibnamefont
  {{Noronha}}}, \bibinfo {author} {\bibfnamefont {M.}~\bibnamefont
  {{Spali{\'n}ski}}}, \ and\ \bibinfo {author} {\bibfnamefont {E.}~\bibnamefont
  {{Speranza}}},\ }\href@noop {} {\bibfield  {journal} {\bibinfo  {journal}
  {arXiv e-prints}\ ,\ \bibinfo {eid} {arXiv:2105.01034}} (\bibinfo {year}
  {2021})},\ \Eprint {http://arxiv.org/abs/2105.01034} {arXiv:2105.01034
  [nucl-th]} \BibitemShut {NoStop}%
\bibitem [{\citenamefont {Landau}\ and\ \citenamefont
  {Lifshitz}(2013{\natexlab{b}})}]{landau6}%
  \BibitemOpen
  \bibfield  {author} {\bibinfo {author} {\bibfnamefont {L.}~\bibnamefont
  {Landau}}\ and\ \bibinfo {author} {\bibfnamefont {E.}~\bibnamefont
  {Lifshitz}},\ }\href@noop {} {\emph {\bibinfo {title} {Fluid Mechanics}}},\
  \bibinfo {number} {v. 6}\ (\bibinfo  {publisher} {Elsevier Science},\
  \bibinfo {year} {2013})\BibitemShut {NoStop}%
\bibitem [{\citenamefont {Gavassino}\ \emph {et~al.}(2022)\citenamefont
  {Gavassino}, \citenamefont {Antonelli},\ and\ \citenamefont
  {Haskell}}]{GavassinoKhalatnikov2022}%
  \BibitemOpen
  \bibfield  {author} {\bibinfo {author} {\bibfnamefont {L.}~\bibnamefont
  {Gavassino}}, \bibinfo {author} {\bibfnamefont {M.}~\bibnamefont
  {Antonelli}}, \ and\ \bibinfo {author} {\bibfnamefont {B.}~\bibnamefont
  {Haskell}},\ }\href {\doibase 10.1103/PhysRevD.105.045011} {\bibfield
  {journal} {\bibinfo  {journal} {Phys. Rev. D}\ }\textbf {\bibinfo {volume}
  {105}},\ \bibinfo {pages} {045011} (\bibinfo {year} {2022})}\BibitemShut
  {NoStop}%
\bibitem [{\citenamefont {{Camelio}}\ \emph
  {et~al.}(2022{\natexlab{b}})\citenamefont {{Camelio}}, \citenamefont
  {{Gavassino}}, \citenamefont {{Antonelli}}, \citenamefont {{Bernuzzi}},\ and\
  \citenamefont {{Haskell}}}]{Camelio2022IIarXiv}%
  \BibitemOpen
  \bibfield  {author} {\bibinfo {author} {\bibfnamefont {G.}~\bibnamefont
  {{Camelio}}}, \bibinfo {author} {\bibfnamefont {L.}~\bibnamefont
  {{Gavassino}}}, \bibinfo {author} {\bibfnamefont {M.}~\bibnamefont
  {{Antonelli}}}, \bibinfo {author} {\bibfnamefont {S.}~\bibnamefont
  {{Bernuzzi}}}, \ and\ \bibinfo {author} {\bibfnamefont {B.}~\bibnamefont
  {{Haskell}}},\ }\href@noop {} {\bibfield  {journal} {\bibinfo  {journal}
  {arXiv e-prints}\ ,\ \bibinfo {eid} {arXiv:2204.11810}} (\bibinfo {year}
  {2022}{\natexlab{b}})},\ \Eprint {http://arxiv.org/abs/2204.11810}
  {arXiv:2204.11810 [gr-qc]} \BibitemShut {NoStop}%
\bibitem [{\citenamefont {{Krotscheck}}\ and\ \citenamefont
  {{Kundt}}(1978)}]{Krotscheck1978}%
  \BibitemOpen
  \bibfield  {author} {\bibinfo {author} {\bibfnamefont {E.}~\bibnamefont
  {{Krotscheck}}}\ and\ \bibinfo {author} {\bibfnamefont {W.}~\bibnamefont
  {{Kundt}}},\ }\href {\doibase 10.1007/BF01609447} {\bibfield  {journal}
  {\bibinfo  {journal} {Communications in Mathematical Physics}\ }\textbf
  {\bibinfo {volume} {60}},\ \bibinfo {pages} {171} (\bibinfo {year}
  {1978})}\BibitemShut {NoStop}%
\bibitem [{\citenamefont {{Heimburg}}(2017)}]{Heimburg2017}%
  \BibitemOpen
  \bibfield  {author} {\bibinfo {author} {\bibfnamefont {T.}~\bibnamefont
  {{Heimburg}}},\ }\href {\doibase 10.1039/C7CP02189E} {\bibfield  {journal}
  {\bibinfo  {journal} {Physical Chemistry Chemical Physics (Incorporating
  Faraday Transactions)}\ }\textbf {\bibinfo {volume} {19}},\ \bibinfo {pages}
  {17331} (\bibinfo {year} {2017})},\ \Eprint {http://arxiv.org/abs/1608.06093}
  {arXiv:1608.06093 [physics.chem-ph]} \BibitemShut {NoStop}%
\bibitem [{\citenamefont {{Moore}}(2018)}]{moore2018cuts}%
  \BibitemOpen
  \bibfield  {author} {\bibinfo {author} {\bibfnamefont {G.~D.}\ \bibnamefont
  {{Moore}}},\ }\href {\doibase 10.1007/JHEP05(2018)084} {\bibfield  {journal}
  {\bibinfo  {journal} {Journal of High Energy Physics}\ }\textbf {\bibinfo
  {volume} {2018}},\ \bibinfo {eid} {84} (\bibinfo {year} {2018})},\ \Eprint
  {http://arxiv.org/abs/1803.00736} {arXiv:1803.00736 [hep-ph]} \BibitemShut
  {NoStop}%
\bibitem [{\citenamefont {{Romatschke}}(2019)}]{Romatschke2019PhRvL}%
  \BibitemOpen
  \bibfield  {author} {\bibinfo {author} {\bibfnamefont {P.}~\bibnamefont
  {{Romatschke}}},\ }\href {\doibase 10.1103/PhysRevLett.122.231603} {\bibfield
   {journal} {\bibinfo  {journal} {\prl}\ }\textbf {\bibinfo {volume} {122}},\
  \bibinfo {eid} {231603} (\bibinfo {year} {2019})}\BibitemShut {NoStop}%
\bibitem [{\citenamefont {Perna}\ and\ \citenamefont
  {Calzetta}(2021)}]{Perna_21_linear_cuts}%
  \BibitemOpen
  \bibfield  {author} {\bibinfo {author} {\bibfnamefont {G.}~\bibnamefont
  {Perna}}\ and\ \bibinfo {author} {\bibfnamefont {E.}~\bibnamefont
  {Calzetta}},\ }\href {\doibase 10.1103/PhysRevD.104.096005} {\bibfield
  {journal} {\bibinfo  {journal} {Phys. Rev. D}\ }\textbf {\bibinfo {volume}
  {104}},\ \bibinfo {pages} {096005} (\bibinfo {year} {2021})}\BibitemShut
  {NoStop}%
\bibitem [{\citenamefont {Hiscock}\ and\ \citenamefont
  {Lindblom}(1988)}]{Hishcock1988}%
  \BibitemOpen
  \bibfield  {author} {\bibinfo {author} {\bibfnamefont {W.~A.}\ \bibnamefont
  {Hiscock}}\ and\ \bibinfo {author} {\bibfnamefont {L.}~\bibnamefont
  {Lindblom}},\ }\href {\doibase https://doi.org/10.1016/0375-9601(88)90679-2}
  {\bibfield  {journal} {\bibinfo  {journal} {Physics Letters A}\ }\textbf
  {\bibinfo {volume} {131}},\ \bibinfo {pages} {509 } (\bibinfo {year}
  {1988})}\BibitemShut {NoStop}%
\end{thebibliography}%

\label{lastpage}

\end{document}